\documentclass[a4paper,11pt]{article}
\pdfoutput=1 

\usepackage{jheppub} 
\usepackage[T1]{fontenc} 
\usepackage{slashed}
\usepackage{arydshln}
\usepackage{caption}
\usepackage{subcaption}
\usepackage{nicematrix}
\usepackage{tikz}
\usepackage{tkz-euclide}
\usepackage[compat=1.1.0]{tikz-feynman}
\usepackage{booktabs}
\usepackage{lipsum}
\usepackage{setspace}
\usepackage{subfiles}
\usepackage{hhline}
\usepackage{stackengine}
\usepackage{multirow}
\usepackage{graphicx} 
\usepackage{array}

\title{\boldmath QCD sum rule analysis of $0^{+}$ four-quark states}

\author[a,1]{Shuang-Hong Li,\note{Corresponding author.}}
\author[b]{Ze-Sheng Chen,}
\author[a]{Yi-Xin Chen,}
\author[a,1]{Hong-Ying Jin}

\affiliation[a]{Zhejiang Institute of Modern Physics, School of Physics, Zhejiang University,\\Hangzhou, 310027, China}
\affiliation[b]{College of Digital Technology and Engineering, Ningbo University of Finance and Economics,\\Ningbo, 315327, China}

\emailAdd{leesh@zju.edu.cn}
\emailAdd{jinhongying@zju.edu.cn}
\emailAdd{yixinchenzimp@zju.edu.cn}
\emailAdd{ventuschen@zju.edu.cn}

\abstract{We present a comprehensive QCD sum rules analysis at next-to-leading order for all types of $J^P=0^{+}$ four-quark states composed of $u$, $d$, and $s$ quarks. The eigenvectors of the renormalization matrix are chosen to be the renormalized four-quark operators, which can be equally interpreted as tetraquark or molecule operators. Meanwhile, the typical nonet masses given by bare tetraquark operators are lower than those given by bare molecule operators. Most of the nonet masses are around $1-2\text{GeV}$, and they can be interpreted as the $0^+$ mesons observed in experiments. We find a category of four-quark nonets with masses $\lesssim1\text{GeV}$, potentially corresponding to the light $0^+$ mesons $f_0(500)$, $K^*_0(700)$, $f_0(980)$, and $a_0(980)$. On the other hand, the possible 27-fold states are heavier than most of the nonets, with masses $\gtrsim 2\text{GeV}$. The main uncertainty arises from the factorization of high-dimensional condensates, which usually underestimates their values. To address this, we introduce deviation factors for the dimension-6, -8, and -10 condensates, and vary them over a wide range to obtain conservative estimates of the $0^+$ four-quark state masses. Some general properties of the $0^+$ light four-quark states can be derived that do not rely on precise numerical values. We also find that the ambiguity in the factorization of the dimension-8 condensate can introduce a larger discrepancy than previously estimated. As a byproduct, we propose a simple trick for renormalizing multi-quark operators at the one-loop level.}

\begin{document} 
\maketitle
\flushbottom

\section{Introduction}
\label{sec:intro}

Scalar mesons with $J^{P}=0^{+}$ are elusive. They can be $q\bar{q}$ mesons, glueballs, hybrids, or multi-quark states. Many $0^{+}$ mesons have been observed in experiments \cite{pdg}, and there are many studies focused on them, but the nonperturbative nature of QCD hinders the theoretical research.

The structures of light $0^+$ mesons with masses $\lesssim1\text{GeV}$ are particularly confusing \cite{f0(500)}. The mass hierarchy $M_\sigma<M_{K^*}<M_{f_0},M_{a_0}$ provides evidence for their tetraquark nature. Many $0^+$ states around $1-2\text{GeV}$ have been observed; some of them are also interpreted as four-quark states. However, $0^+$ glueball and $q\bar{q}$ exist in this range as well. For details, see the reviews ``\textit{Scalar Mesons below 1 GeV}'' and ``\textit{Scalar Mesons below 2 GeV}'' in PDG~\cite{pdg}.

To test the four-quark interpretation of $0^+$ mesons, the QCD sum rules~\cite{qsr} provide a convenient method. Some studies~\cite{prd_2007_tetraquark, prd_2007_tetraquark_mixing, prd_2008_tetraquark, prd_2019_tetraquark} based on QCD sum rules offer different interpretations of the light $0^+$ mesons, owing to the numerous possible $0^+$ four-quark operator structures. We thus enumerate all possible $0^+$ four-quark operators constructed by light quarks to investigate their masses, and find that most of them have masses around $1-2\text{GeV}$, as shown in section~\ref{sec_3_mass}.

Nevertheless, as discussed in Section \ref{sec_2}, the correspondence between four-quark operators and four-quark states is vague, making it difficult to give exact interpretations for the corresponding states. Furthermore, the compact tetraquarks and four-quark molecules may be mixed. In the case of nonets, the renormalized currents involve hybrid-like currents, which can be written as four-quark currents using the equation of motion. However, these two types of current are not exactly equivalent, as discussed in Appendix \ref{A_ward}.

The calculation shows that most four-quark correlators are dominated by the contributions of $\langle\bar{q}q\rangle^2$ and $\langle\bar{q}q\rangle\langle\bar{q}Gq\rangle$ condensates, except in certain cases where these contributions vanish. The exact values of these high-dimensional condensates are not known. The factorization (vacuum saturation) hypothesis typically underestimates their value by factors around $2-5$ \cite{deviation_factors,deviation_factor_d6}. To assess how these underestimations affect mass estimations, we introduce factorization deviation factors $\eta_n$ and assign them different values; precise estimations are therefore unattainable. Our goal is to determine rough mass ranges and draw conclusions that do not rely on precise masses. We also find that the ambiguity in the factorization of dimension-8 condensate leads to a discrepancy larger than previously estimated~\cite{va_sum}, and this discrepancy depends on specific configurations, as shown in Appendix \ref{fac_d8}. The mass estimations of the $0^+$ four-quark states will be discussed in Section \ref{sec_3}.

To complete the extensive calculations, we developed a Mathematica package~\cite{pack} to evaluate the diagrams shown in Appendix~\ref{dias}. Since the light quarks are treated as massless, each diagram is straightforward to calculate using Fourier transformation, except for the next-to-leading-order (NLO) perturbative diagrams that involve a two-loop propagator-type subdiagram. The TARCER~\cite{tarcer,feyncalc_10} package is used to reduce these two-loop diagrams. To avoid redundant calculations for the same diagram with different flavor configurations, we left the flavors unspecified during the calculations. This approach results in $\delta_{f_i f_j}$ terms in the outputs, as listed in Appendix~\ref{ope_results}, where $f_i$ is a flavor label. By specifying flavor labels corresponding to specific states, the required operator product expansion (OPE) results can be obtained. The Laplace sum rules (LSR)\cite{qsr} are then used to estimate the masses based on these correlators. We present all the figures related to the mass estimation in Ref.\cite{results}, which serves as a supplementary material. The necessary figures included in this paper are listed in Appendix~\ref{sec_mass}.

\section{The Construction of $0^{+}$ Four-Quark Currents}\label{sec_2}
\subsection{Renormalization of Four-Quark Currents} 
A general tetraquark current can be written as
\begin{equation}
	\Psi^{a\,T}_{f_1}\mathcal{C} \Gamma_A \Psi^b_{f_2}\, \overline{\Psi}^b_{f_3} \Gamma_B \mathcal{C}\overline{\Psi}^{a\,T}_{f_4}\quad\text{or}\quad\Psi^{a\,T}_{f_1}\mathcal{C} \Gamma_A \Psi^b_{f_2}\, \overline{\Psi}^a_{f_3} \Gamma_B \mathcal{C}\overline{\Psi}^{b\,T}_{f_4},
	\label{tetra_o}
\end{equation} 
where $\mathcal{C}=i\gamma^0\gamma^2$, $f_i$ are the flavor labels, $\Gamma_A$ and $\Gamma_B$ denote arbitrary $\gamma$-matrices. The transpose acts only on spin indices. By Fierz rearrangement, the tetraquark current can also be expressed as a combination of
\begin{equation}
	\overline{\Psi}_{f_1}\Gamma_A \Psi_{f_2}\, \overline{\Psi}_{f_3}\Gamma_B \Psi_{f_4}\quad\text{or}\quad\overline{\Psi}_{f_1}T^a\Gamma_A \Psi_{f_2}\, \overline{\Psi}_{f_3}T^a\Gamma_B \Psi_{f_4}.
	\label{mole_o}
\end{equation}
In the case where the $SU(n)$ generator $T^a$ is not involved, the meson-meson configuration suggests that eq.~\ref{mole_o} may prefer to couple to molecular states. Conversely, the diquark-antidiquark configuration of eq.~\ref{tetra_o} implies it may prefer to couple to compact tetraquark states. We refer to the states corresponding to currents \ref{tetra_o} and \ref{mole_o} as tetraquark and molecule respectively. If no confusion arises, we call both of them four-quark states.

\begin{figure}[t!]
	\centering
	\begin{subfigure}{\textwidth}
		\centering
		\includegraphics{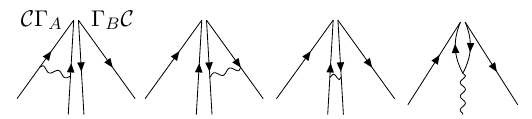}
		\caption{One-loop renormalization of tetraquark operators; diagrams that differ by permuting the legs are omitted.}
		\label{fig_r_a}
	\end{subfigure}
	\begin{subfigure}{\textwidth}
		\centering
		\includegraphics{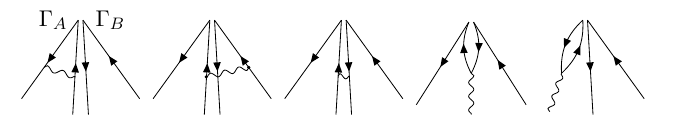}
		\caption{One-loop renormalization of molecule operators; diagrams that differ by exchanging $\Gamma_A\leftrightarrow\Gamma_B$ are omitted.}
		\label{fig_r_b}
	\end{subfigure}
	\caption{Renormalization of four-quark operators at one-loop level. The four-quark vertices have been slightly split to clarify how the quarks are connected; the vertices' structures are shown in the first diagrams of (a) and (b).}
	\label{fig_r}
\end{figure}

The renormalization~\cite{book_r} of four-quark operators at one-loop level involves two Green functions
\begin{subequations}
	\begin{equation}
		\langle T\,\,\mathcal{O}(q)\ \Psi^a_{f_1}(q_1)\Psi^b_{f_2}(q_2)\overline{\Psi}^c_{f_3}(q_3)\overline{\Psi}^d_{f_4}(q_4)\rangle
		\label{green_4q}
	\end{equation}
	and
	\begin{equation}
		\langle T\,\,\mathcal{O}(q)\ \Psi^a_{f_1}(q_1)\overline{\Psi}^b_{f_2}(q_2)A^n_\mu(q_3)\rangle,
		\label{green_qgq}
	\end{equation}
\end{subequations}
where $\mathcal{O}(q)$ is the four-quark operator; the operators and fields here are in momentum space, and it is enough to consider the case with $q=\sum_i q_i=0$. The diagrams are shown in figure~\ref{fig_r}.

To cancel the $\varepsilon$-pole, two types of counterterms are involved. Both can be obtained by a simple trick. The first type involves four-quark operators. In the Feynman gauge, it can be obtained by replacing the (anti)fermion fields in eqs.~\ref{tetra_o} and \ref{mole_o} according to the rule

\begin{equation}
	\Psi\rightarrow \Big(\frac{ig}{8\pi\sqrt{\varepsilon}}\,T^n\gamma^\nu.\gamma^\mu+1\Big)\Psi.
	\label{rep_1}
\end{equation}
Then extract the terms $\propto g^2$, and sum over the indices only for these terms. For the tetraquark operator, to retain the form as eq.~\ref{tetra_o}, the identity $T^{n\,ab}T^{n\,cd}=\frac{1}{2}\delta^{ad}\delta^{bc}-\frac{1}{2C_A}\delta^{ab}\delta^{cd}$ is needed.

The second type refers to hybrid-like operators, obtained by choosing a fermion field and an anti-fermion field in eqs.~\ref{tetra_o} and \ref{mole_o}. If they are not adjacent, commute them, and then proceed by replacing

\begin{equation}
		(\Psi_{f_i})_k^a(\overline{\Psi}_{f_j})_l^b\rightarrow
	-\frac{g\,\delta_{f_if_j}}{32\pi^2\varepsilon}\, m\,T^{n\,ab}(\sigma_{\alpha\beta})_{kl}\, G^{n\,\alpha\beta}+\frac{g\,\delta_{f_if_j}}{48\pi^2\varepsilon}\, T^{n\,ab}\,(\gamma_\nu)_{kl}\, D^{nm}_\mu G^{m\,\mu\nu},
	\label{rep_2}
\end{equation}
where $k$ and $l$ are spin indices; $m$ is the mass of the quark with flavor $f_i$; $\sigma_{\alpha\beta} = \frac{i}{2}[\gamma_\alpha,\gamma_\beta]$; $D^{nm}_\mu$ is covariant derivative in adjoint representation. Here the sign of $g$ is chosen such that the covariant derivative $\nabla_\mu=\partial_\mu -igT^nA^n_\mu$. After enumerating all such possible replacements, the required counterterm is obtained. The presence of the term $\propto mG$ indicates the four-quark operators are not multiplicatively renormalizable. In the massless limit, only the second term is necessary, which defines a hybrid-like operator.

\begin{figure}
	\centering
	\begin{subfigure}{0.34\textwidth}
		\centering
		\includegraphics{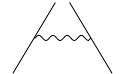}
		\caption{Subdiagram related to four-quark counter term in figure~\ref{fig_r_a}.}
		\label{fig_pole_4q}
	\end{subfigure}
	\hspace*{2.5cm}
	\begin{subfigure}{0.34\textwidth}
		\centering
		\includegraphics{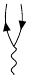}
		\caption{Subdiagram related to hybrid-like counter term in figure~\ref{fig_r_b}.}
		\label{fig_pole_h-like}
	\end{subfigure}
	\caption{One-loop subdiagrams involved in figure~\ref{fig_r}. Here the loops are not "closed" to indicate that the propagators are not contracted with vertices. The directions of the propagators are omitted in the left figure.}
	\label{fig_pole}
\end{figure}

It is also easy to see that this replacement procedure for finding counterterms is applicable to arbitrary multi-quark operators at one-loop level. This is because there is a one-to-one correspondence between the selection of two quark propagators to form a loop and the selection of two quarks for replacement. As a verification, consider the subdiagrams shown in figure~\ref{fig_pole}. In Feynman gauge, the left diagram gives an $\varepsilon$-pole:
\begin{equation}
	\begin{split}
		\int\frac{d^d k}{(2\pi)^d}\ & \bigg(i\frac{igT^n\gamma_\mu.(\slashed{q}+\slashed{k}-m)}{(q+k)^2-m^2}\bigg)\times \bigg(i\frac{igT^m\gamma_\nu.(\slashed{q}+\slashed{k}-m)}{(q+k)^2-m^2}\bigg)\times \frac{g^{\mu\nu}\delta^{nm}}{i k^2}\\&\ \sim -\frac{1}{64\pi^2\varepsilon}g^{\mu\nu}g^{\alpha\beta}\times (igT^n\gamma_\mu.\gamma_\alpha)\times (igT^n\gamma_\nu.\gamma_\beta).
	\end{split}
\end{equation}
The overall sign and hermitian conjugation for the terms in parentheses depend on the direction of fermions. The right diagram in figure~\ref{fig_pole} gives an $\varepsilon$-pole:
\begin{equation}
	\begin{split}
	\int\frac{d^d k}{(2\pi)^d}\ & ig \frac{i(\slashed{q}+\slashed{k}-m)}{q^2-m^2}.T^n.\gamma^\mu \frac{i\slashed{k}-m}{k^2-m^2}\\&\ \sim\ \frac{ig\,m}{16\pi^2\varepsilon} T^n q_\nu\sigma^{\mu\nu}+\frac{g}{48\pi^2\varepsilon}T^n(q^2\gamma^\mu-\slashed{q}q^\mu).
	\end{split}
\end{equation}
These two equations correspond to eqs.~\ref{rep_1} and \ref{rep_2}. The renormalized four-quark operator at $O(\alpha_s)$ can then be expressed as
\begin{equation}
	\mathcal{O}^r=Z_2^{-2}\mathcal{O}+\mathcal{O}_A+\mathcal{O}_B,
	\label{renor_4quark}
\end{equation}
where $Z_2=1-\frac{g^2\,C_F}{16\pi^2 \varepsilon}$ in Feynman gauge; $\mathcal{O}$ is the bare four-quark operator; $\mathcal{O}_A$ and $\mathcal{O}_B$ are the four-quark and hybrid-like counterterms obtained by eq.~\ref{rep_1} and eq.~\ref{rep_2} respectively. The hybrid-like operator is involved only in the singlet and eight-fold states. However, it begins to mix with the four-quark operator at $O(\alpha_s^2)$ through the diagrams
\begin{equation}
	\vcenter{\hbox{\includegraphics[width=0.145\textwidth]{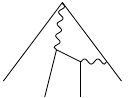}}}\quad\text{and}\quad\vcenter{\hbox{\includegraphics[width=0.145\textwidth]{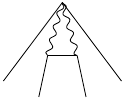}}},
\end{equation}
which correspond to eq.~\ref{green_4q} with $\mathcal{O}(q)$ replaced by a hybrid-like operator. Therefore, the renormalization matrix is upper blocked at $O(\alpha_s)$, and eq.~\ref{renor_4quark} can also be written as
\begin{equation}
	\left[\begin{tikzpicture}[baseline=(X.base)]
	\node[inner sep=0] (X){};
	\node[inner sep=0, above=0.12cm of X](XU){$\mathcal{O}^r_F$};
	\node[inner sep=0, below=0.02cm of X] (Y) {$\mathcal{O}^r_H$};
	\end{tikzpicture}\right]
	=\left[\begin{tikzpicture}[baseline=(X0.base)]
		\node[inner sep=0](X0){};
		\node[inner sep=0, above=0.36cm of X0](XX){};
		\node[inner sep=0, left=0.05cm of XX](XL){$Z_{FF}$};
		\node[inner sep=0, right=0.18cm of XX](XR){$Z_{FH}$};
		\node[inner sep=0, below=0.2cm of X0](YY){};
		\node[inner sep=0, left=0.2cm of YY](YL){{\large 0}};
		\node[inner sep=0, right=0.18cm of YY](YR){$Z_{HH}$};
	\end{tikzpicture}\right]
	 \left[\begin{tikzpicture}[baseline=(X.base)]
	 	\node[inner sep=0](X){};
	 	\node[inner sep=0, above=0.2cm of X](XU){$\mathcal{O}_F$};
	 	\node[inner sep=0, below=0.02cm of X] (Y) {$\mathcal{O}_H$};
	 \end{tikzpicture}\right].
	\label{renor_matrix}
\end{equation}
Here the subscripts indicate the corresponding types of operators, and the superscripts $r$ indicate that the operators are renormalized. The $\mathcal{O}_F$ and $\mathcal{O}_H$ are two classes of operators\textemdash four-quark and hybrid-like operators, respectively.

For the hybrid-like current, the equation of motion gives
\begin{equation}
	g\,\overline{\Psi}_{f_a}T^n\Gamma D_\mu G^{\mu\nu}\gamma_\nu\Psi_{f_b}=-g^2\,\overline{\Psi}_{f_a}T^n\Gamma \Psi_{f_b}\sum_f \overline{\Psi}_{f}T^n\gamma^\nu \Psi_{f}.
	\label{fc_eom}
\end{equation}
\vspace{-0.5cm}

As we have shown in Appendix \ref{A_ward}, this relation is not exact at the quantum level. Nevertheless, for the two-point function of four-quark currents at $O(\alpha_s)$, this relation still holds. However, rewriting eq.~\ref{fc_eom} as tetraquark currents by Fierz rearrangement causes the tetraquark operators to mix with each other. Moreover, the flavors must be written explicitly, due to the $\delta_{f_i f_j}$ in eq.~\ref{rep_2} and the flavor summation in eq.~\ref{fc_eom}. Thus, the renormalization becomes cumbersome, so we keep the hybrid-like currents as they are. Terms similar to eq.~\ref{fc_eom} are also involved in the renormalized molecule currents. The terms
\begin{equation}
	\propto\overline{\Psi}_{f_1}T^n\Gamma_A \Psi_{f_2}\, \overline{\Psi}_{f_3}T^n\Gamma_B \Psi_{f_4},
	\label{mole_t}
\end{equation}
originate from the second to fifth diagrams in figure~\ref{fig_r_b}. The presence of $T^n$ suggests that eq.~\ref{mole_t} does not correctly represent a molecule operator. Nevertheless, by Fierz rearrangement, it can be written as
\begin{equation}
	-\frac{1}{2}\sum_i \overline{\Psi}_{f_1}\Gamma_i \Psi_{f_4}\, \overline{\Psi}_{f_3}\Gamma_B\Gamma_i\Gamma_A \Psi_{f_2} - \frac{1}{2C_A}\overline{\Psi}_{f_1}\Gamma_A \Psi_{f_2}\, \overline{\Psi}_{f_3}\Gamma_B \Psi_{f_4},
\end{equation}
where $\Gamma_i$ are the basis matrices. Instead of involving $T^n$, the flavor exchange ${f_2\leftrightarrow f_4}$ is involved here. Thus, we write the molecule operator as
\begin{equation}
	\overline{\Psi}_{f_1}\Gamma_A \Psi_{f_2}\, \overline{\Psi}_{f_3}\Gamma_B \Psi_{f_4}\pm \{f_2\leftrightarrow f_4\}.
	\label{mole_base}
\end{equation}
Eqs.\ref{tetra_o} and \ref{mole_base} are simply two types of bases for the four-quark operators in this paper.

\begin{figure}[t!]
	\centering
	\begin{subfigure}{\textwidth}
		\centering
		\includegraphics{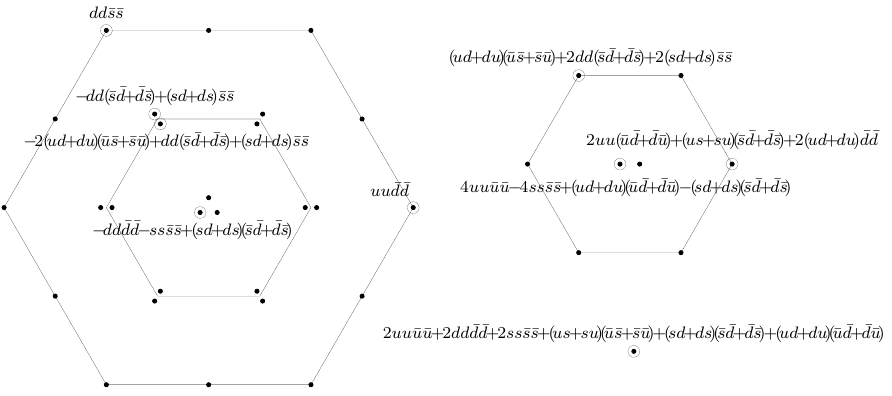}
		\caption{Decomposition of $6\times\bar{6}$, labeled as $27_S$, $8_S$ and $1_S$.}
		\label{su3_6_6}
	\end{subfigure}\\[0.3cm]
	\begin{subfigure}{\textwidth}
		\centering
		\includegraphics{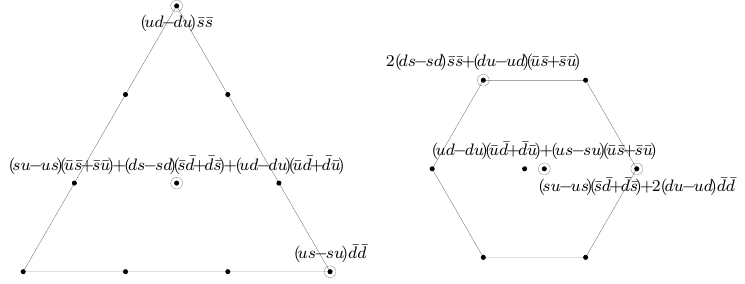}
		\caption{Decomposition of $\bar{3}\times\bar{6}$, labeled as $\overline{10}_M$ and $\overline{8}_M$.}
		\label{su3_6_3}
	\end{subfigure}\\[0.3cm]
	\begin{subfigure}{\textwidth}
		\centering
		\includegraphics{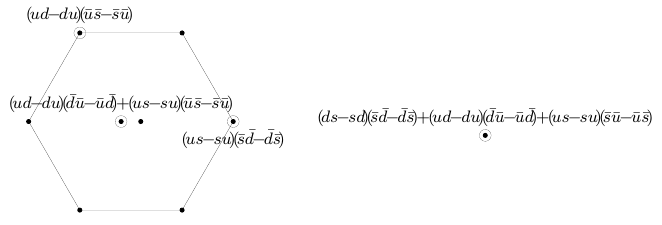}
		\caption{Decomposition of $\bar{3}\times 3$, labeled as $8_A$ and $1_A$.}
		\label{su3_3_3}
	\end{subfigure}
	\caption{Decomposition of $(3\times3)\times(\bar{3}\times\bar{3})=(6+\bar{3})\times(\bar{6}+3)$; only some of the flavors are shown and circled. The charge conjugations of $\overline{10}_M$ and $\overline{8}_M$ are omitted. The subscripts $S$ or $A$ indicate that exchanging the flavors of two quarks (or anti-quarks) is symmetric or antisymmetric, respectively; $M$ indicates that the flavor-exchanging symmetries for quarks and anti-quarks are different. For later convenience, we use square brackets to denote the state with the highest weight in a representation, e.g., $[27_S]=dd\bar{s}\bar{s}$, $[8_A]=(ud-du)(\bar{u}\bar{s}-\bar{s}\bar{u})$. Complete flavor structures can be found in~\cite{results}.}
	\label{su3}
\end{figure}

\subsection{The $0^+$ Four-Quark Currents}\label{sec2_2}

The flavor structure of four-quark states is complex. The decomposition $3\times3\times\bar{3}\times\bar{3}=1+1+8+8+8+8+10+\overline{10}+27$ yields different $su(3)$-flavor representations, as shown in fig~\ref{su3}. To avoid energy scale dependence in the structure of operators, the operator representing a hadron should be multiplicatively renormalizable whenever possible. We thus choose the eigenvectors of $Z_{FF}$ in eq.~\ref{renor_matrix} as the basis of renormalized operators. There are two types of $0^+$ tetraquark operators. The first type is:
\begin{equation}
	\begin{split}
		J_1=\,&\Psi^{a\,T}_{f_1}\,\mathcal{C}\gamma^\mu\,\Psi^b_{f_2}\Big(\overline{\Psi}^b_{f_3}\,\gamma_\mu \mathcal{C}\ \overline{\Psi}^{a\,T}_{f_4}+\{a\leftrightarrow b\}\Big)\\
		J_2=\,&\Psi^{a\,T}_{f_1}\,\mathcal{C}\gamma^\mu\,\Psi^b_{f_2}\Big(\overline{\Psi}^b_{f_3}\,\gamma_\mu \mathcal{C}\ \overline{\Psi}^{a\,T}_{f_4}-\{a\leftrightarrow b\}\Big)\\
		J_3=\,&\Psi^{a\,T}_{f_1}\,\mathcal{C}\gamma^\mu\gamma^5\,\Psi^b_{f_2}\Big(\overline{\Psi}^b_{f_3}\,\gamma_\mu\gamma^5 \mathcal{C}\ \overline{\Psi}^{a\,T}_{f_4}+\{a\leftrightarrow b\}\Big)\\
		J_4=\,&\Psi^{a\,T}_{f_1}\,\mathcal{C}\gamma^\mu\gamma^5\,\Psi^b_{f_2}\Big(\overline{\Psi}^b_{f_3}\,\gamma_\mu\gamma^5 \mathcal{C}\ \overline{\Psi}^{a\,T}_{f_4}-\{a\leftrightarrow b\}\Big)
		\label{tetra_1}
	\end{split}
\end{equation}
The same notations as in eq.~\ref{tetra_o} are adopted here. The renormalized currents can be defined as follows:
\NiceMatrixOptions{cell-space-limits = 0.03cm}
\begin{equation}
	\begin{bNiceMatrix}
		\big(1-\frac{\alpha_s}{\pi\varepsilon}2C_F\big)^{-1}J^r_1\\ \big(1-\frac{\alpha_s}{\pi\varepsilon}2C_F\big)^{-1}J^r_2\\ \big[1-\frac{\alpha_s}{2\pi\varepsilon}(C_F-\frac{3}{2C_A})\big]^{-1}J^r_3\\ \big[1-\frac{\alpha_s}{2\pi\varepsilon}(C_F-\frac{3}{2C_A})\big]^{-1}J^r_4
	\end{bNiceMatrix} = \begin{bNiceMatrix}
		-1 & 0 & \,0 & \,\,1\\
		0 & -1 & \,1 & \,\,0\\
		\frac{C_A-1}{1+C_A} & 0 & \,0 & \,\,1\\
		0 & \frac{1+C_A}{C_A-1} & \,1 & \,\,0\\
	\end{bNiceMatrix} 
	\begin{bNiceMatrix}
		J_1\\J_2\\J_3\\J_4\\
	\end{bNiceMatrix}
	\label{tetra_1_r}
\end{equation}
Here, $C_F$ and $C_A$ are Casimir constants in the fundamental and adjoint
representations, respectively; $J^r_i$ denote the renormalized currents. Note that the $J_i^r$ and $J_i$ with the same subscripts have no direct correspondence. The hybrid-like counterterms will be needed for the singlet and eight-fold states. In this paper, only the hybrid-like counterterm $\propto\overline{\Psi}_{f_i}T^n D^{nm}_\mu G^{m\,\mu\nu}\gamma_\nu\Psi_{f_j}$ is involved; therefore, we use a symbol
\begin{equation}
	H_{f_1f_2,f_3f_4}\equiv\, \delta_{f_1f_2}\overline{\Psi}_{f_3}T^n D^{nm}_\mu G^{m\,\mu\nu}\gamma_\nu\Psi_{f_4}
\end{equation}
to denote the hybrid-like operators. The hybrid-like counterterms for $[J^r_1\ J^r_2\ J^r_3\ J^r_4]^T$ are
\NiceMatrixOptions{cell-space-limits = 0.03cm}
\begin{equation}
	\begin{bNiceMatrix}
		-\frac{g}{\vphantom{a^b}12\pi^2\varepsilon}(H_{f_1f_3,f_4f_2}-H_{f_2f_3,f_4f_1}-H_{f_1f_4,f_3f_2}+H_{f_2f_4,f_3f_1})\\
		-\frac{g}{\vphantom{a^b}12\pi^2\varepsilon}(H_{f_1f_3,f_4f_2}+H_{f_2f_3,f_4f_1}+H_{f_1f_4,f_3f_2}+H_{f_2f_4,f_3f_1})\\
		-\frac{g}{\vphantom{a^b}12(C_A+1)\pi^2\varepsilon}(H_{f_1f_3,f_4f_2}+H_{f_2f_3,f_4f_1}+H_{f_1f_4,f_3f_2}+H_{f_2f_4,f_3f_1})\\
		\frac{g}{\vphantom{a^b}12(C_A-1)\pi^2\varepsilon}(H_{f_1f_3,f_4f_2}-H_{f_2f_3,f_4f_1}-H_{f_1f_4,f_3f_2}+H_{f_2f_4,f_3f_1})
	\end{bNiceMatrix}.
\end{equation}
We write them as vectors to make the equations more compact here and below.

The second type of $0^+$ tetraquark operators is:
\begin{equation}
	\begin{split}
		J_a=\,&\Psi^{a\,T}_{f_1}\,\mathcal{C}\,\Psi^b_{f_2}\Big(\overline{\Psi}^b_{f_3}\,\mathcal{C}\ \overline{\Psi}^{a\,T}_{f_4}+\{a\leftrightarrow b\}\Big)\\
		J_b=\,&\Psi^{a\,T}_{f_1}\,\mathcal{C}\,\Psi^b_{f_2}\Big(\overline{\Psi}^b_{f_3}\,\mathcal{C}\ \overline{\Psi}^{a\,T}_{f_4}-\{a\leftrightarrow b\}\Big)\\
		J_c=\,&\Psi^{a\,T}_{f_1}\,\mathcal{C}\gamma^5\,\Psi^b_{f_2}\Big(\overline{\Psi}^b_{f_3}\,\gamma^5 \mathcal{C}\ \overline{\Psi}^{a\,T}_{f_4}+\{a\leftrightarrow b\}\Big)\\
		J_d=\,&\Psi^{a\,T}_{f_1}\,\mathcal{C}\gamma^5\,\Psi^b_{f_2}\Big(\overline{\Psi}^b_{f_3}\,\gamma^5 \mathcal{C}\ \overline{\Psi}^{a\,T}_{f_4}-\{a\leftrightarrow b\}\Big)\\
		J_e=\,&\Psi^{a\,T}_{f_1}\,\mathcal{C}\sigma^{\mu\nu}\,\Psi^b_{f_2}\Big(\overline{\Psi}^b_{f_3}\,\sigma_{\mu\nu}\mathcal{C}\ \overline{\Psi}^{a\,T}_{f_4}+\{a\leftrightarrow b\}\Big)\\
		J_f=\,&\Psi^{a\,T}_{f_1}\,\mathcal{C}\sigma^{\mu\nu}\,\Psi^b_{f_2}\Big(\overline{\Psi}^b_{f_3}\,\sigma_{\mu\nu}\mathcal{C}\ \overline{\Psi}^{a\,T}_{f_4}-\{a\leftrightarrow b\}\Big)
		\label{tetra_2}
	\end{split}
\end{equation}
We can define the renormalized currents as follows:
\NiceMatrixOptions{cell-space-limits = 0.08cm}
\begin{gather}
	\begin{bNiceMatrix}
		\big[1-\frac{\alpha_s}{\pi\varepsilon}(C_F-\frac{3}{4}+\frac{S}{4C_A})\big]^{-1}J^r_a\\ \big[1-\frac{\alpha_s}{\pi\varepsilon}(C_F+\frac{3}{4}+\frac{S}{4C_A})\big]^{-1}J^r_b\\ \big[1-\frac{\alpha_s}{2\pi\varepsilon}(C_F+\frac{3}{2}+\frac{3}{2C_A})\big]^{-1}J^r_c\\ \big[1-\frac{\alpha_s}{2\pi\varepsilon}(C_F-\frac{3}{2}+\frac{3}{2C_A})\big]^{-1}J^r_d\\ \big[1-\frac{\alpha_s}{\pi\varepsilon}(C_F+\frac{3}{4}-\frac{S}{4C_A})\big]^{-1}J^r_e\\ \big[1-\frac{\alpha_s}{\pi\varepsilon}(C_F-\frac{3}{4}-\frac{S}{4C_A})\big]^{-1}J^r_f
	\end{bNiceMatrix} = 
		\begin{bNiceMatrix}
			A_1&0&A_1&0&0&B+S\\
			0&A_2&0&A_2&B+S&0\\
			0&-1&0&1&0&0\\
			-1&0&1&0&0&0\\
			0&A_2&0&A_2&B-S&0\\
			A_1&0&A_1&0&0&B-S
		\end{bNiceMatrix}
	\begin{bNiceMatrix}
			J_a\\J_b\\J_c\\J_d\\J_e\\J_f\\
	\end{bNiceMatrix}.
	\label{tetra_2_r}
\end{gather}
Here, $A_1=6(C_A-C^2_A)$, $A_2=6(C_A+C^2_A)$, $B=C_A^2-4$, $S=\sqrt{4C_A^4-11C_A^2+16}$. The hybrid-like counterterms for $[J^r_a\ J^r_b\ J^r_c\ J^r_d\ J^r_e\ J^r_f]^T$ are
\NiceMatrixOptions{cell-space-limits = 0.05cm}
\begin{equation}
	\begin{bNiceMatrix}
		0\\
		0\\
		-\frac{g}{24\pi^2\varepsilon}(H_{f_1f_3,f_4f_2}-H_{f_2f_3,f_4f_1}-H_{f_1f_4,f_3f_2}+H_{f_2f_4,f_3f_1})\\
		-\frac{g}{24\pi^2\varepsilon}(H_{f_1f_3,f_4f_2}+H_{f_2f_3,f_4f_1}+H_{f_1f_4,f_3f_2}+H_{f_2f_4,f_3f_1})\\
		0\\
		0
	\end{bNiceMatrix}.
	\label{hybrid_tetra_2}
\end{equation}

The $0^+$ molecule currents can also be categorized into two types. The first type is:
\begin{equation}
	\begin{split}
		I_1=&\,\overline{\Psi}_{f_1}\gamma^\mu\Psi_{f_2}\,\overline{\Psi}_{f_3}\gamma_\mu\Psi_{f_4}+\{f_2\leftrightarrow f_4\}\\[1pt]
		I_2=&\,\overline{\Psi}_{f_1}\gamma^\mu.\gamma^5\Psi_{f_2}\,\overline{\Psi}_{f_3}\gamma_\mu.\gamma^5\Psi_{f_4}+\{f_2\leftrightarrow f_4\}\\[1pt]
		I_3=&\,\overline{\Psi}_{f_1}\sigma^{\mu\nu}\Psi_{f_2}\,\overline{\Psi}_{f_3}\sigma_{\mu\nu}\Psi_{f_4}+\{f_2\leftrightarrow f_4\}\\[1pt]
		I_4=&\,\overline{\Psi}_{f_1}\gamma^5\Psi_{f_2}\,\overline{\Psi}_{f_3}\gamma^5\Psi_{f_4}+\{f_2\leftrightarrow f_4\}\\[1pt]
		I_5=&\,\overline{\Psi}_{f_1}\Psi_{f_2}\,\overline{\Psi}_{f_3}\Psi_{f_4}+\{f_2\leftrightarrow f_4\}.
	\end{split}
	\label{mole_s_basis}
\end{equation}
We can define the renormalized currents as follows:
\NiceMatrixOptions{cell-space-limits = 0.01cm}
\begin{equation}
	\begin{bNiceMatrix}
		\big(1-\frac{\alpha_s}{\pi\varepsilon}2C_F\big)^{-1}I^r_1\\ \big[1-\frac{\alpha_s}{2\pi\varepsilon}(C_F-\frac{3}{2C_A})\big]^{-1}I^r_2\\ \big[\vphantom{\frac{|}{B}}1-\frac{\alpha_s}{2\pi\varepsilon}(C_F-\frac{3}{2}+\frac{3}{2C_A})\big]^{-1}I^r_3\\ \big[\vphantom{\frac{|}{B}}1-\frac{\alpha_s}{\pi\varepsilon}(C_F-\frac{3}{4}+\frac{S}{4C_A})\big]^{-1}I^r_4\\
		\big[\vphantom{\frac{|}{B}}1-\frac{\alpha_s}{\pi\varepsilon}(C_F-\frac{3}{4}-\frac{S}{4C_A})\big]^{-1}I^r_5
	\end{bNiceMatrix} = \begin{bNiceMatrix}
		0 & 0 & 0 & \,-1 & \,\,1 \\
		\frac{C_A}{2} & -\frac{C_A}{2} & 0 & \,-1 &\,\, 1 \\
		1 & 1 & 0 & \,0 & \,\,0 \\
		0 & 0 & \frac{2(C_A^2-1)-S}{6(C_A+2)} & \,1 &\,\, 1\\
		0 & 0 & \frac{2(C_A^2-1)+S}{6(C_A+2)} & \,1 & \,\,1
	\end{bNiceMatrix} 
	\begin{bNiceMatrix}
		I_1\\[3pt]I_2\\[3pt]I_3\\[3pt]I_4\\[3pt]I_5\\
	\end{bNiceMatrix}.
	\label{mole_s_r}
\end{equation}
Here and below, as in eq.~\ref{tetra_2_r}, we have $S=\sqrt{4C_A^4-11C_A^2+16}$. In practice, $I^r_4\propto I_4+I_5$ approximately since $C_A=3$. The hybrid-like counterterms for $[I^r_1\ I^r_2\ I^r_3\ I^r_4\ I^r_5]^T$ are:
\begin{equation}
	\begin{bNiceMatrix}
		-\frac{g}{\vphantom{a^b}24\pi^2\varepsilon}(H_{f_1f_4,f_3f_2}+H_{f_1f_2,f_3f_4}+H_{f_4f_3,f_1f_2}+H_{f_2f_3,f_1f_4})\\
		-\frac{g}{\vphantom{a^b}24\pi^2\varepsilon}(H_{f_1f_4,f_3f_2}+H_{f_1f_2,f_3f_4}+H_{f_4f_3,f_1f_2}+H_{f_2f_3,f_1f_4})\\
		\frac{g}{\vphantom{a^b}12\pi^2\varepsilon}(H_{f_1f_4,f_3f_2}+H_{f_1f_2,f_3f_4}+H_{f_4f_3,f_1f_2}+H_{f_2f_3,f_1f_4})\\
		0\\
		0
	\end{bNiceMatrix}.
	\label{hybrid_mole_S}
\end{equation}

The second type of $0^+$ molecule currents is:
\begin{equation}
	\begin{split}
		I_a=&\,\overline{\Psi}_{f_1}\gamma^\mu\Psi_{f_2}\,\overline{\Psi}_{f_3}\gamma_\mu\Psi_{f_4}-\{f_2\leftrightarrow f_4\}\\[2pt]
		I_b=&\,\overline{\Psi}_{f_1}\gamma^\mu.\gamma^5\Psi_{f_2}\,\overline{\Psi}_{f_3}\gamma_\mu.\gamma^5\Psi_{f_4}-\{f_2\leftrightarrow f_4\}\\[2pt]
		I_c=&\,\overline{\Psi}_{f_1}\sigma^{\mu\nu}\Psi_{f_2}\,\overline{\Psi}_{f_3}\sigma_{\mu\nu}\Psi_{f_4}-\{f_2\leftrightarrow f_4\}\\[2pt]
		I_d=&\,\overline{\Psi}_{f_1}\gamma^5\Psi_{f_2}\,\overline{\Psi}_{f_3}\gamma^5\Psi_{f_4}-\{f_2\leftrightarrow f_4\}\\[2pt]
		I_e=&\,\overline{\Psi}_{f_1}\Psi_{f_2}\,\overline{\Psi}_{f_3}\Psi_{f_4}-\{f_2\leftrightarrow f_4\}.
	\end{split}
	\label{mole_a_basis}
\end{equation}
We can define the renormalized currents as follows
\NiceMatrixOptions{cell-space-limits = 0.01cm}
\begin{equation}
	\begin{bNiceMatrix}
		\big[1-\frac{\alpha_s}{\pi\varepsilon}2C_F\big]^{-1}I^r_a\\ \big[1-\frac{\alpha_s}{2\pi\varepsilon}(C_F-\frac{3}{2C_A})\big]^{-1}I^r_b\\ \big[\vphantom{\frac{|}{B}}1-\frac{\alpha_s}{2\pi\varepsilon}(C_F+\frac{3}{2}+\frac{3}{2C_A})\big]^{-1}I^r_c\\ \big[\vphantom{\frac{|}{B}}1-\frac{\alpha_s}{\pi\varepsilon}(C_F+\frac{3}{4}+\frac{S}{4C_A})\big]^{-1}I^r_d\\
		\big[\vphantom{\frac{|}{B}}1-\frac{\alpha_s}{\pi\varepsilon}(C_F+\frac{3}{4}-\frac{S}{4C_A})\big]^{-1}I^r_e
	\end{bNiceMatrix} = \begin{bNiceMatrix}
		0 & 0 & 0 & \,-1 & \,\,1 \\
		-\frac{C_A}{2} & \frac{C_A}{2} & 0 & \,-1 & \,\,1 \\
		1 & 1 & 0 & \,0 & \,\,0 \\
		0 & 0 & \frac{C_A+2}{2S+4(C_A^2-1)} & \,-1 & \,\,-1 \\
		0 & 0 & \frac{C_A+2}{2S-4(C_A^2-1)} & \,1 & \,\,1 
	\end{bNiceMatrix} 
	\begin{bNiceMatrix}
		I_a\\[3pt]I_b\\[3pt]I_c\\[3pt]I_d\\[3pt]I_e\\
	\end{bNiceMatrix}.
	\label{mole_a_r}
\end{equation}
In practice, $I^r_e\propto I_c$ approximately since $C_A=3$. And the hybrid-like counterterms for $[I^r_a\ I^r_b\ I^r_c\ I^r_d\ I^r_e]^T$ are:
\NiceMatrixOptions{cell-space-limits = 0.05cm}
\begin{equation}
	\begin{bNiceMatrix}
		-\frac{g}{\vphantom{a^b}24\pi^2\varepsilon}(H_{f_1f_4,f_3f_2}-H_{f_1f_2,f_3f_4}-H_{f_4f_3,f_1f_2}+H_{f_2f_3,f_1f_4})\\
		-\frac{g}{\vphantom{a^b}24\pi^2\varepsilon}(H_{f_1f_4,f_3f_2}-H_{f_1f_2,f_3f_4}-H_{f_4f_3,f_1f_2}+H_{f_2f_3,f_1f_4})\\
		\frac{g}{\vphantom{a^b}12\pi^2\varepsilon}(H_{f_1f_4,f_3f_2}-H_{f_1f_2,f_3f_4}-H_{f_4f_3,f_1f_2}+H_{f_2f_3,f_1f_4})\\
		0\\
		0
	\end{bNiceMatrix}.
	\label{hybrid_mole_A}
\end{equation}

To evaluate the correlator corresponding to a specific state, we must specify the flavor structures for the currents. We adopt some notations here for clarity. First, an $su(3)$-flavor state can be written as
\begin{equation}
	\mathcal{F}=\sum_i c_i\, f^i_a f^i_b \bar{f}^i_c \bar{f}^i_d,
	\label{su3-f}
\end{equation}
where $c_i$ are coefficients, $i$ labels different terms, and the subscripts $a$, $b$, $c$, and $d$ label the flavors. The $J_\mathcal{F}$ or $I_\mathcal{F}$ represent the currents $J$ or $I$ with flavors specified as the $su(3)$-flavor state $\mathcal{F}$. Specifically, by keeping only the flavor indices and abbreviating the tetraquark current as $J_{f_1f_2\bar{f}_3\bar{f}_4}$, we have
\begin{equation}
	J_\mathcal{F}=\sum_i c_i\, J_{f^i_a f^i_b \bar{f}^i_c \bar{f}^i_d}.
\end{equation}
For example, consider $[8_A]=(ud-du)(\bar{u}\bar{s}-\bar{s}\bar{u})$ as shown in figure~\ref{su3_3_3} and
\begin{equation}
	J_1=\Psi^{a\,T}_{f_1}\,\mathcal{C}\gamma^\mu\,\Psi^b_{f_2}\big(\overline{\Psi}^b_{f_3}\,\gamma_\mu \mathcal{C}\ \overline{\Psi}^{a\,T}_{f_4}+\{a\leftrightarrow b\}\big),
\end{equation}
we have
\begin{equation}
	\begin{split}
	J_{1,[8_A]}=&\quad u^{aT}\mathcal{C}\gamma^\mu d^b\big(\bar{u}^b\gamma_\mu\mathcal{C}\,\bar{s}^{aT}+\{a\leftrightarrow b\}\big)\\
	&-u^{aT}\mathcal{C}\gamma^\mu d^b\big(\bar{s}^b\gamma_\mu\mathcal{C}\, \bar{u}^{aT}+\{a\leftrightarrow b\}\big)\\
	&-d^{aT}\mathcal{C}\gamma^\mu u^b\big(\bar{u}^b\gamma_\mu\mathcal{C}\, \bar{s}^{aT}+\{a\leftrightarrow b\}\big)\\
	&+d^{aT}\mathcal{C}\gamma^\mu u^b\big(\bar{s}^b\gamma_\mu\mathcal{C}\, \bar{u}^{aT}+\{a\leftrightarrow b\}\big)
\end{split}
\end{equation}
Here, $\Psi_f$ is abbreviated as $f$. For the molecule operator, we rewrite eq.~\ref{su3-f} as
\begin{equation}
	\mathcal{F}=\sum_i c_i\, \bar{f}^i_cf^i_b\bar{f}^i_df^i_a.
\end{equation}
By keeping only the flavor indices and abbreviating the molecule current as $I_{\bar{f}_1f_2\bar{f}_3f_4}$, we have 
\begin{equation}
	I_\mathcal{F}=\sum_i c_i\, I_{\bar{f}^i_c  f^i_b \bar{f}^i_d f^i_a}.
\end{equation}
For example,
\begin{equation}
	I_a=\overline{\Psi}_{f_1}\gamma^\mu\Psi_{f_2}\,\overline{\Psi}_{f_3}\gamma_\mu\Psi_{f_4}-\{f_2\leftrightarrow f_4\},
\end{equation}
and we have
\begin{equation}
	\begin{split}
	I_{a,[8_A]}=&\quad \big(\bar{u}\gamma^\mu d \,\bar{s}\gamma_\mu u - \{d\leftrightarrow u\}\big)\\
	&- \big(\bar{s}\gamma^\mu d \,\bar{u}\gamma_\mu u - \{d\leftrightarrow u\}\big)\\
	&- \big(\bar{u}\gamma^\mu u \,\bar{s}\gamma_\mu d - \{u\leftrightarrow d\}\big)\\
	&+ \big(\bar{s}\gamma^\mu u \,\bar{u}\gamma_\mu d - \{u\leftrightarrow d\}\big).
	\end{split}
\end{equation}
We also use $J_R$ or $I_R$ to denote currents $J$ or $I$ whose flavors are specified within the representation $R$. For example, $J^r_{4,8_S}$ denotes eight $J^r_4$ currents, where each current has flavor structure corresponding to an $su(3)$-flavor state in the $8_S$ representation.

Note that for any four-quark current in eqs.~\ref{tetra_1}, \ref{tetra_2}, \ref{mole_s_basis}, and \ref{mole_a_basis}, exchanging the flavors of two fermions is equivalent to exchanging the flavors of two antifermions, which may change the sign of the current. Consequently, the current may vanish after specifying flavors. For example, $I_{1,[8_A]}=0$ because $I_1$ is symmetric under the exchange of two fermions' flavors. Specifically, the currents in eq.~\ref{mole_s_r} vanish when the flavors are specified as $8_A$ or $1_A$; similarly, the currents in eq.~\ref{mole_a_r} vanish when the flavors are specified as $27_S$, $8_S$, or $1_S$. Moreover, when the flavors are specified as $10_M$ and $8_M$, all of the four-quark currents vanish. For these states, exchanging two fermions' flavors differs from exchanging two antifermions' flavors by a sign, as shown in fig~\ref{su3_6_3}.

Unlike the bare operators, there is a one-to-one correspondence between the renormalized tetraquark operators and the renormalized molecular operators, as shown in table~\ref{bases}. After Fierz rearrangement, the $J_i^r$ in eqs.~\ref{tetra_1_r} and \ref{tetra_2_r} are just the $I_i^r$ in eqs.~\ref{mole_s_r} and \ref{mole_a_r}, up to overall factors and the exchange $f_1\leftrightarrow f_4$ to match their flavor conventions.


\begin{table}[t!]
	\centering
	\caption{The one-to-one correspondence between renormalized tetraquark operators and renormalized molecule operators. The $\gamma^5$-scheme is related to this correspondence if $d\neq4$, and terms involving $\hat{\gamma}^5$ may be introduced, which break the correspondence slightly.\label{bases}}
	\renewcommand{\arraystretch}{1.2}
	\begin{tabular}{|*{10}{w{c}{0.06\textwidth}|}}
		\noalign{\hrule height 1pt}
		$J_1^r$&$J_2^r$&$J_3^r$&$J_4^r$&$J_a^r$&$J_b^r$&$J_c^r$&$J_d^r$&$J_e^r$&$J_f^r$\\
		\noalign{\hrule height 0.2pt}
		$I_a^r$&$I_1^r$&$I_b^r$&$I_2^r$&$I_4^r$&$I_d^r$&$I_c^r$&$I_3^r$&$I_e^r$&$I_5^r$\\
		\noalign{\hrule height 1pt}
	\end{tabular}
\end{table}

While this correspondence may seem surprising, it arises because tetraquark operators and molecular operators are related by Fierz rearrangement; they are simply two different bases of the same operator space. On the other hand, the renormalization matrix provides information about how an operator transforms under the changes of energy scale; its eigenvectors are determined by physics. Such a correspondence between different sets of renormalized operators should always exist, as long as they span the same operator space. Once renormalization is taken into account, constructing an operator that accurately represents a compact tetraquark or four-quark molecule becomes subtle. For the renormalization matrix of four-quark operators, since the eigenvectors can equally be interpreted as tetraquark or molecule operators, the physical states may be mixed states of compact tetraquark and four-quark molecule. This conclusion agrees with ref.~\cite{tetra_mole,mixed_tetra} but from a different perspective.

The selection of diagrams associated with four-quark states remains an unsettled issue~\cite{fourquark_criteria,tetra_Nc_QCD,landau_pole_wang}, and there is no established criterion for distinguishing tetraquark diagrams from molecular diagrams. Commonly, the four-quark molecule is identified as two mesons bound by exchanging color-neutral states. Therefore, a naive criterion for selecting the molecular correlator is to discard diagrams involving gluon exchange between the two mesons, as well as diagrams where quarks annihilate into a gluon---for example, the second to fifth diagrams in figure~\ref{fig_r_b}. These diagrams give rise to terms that do not correctly represent molecular operators. However, as discussed below eq.~\ref{mole_t}, the color matrix $T^n$ can be eliminated by simply rewriting the operator in a different form. This issue is beyond the scope of this paper, and we will not distinguish between tetraquarks and molecules in this analysis.


\section{Mass Estimations}\label{sec_3}
\subsection{The OPE of Four-Quark Correlators}\label{sec_3_ope}
The two-point function $\langle T J_A(x)J^\dagger_B(0)\rangle$ contains information about the position of the mass-pole in the $s$-plane. To evaluate the two-point function in the nonperturbative region, the operator product expansion (OPE)~\cite{qcd_sr_book} is a convenient technique. The two-point function can be expressed as
\begin{equation}
	\Pi_{AB}(q^2)=i\int dx e^{iqx}\langle T J_A(x)J^\dagger_B(0)\rangle = C^n_{AB}(q^2)\langle O_n\rangle,
\end{equation}
where $C^n_{AB}(q^2)$ are the coefficients that can be calculated perturbatively; $\langle O_n\rangle$ are condensates (vacuum expectation values of local operators) which encode the nonperturbative information. Our calculation includes condensates up to mass-dimension 10:
\begin{equation}
	\begin{split}
	\Pi_{AB}(q^2)=&\,C^0_{AB} + C^{4_q}_{AB}m\langle \bar{q}q\rangle + C^{4_g}_{AB}\langle GG\rangle + C^{6_q}_{AB}\langle \bar{q}q\rangle^2 +C^{6_g}_{AB}\langle G^3\rangle \\&+C^8_{AB}\langle \bar{q}q\rangle\langle \bar{q}Gq\rangle+C^{10_m}_{AB}m\langle \bar{q}q\rangle^3+C^{10_q}_{AB}\langle \bar{q}Gq\rangle^2+ C^{10_g}_{AB}\langle GG\rangle\langle\bar{q}q\rangle^2,
	\end{split}
	\label{pi_AB}
\end{equation}
where $\langle GG\rangle=\langle \alpha_s G^n_{\mu\nu}G^{n\,\mu\nu}\rangle$; $\langle \bar{q}Gq\rangle=\langle \bar{q}\,g_s T^nG^n_{\mu\nu}\sigma^{\mu\nu}q\rangle$; $\langle G^3\rangle=\langle g_s^3 f^{abc}G^{a\ \nu}_\mu G^{b\ \rho}_\nu G^{c\ \mu}_\rho\rangle$. The superscripts of $C_{AB}$ denote the corresponding types of condensates. The related diagrams are shown in Appendix~\ref{dias}. We have built a Mathematica package~\cite{pack} to calculate these diagrams. For the calculation technique, refer to ref~\cite{condensates_high_order}. The quark propagator in background gluon fields under Fock-Schwinger gauge~\cite{qsr,condensates_high_order} can be written as
\begin{equation}
	\begin{split}
	S(q) =&\,\, S_0(q) + \frac{ig}{2}G_{\mu\nu}^a S_0(q)T^a\partial^\mu\gamma^\nu S_0(q) + \frac{g}{3}D^{ab}_\alpha G^b_{\mu\nu}S_0(q)T^a\partial^\alpha\partial^\mu\gamma^\nu S_0(q)\\
	&-\!\frac{ig}{8}D^{ab}_\alpha D^{bc}_\beta G^c_{\mu\nu}S_0(q)T^a\partial^\alpha\partial^\beta\partial^\mu\gamma^\nu S_0(q)-\!\frac{g^2}{4}G^a_{\rho\sigma}G^b_{\mu\nu}S_0(q)T^a\partial^\rho\gamma^\sigma S_0(q)T^b\partial^\mu\gamma^\nu S_0(q)\\
	&-\frac{ig^3}{8}G^a_{\alpha\beta}G^b_{\rho\sigma}G^c_{\mu\nu}\, S_0(q)T^a\partial^\alpha\gamma^\beta S_0(q)T^b\partial^\rho\gamma^\sigma S_0(q)T^c\partial^\mu\gamma^\nu S_0(q) + \cdots,
	\label{q_propagator}
	\end{split}
\end{equation}
where $S_0(q)=(\slashed{q}+m)/q^2$ is the free quark propagator in the $m^2\rightarrow0$ limit, and the terms irrelevant to our calculation are omitted. The gluon propagator $\Delta(q)_{\mu\nu}^{nm}=-\delta^{nm}g_{\mu\nu}/q^2$ is involved only in the NLO perturbative diagrams. The evaluation of eq.~\ref{pi_AB} is straightforward in coordinate space by Fourier transformation, since only $S(x)=\int \frac{d^4q}{(2\pi)^4} e^{-iqx} i S(q)$ is involved. The only exceptions are the NLO perturbative diagrams, which involve two-loop propagator-type subdiagrams. Keeping only the terms relevant to momentum, it can be written as
\begin{equation}
	\int\frac{d^dk_1}{(2\pi)^d}\frac{d^dk_2}{(2\pi)^d}\,\frac{k_1^\alpha k_2^\beta(k_1+q)^\mu(k_2+q)^\nu}{k_1^2k_2^2(k_1-k_2)^2(k_1+q)^2(k_2+q)^2}.
\end{equation}
This integral can be reduced as
\begin{equation}
\begin{tikzpicture}[baseline=-\the\dimexpr\fontdimen22\textfont2\relax]
	\begin{feynman}[inline=(a0)]
		\vertex (a0);
		\vertex [right=2.4cm of a0](a2);
		\vertex [right=1.2cm of a0](c0);
		\vertex [above=0.33cm of c0](cu);
		\vertex [below=0.33cm of c0](cd);
		
		\diagram*[small]{
			(a2)--[bend left=28](a0)--[bend left=28](a2),
			(cu)--(cd)
		};
	\end{feynman}
\end{tikzpicture}\sim
\begin{tikzpicture}[baseline=-\the\dimexpr\fontdimen22\textfont2\relax]
	\begin{feynman}[inline=(a0)]
		\vertex (a0);
		\vertex [right=2cm of a0](a2);
		\diagram*[small]{
			(a2)--[bend left=30](a0)--[bend left=30](a2),
			(a2)--(a0)
		};
	\end{feynman}
\end{tikzpicture}\,+\,
\begin{tikzpicture}[baseline=-\the\dimexpr\fontdimen22\textfont2\relax]
	\begin{feynman}[inline=(a0)]
		\vertex (a0);
		\vertex [right=2.4cm of a0](a2);
		\vertex [right=1.2cm of a0](a1);
		\diagram*[small]{
			(a0)--[bend left=30](a1)--[bend left=30](a2),
			(a2)--[bend left=30](a1)--[bend left=30](a0)
		};
	\end{feynman}
\end{tikzpicture}\,+\,
\begin{tikzpicture}[baseline=-\the\dimexpr\fontdimen22\textfont2\relax]
	\begin{feynman}[inline=(a0)]
		\vertex (a0);
		\vertex [right=2cm of a0](a2);
		\vertex [right=1.2cm of a0](b0);
		\vertex [above=0.286cm of b0](bu);
		\diagram*[small]{
			(a2)--[bend left=30](a0)--[bend left=30](a2),
			(a0)--[bend right=20](bu)
		};
	\end{feynman}
\end{tikzpicture}\,+\,
\begin{tikzpicture}[baseline=-\the\dimexpr\fontdimen22\textfont2\relax]
	\begin{feynman}[inline=(a0)]
		\vertex (a0);
		\vertex [right=2cm of a0](a2);
		\vertex [right=0.8cm of a0](b0);
		\vertex [above=0.286cm of b0](bu);
		\diagram*[small]{
			(a2)--[bend left=30](a0)--[bend left=30](a2),
			(a2)--[bend left=20](bu)
		};
	\end{feynman}
\end{tikzpicture}.
\label{2-loop_reduce}
\end{equation}
All types of diagrams on the right-hand side of eq.~\ref{2-loop_reduce}, as well as the subdiagrams in other NLO perturbative diagrams in Appendix~\ref{dias} can recursively integrated and are equivalent to propagators, so the entire diagram can also be evaluated via Fourier transformation.

The results of eq.~\ref{pi_AB}, however, are not scheme-independent. The unrenormalized NLO perturbative diagrams involve an $1/\varepsilon^2$-pole, so different $\gamma^5$ schemes and subtraction schemes lead to a difference $\propto \alpha_s\, \text{Log}(-q^2/\mu^2)$. Fortunately, this difference is tiny and can be neglected in mass estimation. For nonperturbative contributions, the $\langle G^3\rangle$ diagrams involve infrared divergence, but they are canceled after summing up all diagrams, leaving only an $\varepsilon$-pole. As a result, for all nonperturbative contributions, the terms $\propto\text{Log}(-q^2/\mu^2)$ are scheme independent. We choose the commonly adopted $\overline{\text{MS}}$ scheme for subtraction. To minimize ambiguity related to $\gamma^5$, we adopt the BMHV scheme~\cite{feyncalc_10} for the perturbative diagrams, and replace $\gamma^\mu.\gamma^5$ with$(\gamma^\mu.\gamma^5 - \gamma^5.\gamma^\mu)/2$ in the four-quark currents. For the nonperturbative diagrams, we adopt NDR scheme~\cite{feyncalc_10} to avoid unnecessary and cumbersome calculations.

For {\small$J^r_d$}, {\small$J^r_c$}, {\small$J^r_{a,[27_S]}$}, and {\small$J^r_{f,[27_S]}$} ({\small$I_3^r$}, {\small$I_c^r$}, {\small$I^r_{4,27_S}$}, and {\small$I^r_{5,27_S}$}), the contributions from {\small$\langle \bar{q}q\rangle^2$}, {\small$\langle \bar{q}q\rangle\langle \bar{q}Gq\rangle$}, {\small$\langle \bar{q}Gq\rangle^2$}, and {\small$\langle GG\rangle\langle \bar{q}q\rangle^2$} (i.e., the condensates involving four quarks) vanish in the correlators. This occurs because the involved bare currents differ by $\gamma^5$, leading to the cancellation between relevant diagrams. Notably, for {\small$I_3^r$} and {\small$I_c^r$}, this cancellation is independent of the flavor structures. In some cases, the cancellation causes the {\small$\langle \bar{q}q\rangle^2$} and {\small$\langle \bar{q}q\rangle\langle \bar{q}Gq\rangle$} contributions to become relatively small, making the {\small$\langle G^3\rangle$} contribution comparable to the {\small$\langle \bar{q}q\rangle^2$} contribution.

For the dimension-8 quark condensate, the factorization procedure (vacuum saturation hypothesis) is ambiguous. Applying the equation of motion before or after factorization (EM-first or VS-first) yields different results, as discussed in Appendix~\ref{fac_d8}. Fortunately, in this study, using different procedures does not lead to significant differences in the mass estimations. Comparing the masses of four-quark states listed in this section and Appendix~\ref{em-first}, the differences are negligible.

For dimension-10 condensates, a similar ambiguity is also involved but is more complicated, and the EM-first procedure is difficult to apply. We simply choose the VS-first procedure for dimension-10 condensates in this paper, since the correlators are dominated by dimension-6 and -8 condensates except in certain cases where $\langle \bar{q}q\rangle^2$ and $\langle \bar{q}q\rangle\langle \bar{q}Gq\rangle$ contributions vanish.

The results of the correlators are quite tedious. We present them for $C_A=3$ with unspecified flavors in Appendix~\ref{ope_results}, where the VS-first procedure is adopted for dimension-8 and -10 condensates. The complete results are given in ref.~\cite{results}, which serves as a supplemental material.

\subsection{Mass Estimations for the Renormalized Currents}\label{sec_3_mass}
To extract mass information from the two-point function, QCD sum rules utilize the dispersion relation along with integral transformations such as Borel (Laplace) transformation~\cite{qsr} or Gauss-Weierstrass transform~\cite{gaussian}. Under the ``pole $+$ continuum'' ansatz, the spectrum is approximated as $\delta(s-m^2)+\theta(s-s_0)\rho(s)$, where $s_0$ is the continuum threshold and $\rho(s)$ is the continuum spectrum. Consequently, for the moments, we have 
\begin{equation}
	\mathcal{M}^n(\tau,s_0)=\frac{1}{\pi}\int_0^{s_0}ds\ s^n e^{-\tau s}\, \text{Im}\Pi_{AB}(s)\propto m^{2n}e^{-\tau m^2},
	\label{moment}
\end{equation}
where $\text{Im}\Pi_{AB}(s)$ is the imaginary part of $\Pi_{AB}(s)$, which contains information about the spectrum. The mass of the lowest resonance can then be derived from the ratio
\begin{equation}
	\mathcal{R}^n(\tau,s_0)=\frac{\mathcal{M}^{n+1}(\tau,s_0)}{\mathcal{M}^n(\tau,s_0)}=m^2.
	\label{ratio}
\end{equation}
Additionally, the renormalization group improved OPE is obtained by setting $\mu^2=1/\tau$~\cite{qsr_laplace}.

For dimension-6, -8, and -10 four-quark condensates, the factorization procedure (vacuum saturation hypothesis) introduces large uncertainties. Thus, the replacements $\langle \bar{q}q\rangle^2\rightarrow \eta_6 \langle \bar{q}q\rangle^2$, $\langle \bar{q}q\rangle\langle \bar{q}Gq\rangle \rightarrow \eta_8 \langle \bar{q}q\rangle\langle \bar{q}Gq\rangle$, and $\langle \mathcal{O}_{10}\rangle \rightarrow \eta_{10} \langle \mathcal{O}_{10}\rangle$ are applied when estimating the mass, where the $\eta_n$ are factorization deviation factors. The precise values of $\eta_6$, $\eta_8$, and $\eta_{10}$ are unclear; they are estimated to be around $2-5$~\cite{deviation_factors}. We consider the case that $\eta_n=1$, as well as the case that $\eta_6=3$~\cite{deviation_factor_d6} and $\eta_8=\eta_{10}=5$, to derive a conservative range of the masses. The QCD parameters used in the calculation are listed in Table~\ref{con_v}, with $\alpha_s=\frac{4\pi}{9\text{Log}(\mu^2/\Lambda^2_\text{QCD})}$ for $n_f=3$ and $\Lambda_\text{QCD}=0.353\text{GeV}$~\cite{1-+}.


\begin{table}[t!]
	\centering
	\caption{Quark masses~\cite{pdg} and the values of condensates~\cite{qcd_sum_review} at $\mu=1\text{GeV}$; $q=u,d$ here.\label{con_v}}
	\renewcommand{\arraystretch}{1.2}
	\stackengine{0pt}{
	\begin{tabular}{|w{c}{1.4cm}|w{c}{2.9cm}|w{c}{3.05cm}|w{c}{2.4cm}|w{c}{2.9cm}|}
		\noalign{\hrule height 1pt}
		$m_u $&$m_d$&$m_s$&$\langle \bar{q}q\rangle$&$\langle\bar{s}s\rangle$\\
		\noalign{\hrule height 0.2pt}
		$2.2\text{MeV}$&$4.7\text{MeV}$&$93\text{MeV}$&$(-0.276)^3\text{GeV}^3$&$0.8(-0.276)^3\text{GeV}^3$\\
		\noalign{\hrule height 1pt}
	\end{tabular}
	}{
	\begin{tabular}{|w{c}{1.4cm}|w{c}{2.9cm}|w{c}{3.05cm}|w{c}{2.4cm}|}
		$\langle GG\rangle$&$\langle \bar{q}Gq\rangle$&$\langle\bar{s}Gs\rangle$&$\langle G^3\rangle$\\
		\noalign{\hrule height 0.2pt}
		$0.07\text{GeV}^4$&$0.8(-0.276)^3\text{GeV}^5$&$0.8^2(-0.276)^3\text{GeV}^5$&$8\times 0.07 \text{GeV}^6$\\
		\noalign{\hrule height 1pt}
	\end{tabular}
	}{U}{l}{F}{F}{S}
	\renewcommand{\arraystretch}{1}
\end{table}

Within each current, states belonging to the same $su(3)$-flavor representation exhibit similar masses. This is because only the $s$ quark contributes a non-negligible difference, but it is small compared to the masses of the hadrons. After taking the ratio as in eq.~\ref{ratio}, the difference becomes tiny and is overwhelmed by the factorization deviation factors. A comparison of the masses for $J^r_{4,8_S}$ is presented at the beginning of Appendix~\ref{sec_mass}. For each current, tables~\ref{mass} and \ref{mass_eta} list the masses corresponding to the flavor configurations with the highest weights in each $su(3)$-flavor representation. 

Two extra parameters, $\tau$ and $s_0$, are introduced in eq.~\ref{moment}. They are commonly constrained within the Borel Window~\cite{qcd_book}. The first constraint is $\mathcal{M}^0(\tau,s_0)/\mathcal{M}^0(\tau,\infty)\geq50\%$, so that the resonance contribution dominates; the second constraint is that the contribution received from the highest dimensional condensate is less than $10\%$, to ensure the OPE converges.

However, the Borel Window always exists. For sufficiently large $s_0$, the first constraint must be satisfied. Since $\int_0^{s_0}ds\ s^n e^{-\tau s}\sim1/\tau^{(n+1)}$, for sufficiently small $\tau$, the OPE is dominated by the perturbative contribution, and the second constraint will also be satisfied in general.

\begin{figure}[t!]
	\begin{subfigure}{0.45\textwidth}
		\includegraphics[width=\textwidth]{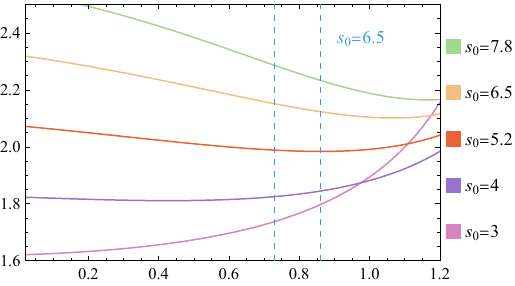}
		\caption{$\sqrt{\mathcal{R}^0}$ for $J_{4,1_S}^r$.}
		\label{r0_t_j4_1_s_no_factors}
	\end{subfigure}
	\hspace*{0.1\textwidth}	
	\begin{subfigure}{0.39\textwidth}
		\includegraphics[width=\textwidth]{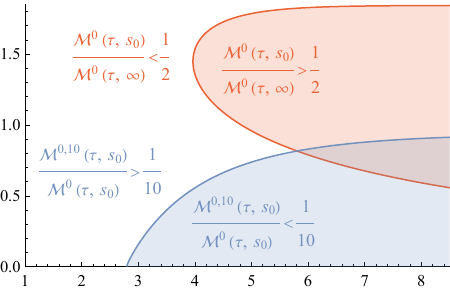}
		\caption{Borel Window for $J_{4,1_S}^r$.}
		\label{borel_window_j4_1_s}
	\end{subfigure}
	\caption{Mass estimation and Borel Window for {\small$J_{4,1_S}^r$}. The left figure shows {\small$\sqrt{\mathcal{R}^0}(\text{GeV}^2)$} versus {\small$\tau(\text{GeV}^{-2})$} for different values of {\small$s_0(\text{GeV}^2)$}, with vertical dashed lines indicating the Borel Window when $s_0=6.5$. The overlap region in the right figure gives the Borel Window; the vertical and horizontal axes corresponding to {\small$\tau(\text{GeV}^{-2})$} and {\small$s_0(\text{GeV}^2)$} respectively. The {\small$\mathcal{M}^{n,10}$} means the contribution of dimension-10 condensates in {\small$\mathcal{M}^n$}, which are the highest-dimensional condensates in our calculation.}
	\label{fig_borel_window}
\end{figure}

\begin{figure}[t!]
	\begin{subfigure}{0.45\textwidth}
		\includegraphics[width=\textwidth]{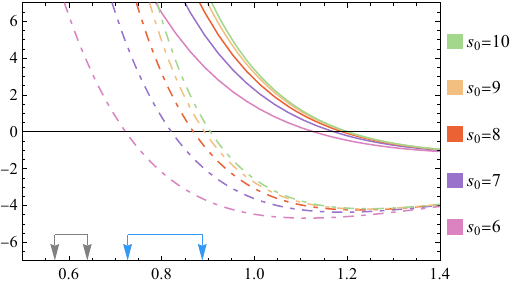}
		\caption{$\mathcal{M}^0$.}
		\label{m0_tau_j2_27_s}
	\end{subfigure}
	\hspace*{\fill}	
	\begin{subfigure}{0.45\textwidth}
		\includegraphics[width=\textwidth]{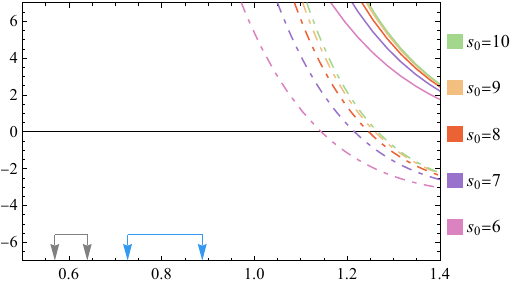}
		\caption{$\mathcal{M}^1$.}
		\label{m1_tau_j2_27_s}
	\end{subfigure}\\[0cm]	
	\begin{subfigure}{0.45\textwidth}
		\includegraphics[width=\textwidth]{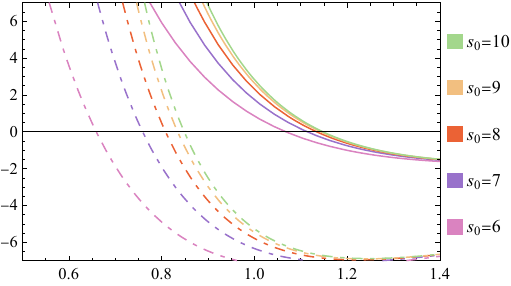}
		\caption{$\mathcal{M}^0$ without dimension-10 condensates.}
		\label{m0_tau_j2_27_s_t8}
	\end{subfigure}
	\hspace*{\fill}	
	\parbox[b][4.8cm][t]{7.09cm}{\caption{{\small$\mathcal{M}^n(10^5\text{GeV}^8)$} versus $\tau(\text{GeV}^{-2})$ for $J^r_{2,[27_S]}$ with different values of {\small$s_0(\text{GeV}^2)$}. The dot-dashed and solid lines represent cases with and without including $\eta_n$, respectively. The arrows indicate the Borel Windows: gray for including $\eta_n$ with {\small$s_0=9\text{GeV}^2$}; blue for {\small$s_0=7\text{GeV}^2$} without $\eta_n$. The zeros of {\small$\mathcal{M}^1$} are farther from the Borel Window than the zeros of {\small$\mathcal{M}^0$}.}
	\label{fig_m01_tau}}
\end{figure}

\begin{figure}[ht!]
	\begin{subfigure}{0.45\textwidth}
		\includegraphics[width=\textwidth]{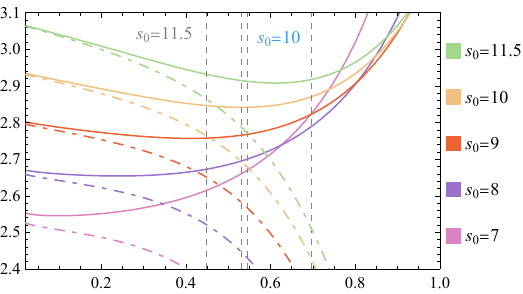}
		\caption{$\sqrt{\mathcal{R}^0}$ for $J_{d,[27_S]}^r$.}
		\label{r0_tau_jd_27_s}
	\end{subfigure}
	\hspace*{\fill}	
	\begin{subfigure}{0.45\textwidth}
		\includegraphics[width=\textwidth]{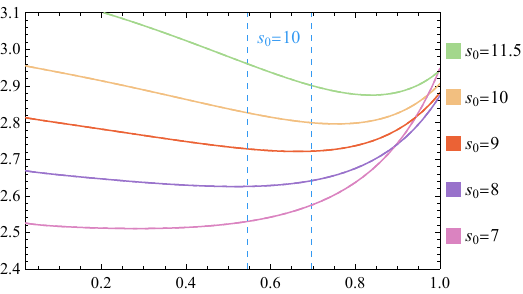}
		\caption{$\sqrt{\mathcal{R}^1}$ for $J_{d,[27_S]}^r$.}
		\label{r1_tau_jd_27_s}
	\end{subfigure}	\\[0.1cm]
	\begin{subfigure}{0.45\textwidth}
		\includegraphics[width=\textwidth]{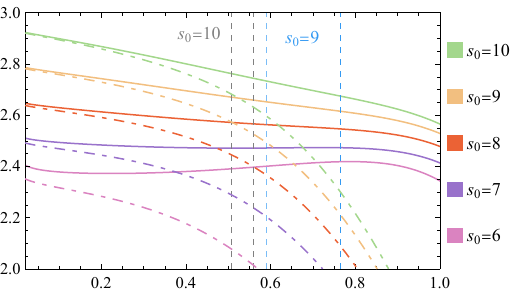}
		\caption{$\sqrt{\mathcal{R}^0}$ for $I_{3,[27_S]}^r$.}
		\label{r0_tau_i3_27_s}
	\end{subfigure}
	\hspace*{\fill}	
	\begin{subfigure}{0.45\textwidth}
		\includegraphics[width=\textwidth]{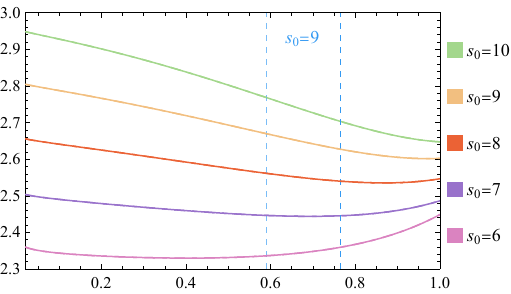}
		\caption{$\sqrt{\mathcal{R}^1}$ for $I_{3,[27_S]}^r$.}
		\label{r1_tau_i3_27_s}
	\end{subfigure}
	\caption{{\small$\sqrt{\mathcal{R}^n}(\text{GeV})$} versus {\small$\tau(\text{GeV}^{-2})$} for {\small$J_{d,[27_S]}^r$} and {\small$I_{3,[27_S]}^r$} with different values of {\small$s_0(\text{GeV}^2)$}. The dot-dashed and solid lines represent the cases with and without including $\eta_n$, respectively. The vertical dashed lines indicate the Borel Window, with gray and blue referring to including and not including $\eta_n$, respectively. The $s_0$ for Borel windows are shown beside their corresponding vertical lines in matching colors. The four-quark condensates contribution is absent in these cases, while the $m\langle\bar{q}q\rangle^3$ contribution only affects {\small$\mathcal{M}^0$} and {\small$\mathcal{R}^0$}. The {\small$\mathcal{M}^0$} suffers from the same problem as in figure~\ref{fig_m01_tau}.}
	\label{fig_r_tau}
\end{figure}

As the case shown in figure~\ref{fig_borel_window}, the Borel Window exists starting from {\small$s_0\geq5.8$}. The {\small$\sqrt{\mathcal{R}^0}(\tau)$} is stable when $s_0\simeq4$. We find that for all the cases in this paper, the $s_0$ that makes {\small$\sqrt{\mathcal{R}^0}$} over $\tau$ stable is smaller than the threshold $s_0$ where the Borel Window begins. Therefore, we choose the $s_0$ close to this threshold to give the corresponding range of $\tau$ as the Borel Window. To reduce the bias from parameter selection, the stability criteria~\cite{qcd_sum_review} also adopted, allowing mass estimation without parameter adjustment.

\begin{figure}[t!]
	\begin{subfigure}{0.45\textwidth}
		\includegraphics[width=\textwidth]{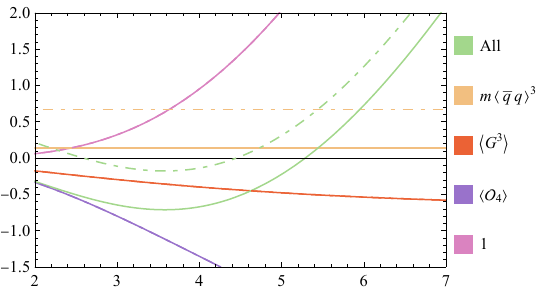}
		\caption{$J_{d,[27_S]}^r$, scaled up by $10^5$.}
		\label{c_ratio_jd_27_s_0.2}
	\end{subfigure}
	\hspace*{\fill}	
	\begin{subfigure}{0.45\textwidth}
		\includegraphics[width=\textwidth]{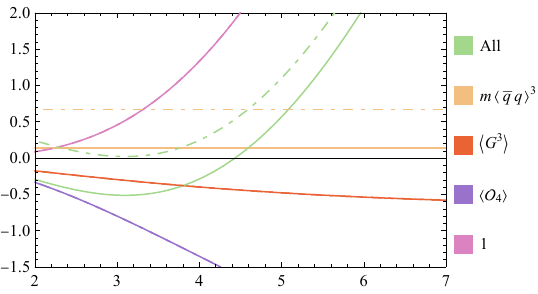}
		\caption{$I_{3,[27_S]}^r$, scaled up by $\frac{5}{2}\times10^4$.}
		\label{c_ratio_jd_27_s_0.5}
	\end{subfigure}
	\caption{Contribution of each condensate in {\small$\mathcal{M}^0(\text{GeV}^8)$} as a function of {\small$s_0(\text{GeV}^2)$} for {\small$\tau=0.5\text{GeV}^{-2}$}. Here the ``1'' denotes the perturbative contribution. The dot-dashed and solid lines represent the cases with and without $\eta_n$, respectively. In both cases, condensates involving four quarks do not contribute. The perturbative contribution is affected by the $\gamma^5$-scheme, leading to significant differences in the zeros of {\small$\mathcal{M}^0$} for {\small$J_{d,[27_S]}^r$} and {\small$I_{3,[27_S]}^r$}.}
	\label{fig_c_list}
\end{figure}

\begin{figure}[ht!]
	\begin{subfigure}{0.42\textwidth}
		\includegraphics[width=\textwidth]{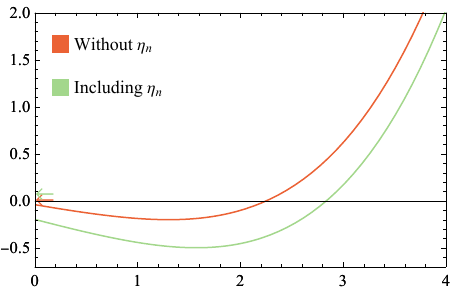}
		\caption{$J_{2,[27_S]}^r$, scale up by $10^3$.}
		\label{im_j2_27_s}
	\end{subfigure}
	\hspace*{1.5cm}	
	\begin{subfigure}{0.42\textwidth}
		\includegraphics[width=\textwidth]{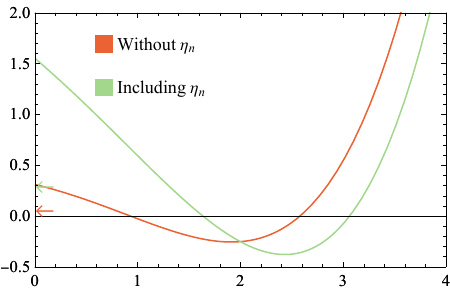}
		\caption{$J_{3,1_A}^r$, scale up by $10^3$.}
		\label{im_j3_1_a}
	\end{subfigure}\\[0cm]
	\begin{subfigure}{0.42\textwidth}
		\includegraphics[width=\textwidth]{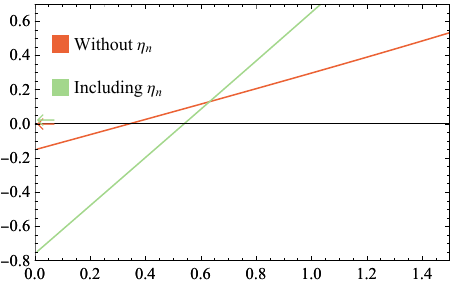}
		\caption{$J_{a,[8_S]}^r$, scale up by $10^{-2}$.}
		\label{im_ja_8_s}
	\end{subfigure}
	\hspace*{1.5cm}	
	\parbox[b][5.2cm][t]{7.06cm}{
		\caption{{\small$\text{Im}\Pi(s)(\text{GeV}^8)$} versus {\small$s(\text{GeV}^2)$} when {\small $\mu=1\text{GeV}$}. The term {\small$\propto\delta(s)$} cannot be plotted; the arrows indicate the coefficient $c$ in {\small$c\,\delta(s)$}. The property of {\small$\text{Im}\Pi(s)$} strongly affects the mass estimation, even though it does not directly give valid spectrum information. The currents {\small$J_{2,[27_S]}^r$} and {\small$J_{3,1_A}^r$} are hard to couple to states with masses {\small$\lesssim \sqrt{3}\text{GeV}$}, while {\small$J_{a,[8_S]}^r$} is hard to couple to a state with mass {\small$\lesssim \sqrt{0.5}\text{GeV}$}.\label{fig_im}}}
\end{figure}

In some cases, negative contributions from certain condensates cause $\mathcal{M}^0(\tau,s_0)$ to have a zero near the Borel Window. Consequently, the $\mathcal{R}^0(\tau)$ develops a pole and the mass estimation becomes delicate. We find that the zero of $\mathcal{M}^1$ appears at a larger $\tau$ and farther from the Borel Window compared to the zero of $\mathcal{M}^0$. In these cases, $\mathcal{R}^1$ gives more reliable mass estimations. The typical figures of $\mathcal{M}^n$ are shown in figure~\ref{fig_m01_tau}. This property appears related to the dimension-10 condensate, since the dimension-10 condensate only contributes to $\mathcal{M}^0(\tau,s_0)$. However, as shown in figure~\ref{m0_tau_j2_27_s_t8}, the zero of $\mathcal{M}^0(\tau,s_0)$ is still closer to the Borel Window even if we discard the contribution of dimension-10 condensates.

\begin{figure}[ht!]
	\begin{subfigure}{0.45\textwidth}
		\includegraphics[width=\textwidth]{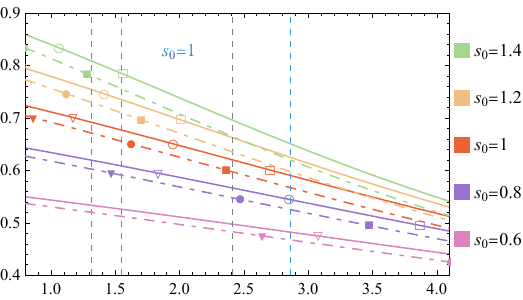}
		\caption{$\sqrt{\mathcal{R}^0}(\tau)$ for $J^r_{b,[8_A]}$.}
		\label{r0_t_jb_8_a}
	\end{subfigure}
	\hspace*{\fill}	
	\begin{subfigure}{0.45\textwidth}
		\includegraphics[width=\textwidth]{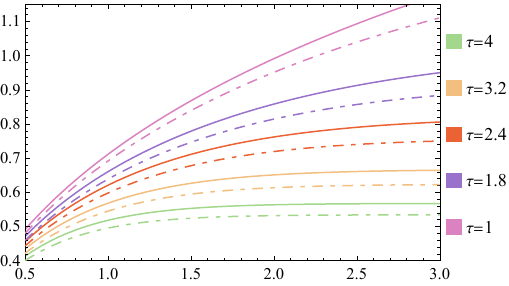}
		\caption{$\sqrt{\mathcal{R}^0}(s_0)$ for $J^r_{b,[8_A]}$.}
		\label{r0_s0_jb_8_a}
	\end{subfigure}\\[0.3cm]
	\begin{subfigure}{0.45\textwidth}
		\includegraphics[width=\textwidth]{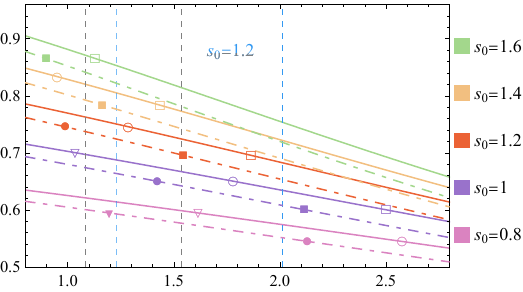}
		\caption{$\sqrt{\mathcal{R}^0}(\tau)$ for $J^r_{b,1_A}$.}
		\label{r0_t_jb_1_a}
	\end{subfigure}
	\hspace*{\fill}	
	\begin{subfigure}{0.45\textwidth}
		\includegraphics[width=\textwidth]{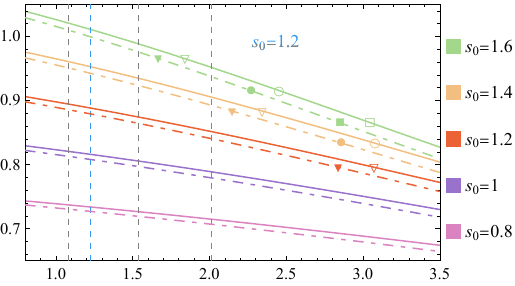}
		\caption{$\sqrt{\mathcal{R}^1}(\tau)$ for $J^r_{b,1_A}$.}
		\label{r1_t_jb_1_a}
	\end{subfigure}
	\caption{Mass estimation for $J^r_b$. Figure~\ref{r0_s0_jb_8_a} shows {\small$\sqrt{\mathcal{R}^n}(\text{GeV})$} versus {\small$s_0(\text{GeV}^2)$} for different values of {\small$\tau(\text{GeV}^{-2})$}, while other figures show {\small$\sqrt{\mathcal{R}^n}(\text{GeV})$} versus {\small$\tau(\text{GeV}^{-2})$} for different values of {\small$s_0(\text{GeV}^2)$}. The lines follow the same conventions as in figure~\ref{fig_r_tau}. The corresponding $s_0$ for Borel windows are overlapped here to indicate they are identical. The markers on the curves correspond to the constraint {\small$s_0=(\sqrt{\mathcal{R}^n}+\Lambda)^2$}, with filled and empty marks corresponding to including and not including $\eta_n$, respectively. The triangle, circle, and square markers refer to choosing $\Lambda=0.3$, $0.35$, and $0.4\text{GeV}$ respectively. The markers outside the visible range are omitted.}
	\label{fig_r0_jb}
\end{figure}

For these cases, the negativity of $\mathcal{M}^n(\tau,s_0)$ in a certain range of $s_0$ indicates that the $\text{Im}\Pi(s)$ is negative in a certain range of $s$, as shown in figure~\ref{fig_im}. Consequently, the currents cannot couple to the state with $m^2$ in this range. Nevertheless, for sufficiently large $s_0$, $\mathcal{M}^n(\tau,s_0)$ is positive and the Borel Window exists. Since $\text{Im}\Pi(s)$ does not directly provide valid spectrum information, we still keep the corresponding results.

As previously discussed,  there is a one-to-one correspondence between the renormalized tetraquark and molecule operators, which is confirmed by the masses listed in Tables~\ref{mass} and~\ref{mass_eta}. The only exceptions are $J_d^r$/$I_3^r$ and $J^r_c$/$I^r_c$, because for these correlators, the perturbative and nonperturbative contributions cancel each other out when $s_0$ is around $4-5\text{GeV}^2$, as shown in figure~\ref{fig_c_list}. The difference $\propto d-4$ between these tetraquark and molecule operators then leads to a non-negligible difference in their masses. Typical plots for $J_d^r$ and $I_3^r$ are shown in figure~\ref{fig_r_tau}. For other four-quark currents, we present only the figures related to tetraquarks. All figures related to mass estimation are provided in supplementary material~\cite{results}.

Based on the properties of $\text{Im}\Pi(s)$, the currents can be sorted into six categories:
\begin{enumerate}
	\item $J^r_{2,27_S}$, $J^r_3$, and $J^r_4$ ($I^r_{1,27_S}$, $I^r_b$, and $I^r_2$): $\text{Im}\Pi(s)$ is negative in a certain range of $s$, as the cases shown in figures~\ref{im_j2_27_s} and \ref{im_j3_1_a}. The typical masses $\gtrsim2\text{GeV}$.
	\item $J^r_{a,27_S}$ and $J^r_d$ ($I^r_{4,27_S}$ and $I^r_3$): Similar to category 1, but the condensates involving four quarks are absent.
	\item $J^r_{a,8_S}$, $J^r_{a,1_S}$ and $J^r_e$ ($I^r_{4,8_S}$, $I^r_{4,1_S}$, and $I^r_e$): $\text{Im}\Pi(s)$ is positive, or negative only in a tiny range of $s$ around $0$, as the case shown in figure~\ref{im_ja_8_s}. The typical masses $\lesssim2\text{GeV}$.
	\item $J^r_1$, $J^r_{2,8_S}$, $J^r_{2,1_S}$, and $J^r_b$ ($I^r_a$, $I^r_{1,8_S}$, $I^r_{1,1_S}$, and $I^r_d$): $\text{Im}\Pi(s)$ is positive; $\mathcal{R}^n$ changes monotonically with respect to $\tau$ and $s_0$, and no plateaus exist. The stability criterion is not applicable here. We choose $\tau$ and $s_0$ such that $s_0=(m+\Lambda_\text{QCD})^2$, where $m=\sqrt{\mathcal{R}^n}$, since for category 3, the Borel Window gives $\sqrt{s_0}\sim m+\Lambda_\text{QCD}$. The states in this category correspond to nonets with masses $\lesssim1\text{GeV}$; the typical figures are shown in figure~\ref{fig_r0_jb}. However, the results rely on the choice of $s_0$; bias may be introduced.
	\item $J^r_c$/$I^r_c$: Similar to category 4, but the condensates involving four quarks are absent.
	\item $J^r_f$/$I^r_5$: $\text{Im}\Pi(s)$ is negative in a large range of $s$. No reasonable results can be derived. Blindly applying Laplace sum rules obtains masses $\gtrsim4\text{GeV}$.
\end{enumerate}
The division of these categories is not strict. A current belonging to one category may shift to another if the factorization deviation factors $\eta_n$ are changed.


\renewcommand{\arraystretch}{1.3}
\begin{table}[t!]
	\centering
	\caption{The four-quark mass estimations correspond to $\eta_n=1$. The header row specifies the flavors, while the first column lists the currents. A blank cell indicates the corresponding current vanishes, whereas a cell marked with ``$\times$'' means no reasonable result can be derived, corresponds to category 6. Values enclosed in parentheses are determined using stability criteria, where $\sqrt{\mathcal{R}^n}(\tau)$ is most stable. Other values are derived from the values in the Borel Window or by the constraint {\small$s_0=(\sqrt{\mathcal{R}^n}+\Lambda_\text{QCD})^2$}. In some cases, the values are hard to ascertain based on the curves of $\sqrt{\mathcal{R}^n}(\tau)$; refer to the figures in Appendix~\ref{sec_mass_r} for details. \label{mass}}
	\stackengine{0pt}{
		\begin{tabular}{|w{c}{0.7cm}!{\vrule width 1pt}w{c}{2.2cm}|w{c}{2.2cm}|w{c}{2.2cm}|w{c}{2.2cm}|w{c}{2.2cm}|}
			\noalign{\hrule height 1pt}
			$J_1^r$&&&&$0.7-0.8$&$0.8-1$\\
			\noalign{\hrule height 0.2pt}
			$J_2^r$&$2.3\,(2.2)$&$0.6-0.8$&$0.9-1.1$&&\\
			\noalign{\hrule height 0.2pt}
			$J_3^r$&&&&$2.3$&$2.3-2.4$\\
			\noalign{\hrule height 0.2pt}
			$J_4^r$&$2.4\,(1.7)$&$2.3\,(1.6)$&$2.2\,(1.6)$&&\\
			\noalign{\hrule height 1pt}
			$J_a^r$&{\small2.4-2.6\,(2.2-2.5)}&$1.2\,(1-1.2)$&$1.2-1.3\,(0.9)$&&\\
			\noalign{\hrule height 0.2pt}
			$J_b^r$&&&&$0.6-0.8$&$0.7-0.9$\\
			\noalign{\hrule height 0.2pt}
			$J_c^r$&&&&$1.6-1.8$&$1.6-1.8$\\
			\noalign{\hrule height 0.2pt}
			$J_d^r$&{\small2.8-2.9\,(2.6-2.8)}&{\small2.7-3\,(2.7-2.8)}&{\small2.8-3\,(2.7-2.8)}&&\\
			\noalign{\hrule height 0.2pt}
			$J_e^r$&&&&$1.9-2.1\,(1.3)$&{\small2-2.1\,(1.3-1.4)}\\
			\noalign{\hrule height 0.2pt}
			$J_f^r$&$\times$&$\times$&$\times$&&\\
			\noalign{\hrule height 1pt}
			$I_1^r$&$2.2\,(2.1)$&$0.6-0.8$&$0.9-1$&&\\
			\noalign{\hrule height 0.2pt}
			$I_2^r$&$2.3\,(1.7)$&$2.3\,(1.7)$&$2.1\,(1.6)$&&\\
			\noalign{\hrule height 0.2pt}
			$I_3^r$&$2.7\,(2.4-2.5)$&$2.6\,(2.4-2.6)$&$2.6\,(2.3-2.6)$&&\\
			\noalign{\hrule height 0.2pt}
			$I_4^r$&{\small2.4-2.5\,(2.2\,-2.4)}&$1.2\,(1)$&$1.4\,(1)$&&\\
			\noalign{\hrule height 0.2pt}
			$I_5^r$&$\times$&$\times$&$\times$&&\\
			\noalign{\hrule height 1pt}
			$I_a^r$&&&&$0.7-0.8$&$0.8-1$\\
			\noalign{\hrule height 0.2pt}
			$I_b^r$&&&&$2.1-2.3$&$1.6-1.8$\\
			\noalign{\hrule height 0.2pt}
			$I_c^r$&&&&$1.6-1.8$&$1.7-1.9$\\
			\noalign{\hrule height 0.2pt}
			$I_d^r$&&&&$0.6-0.8$&$0.6-0.9$\\
			\noalign{\hrule height 0.2pt}
			$I_e^r$&&&&$2.1\,(1.3)$&$2\,(1.4)$\\
			\noalign{\hrule height 1pt}
		\end{tabular}
	}{
		\begin{tabular}{!{\vrule width 1pt}w{c}{2.2cm}|w{c}{2.2cm}|w{c}{2.2cm}|w{c}{2.2cm}|w{c}{2.2cm}|}
			\noalign{\hrule height 1pt}
			$[27_S]$&$[8_S]$&$1_S$&$[8_A]$&$1_A$
		\end{tabular}
	}{O}{r}{F}{F}{S}
\end{table}

\renewcommand{\arraystretch}{1.3}
\begin{table}[t!]
	\centering
	\caption{The four-quark mass estimations correspond to $\eta_6=3$ and $\eta_8=\eta_{10}=5$. The same conventions as in table~\ref{mass} are adopted here. In some cases, the values are hard to ascertain based on the curves of $\sqrt{\mathcal{R}^n}(\tau)$; refer to the figures in Appendix~\ref{sec_mass_r} for details.\label{mass_eta}}
	\stackengine{0pt}{
		\begin{tabular}{|w{c}{0.7cm}!{\vrule width 1pt}w{c}{2.2cm}|w{c}{2.2cm}|w{c}{2.2cm}|w{c}{2.2cm}|w{c}{2.2cm}|}
			\noalign{\hrule height 1pt}
			$J_1^r$&&&&$0.7-0.8$&$1-1.1$\\
			\noalign{\hrule height 0.2pt}
			$J_2^r$&$2.7\,(2.5)$&$0.6-0.8$&$1-1.1$&&\\
			\noalign{\hrule height 0.2pt}
			$J_3^r$&&&&$2.5$&$2.6$\\
			\noalign{\hrule height 0.2pt}
			$J_4^r$&$2.7\,(2.2)$&$2.7\,(2.2)$&$2.6\,(2.2)$&&\\
			\noalign{\hrule height 1pt}
			$J_a^r$&$2.6$&{\small1.4-1.5\,(1.3-1.5)}&$1.5\,(1.2-1.4)$&&\\
			\noalign{\hrule height 0.2pt}
			$J_b^r$&&&&$0.6-0.8$&$0.7-0.9$\\
			\noalign{\hrule height 0.2pt}
			$J_c^r$&&&&$2-2.2$&$2-2.2$\\
			\noalign{\hrule height 0.2pt}
			$J_d^r$&$2.8-3$&$2.9-3$&$2.9-3$&&\\
			\noalign{\hrule height 0.2pt}
			$J_e^r$&&&&$2.4\,(1.9)$&$2.4\,(2)$\\
			\noalign{\hrule height 0.2pt}
			$J_f^r$&$\times$&$\times$&$\times$&&\\
			\noalign{\hrule height 1pt}
			$I_1^r$&$2.7\,(2.4)$&$0.7-0.8$&$1.1-1.2$&&\\
			\noalign{\hrule height 0.2pt}
			$I_2^r$&$2.7\,(2.2)$&$2.7\,(2.2)$&$2.6\,(2.1)$&&\\
			\noalign{\hrule height 0.2pt}
			$I_3^r$&$2.7-2.8$&$2.7-2.9$&$2.7-2.8$&&\\
			\noalign{\hrule height 0.2pt}
			$I_4^r$&$2.4-2.5$&$1.4\,(1.3)$&$1.6\,(1.2)$&&\\
			\noalign{\hrule height 0.2pt}
			$I_5^r$&$\times$&$\times$&$\times$&&\\
			\noalign{\hrule height 1pt}
			$I_a^r$&&&&$0.7-0.9$&$0.9-1.1$\\
			\noalign{\hrule height 0.2pt}
			$I_b^r$&&&&$2.1-2.6$&$2-2.2$\\
			\noalign{\hrule height 0.2pt}
			$I_c^r$&&&&$2-2.2$&$2-2.3$\\
			\noalign{\hrule height 0.2pt}
			$I_d^r$&&&&$0.6-0.8$&$0.7-0.9$\\
			\noalign{\hrule height 0.2pt}
			$I_e^r$&&&&$2.4\,(1.8)$&$2.4\,(1.9-2.2)$\\
			\noalign{\hrule height 1pt}
		\end{tabular}
	}{
		\begin{tabular}{!{\vrule width 1pt}w{c}{2.2cm}|w{c}{2.2cm}|w{c}{2.2cm}|w{c}{2.2cm}|w{c}{2.2cm}|}
			\noalign{\hrule height 1pt}
			$[27_S]$&$[8_S]$&$1_S$&$[8_A]$&$1_A$
		\end{tabular}
	}{O}{r}{F}{F}{S}
\end{table}

\renewcommand{\arraystretch}{1.2}
\begin{table}[ht!]
	\centering
	\caption{The allowed decay channels of four-quark states. For specific flavor structures, some channels may not be allowed. \label{decay_r}}
	\begin{tabular}{|w{c}{0.035\textwidth}|*{10}{w{c}{0.063\textwidth}|}}
		\noalign{\hrule height 1pt}
		&$J_1^r(I_a^r)$&$J_2^r(I_1^r)$&$J_3^r(I_b^r)$&$J_4^r(I_2^r)$&$J_a^r(I_4^r)$&$J_b^r(I_d^r)$&$J_c^r(I_c^r)$&$J_d^r(I_3^r)$&$J_e^r(I_e^r)$&$J_f^r(I_5^r)$\\[2pt]
		\noalign{\hrule height 0.2pt}
		\multirow{2}*{$J^P$}&\multirow{2}*{$0^-,0^+$}&\multirow{2}*{$0^-,0^+$}&$0^-,0^+$&$0^-,0^+$&$0^-,0^+$&$0^-,0^+$&$0^-,0^+$&$0^-,0^+$&$0^-,0^+$&$0^-,0^+$\\
		&&&$1^-,1^+$&$1^-,1^+$&$1^-,2^+$&$1^-,2^+$&$1^-,1^+$&$1^-,1^+$&$1^-,2^+$&$1^-,2^+$\\
		\noalign{\hrule height 1pt}			
	\end{tabular}
\end{table}
\renewcommand{\arraystretch}{1}


A precise mass estimation based solely on the curves of $\sqrt{\mathcal{R}^n}(\tau)$ is not feasible; the results listed in tables.~\ref{mass} and \ref{mass_eta} are thus not decisive. For category 1, the $\sqrt{\mathcal{R}^0}(\tau)$ exhibits a pole near the Borel Window as discussed previously. The $\sqrt{\mathcal{R}^0}$ and $\sqrt{\mathcal{R}^1}$ are nearly identical at $\tau\sim0$ as shown in figure~\ref{fig_r_tau}, but $\sqrt{\mathcal{R}^0}$ becomes steep at $\tau\sim1$, which increases the uncertainty. Therefore, we choose $\sqrt{\mathcal{R}^1}$ to derive the masses for this category. For $J_3^r$/$I_b^r$, the $\sqrt{\mathcal{R}^n}(\tau)$ has no plateau, so we choose $\tau$ and $s_0$ such that $s_0=(m+\Lambda_\text{QCD})^2$, the same as in categories 4 and 5. 

For categories 2 and 5, we keep the results derived from $\sqrt{\mathcal{R}^0}$ even though some of them suffer from the same problem as in category 1. Since the $m\langle\bar{q}q\rangle^3$ contribution is absent in the $\sqrt{\mathcal{R}^1}$, it is necessary to keep $\sqrt{\mathcal{R}^0}$ to investigate the effect of factorization deviation.

The values in tables~\ref{mass} and \ref{mass_eta} give wide ranges of masses for some currents. The first source of uncertainty arises from the factorization deviation factors $\eta_n$. The second arises from the method used to estimate the mass, as stability criteria and values within the Borel Window can yield different results. Despite these uncertainties, we can still draw some conclusions.

Naively, a four-quark molecule state denoted by $\overline{\Psi}_{f_1}\Gamma_A \Psi_{f_2}\, \overline{\Psi}_{f_3}\Gamma_B \Psi_{f_4}$ mainly decays into $\bar{f_1}f_2$ and $\bar{f_3}f_4$ mesons. Based on the operators listed in Section~\ref{sec2_2}, we can list the $J^P$ for the decay channels of four-quark states, as shown in table~\ref{decay_r}. In practice, $C_A=3$, eqs.~\ref{mole_s_basis} and \ref{mole_a_basis} imply that the $1^-$ and $2^+$ decay channels are rare for $J_a^r$/$I_4^r$, while for $J_e^r$/$I_e^r$, the $0^-$ and $0^+$ decay channels are rare. On the other hand, $I_e^r\propto I_c$ approximately and does not mix with hybrid-like operators, indicating that $J_e^r$ may primarily be a four-quark molecule.

The $f_0(500)$, $K^*_0(700)$, $f_0(980)$, and $a_0(980)$ are commonly identified as a four-quark nonet. Based on table~\ref{mass}, the $J_1^r$/$I_a^r$, $J_b^r$/$I_d^r$, $J^r_{2,8_S}$/$I^r_{1,8_S}$, and $J^r_{2,1_S}$/$I^r_{1,1_S}$ are all possible interpretations. Additionally, table~\ref{decay_r} provides further support for $J^r_1$/$I_a^r$, $J^r_{2,8_S}$/$I^r_{1,8_S}$, and $J^r_{2,1_S}$/$I^r_{1,1_S}$, since only the $0^-$ decay channel has been observed for these mesons~\cite{pdg}. For other four-quark states, reliable interpretations cannot be derived due to significant uncertainties in their masses. Nevertheless, some properties can still be listed:

\begin{enumerate}
	\item The $a_0(1450)$, $K^*_0(1430)$, and two of the $f_0(1370)$, $f_0(1500)$, and $f_0(1710)$ are commonly identified as another four-quark nonet~\cite{pdg}. The masses of $J^r_{a,8_S}$/$I^r_{4,8_S}$ and $J^r_{a,1_S}$/$I^r_{4,1_S}$, as well as the lower-bound masses of $J^r_e$/$I^r_e$ are around this range.

	\item Since the $a_0(980)$, $a_0(1450)$, and $a_0(1950)$ have nearly equal mass differences, it is possible that $a_0(1950)$, $K^*_0(1950)$, and two of the $f_0$ around $2\text{GeV}$ listed in PDG~\cite{pdg} form a nonet with masses around $2\text{GeV}$. The $J^r_c$, $J^r_e$, $J^r_{4,8_S}$, and $J^r_{4,1_S}$ ($I^r_c$, $I^r_e$, $I^r_{2,8_S}$, and $I^r_{2,1_S}$) are all possible interpretations.
	
	\item $J^r_3$/$I^r_b$ could be another nonet with masses above $2\text{GeV}$, but currently, no observed mesons in this range can be identified as part of the nonet.
	
	\item For all flavor configurations, the $J^r_d$/$I^r_3$ give masses close to $3\text{GeV}$, but currently, no observed $0^+$ mesons are in this range. 
\end{enumerate}

Unlike the nonets, all the 27-fold states in table~\ref{mass} have masses $\gtrsim2\text{GeV}$. The typical masses of $J^r_{2,27_S}$/$I^r_{1,27_S}$ and $J^r_{a,27_S}$/$I^r_{4,27_S}$ are around $2.5\text{GeV}$. This property may be relevant to the mixing between four-quark and hybrid states, since only the nonets can mix with hybrid states. 

\begin{figure}[t!]
	\begin{subfigure}[t]{0.39\textwidth}
		\includegraphics[width=\textwidth]{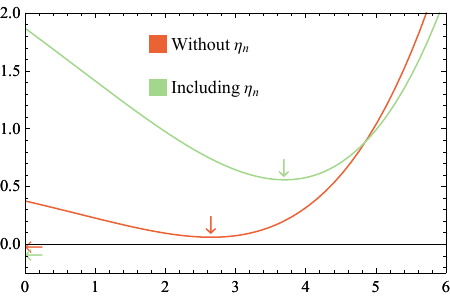}
		\caption{$\text{Im}\Pi(s)(10^3\text{GeV}^8)$ versus $s(\text{GeV}^2)$ with $\mu=1\text{GeV}$.}
		\label{rho_s_jb_1_a_10^3}
	\end{subfigure}
	\hspace*{\fill}	
	\begin{subfigure}[t]{0.45\textwidth}
		\includegraphics[width=\textwidth]{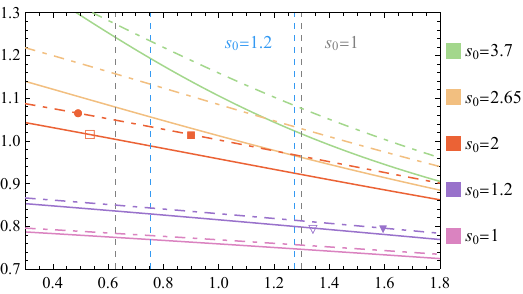}
		\caption{$\sqrt{\mathcal{R}^1}(\text{GeV})$ versus $\tau(\text{GeV}^{-2})$ with different values of $s_0(\text{GeV}^2)$.}
		\label{r0_t_jb_1_a_m}
	\end{subfigure}
	\caption{Mass estimation and {\small$\text{Im}\Pi(s)$} for $J_{b,1_A}$. The left figure shows the {\small$\text{Im}\Pi(s)$} when {\small$\mu=1\text{GeV}$}; the term {\small$\propto\delta(s)$} cannot be plotted. The left-pointing arrows indicate the coefficient $c$ in {\small$c\,\delta(s)$}; the minima are marked by downward arrows, occurring at $s=2.65$ and $3.7$ for the cases with and without including $\eta_n$, respectively. The lines and markers in the right figure follow the same conventions as in figure~\ref{fig_r0_jb}.}
	\label{fig_s_min}
\end{figure}


\subsection{Mass Estimations for the Bare Currents }\label{sec_3_mass_bare}

As we mentioned at the end of section~\ref{sec2_2}, the renormalized molecule currents may not properly represent the molecular states. It is therefore worthwhile to compare the results derived from the bare currents. We present the masses derived from the bare currents at leading order in tables~\ref{mass_bare} and \ref{mass_bare_eta}. The bare currents provide masses similar to those of the renormalized ones. Based on the properties of $\text{Im}\Pi(s)$, they can be sorted into 7 categories:
\begin{enumerate}
	\item $I_{1,27_S}$, $I_4$
	\item $J_{f,27_S}$
	\item $J_{2,27_S}$, $J_c$, $J_d$, $J_{f,8_S}$, $J_{f,1_S}$, $I_{1,8_S}$, $I_{1,1_S}$, $I_{3,8_S}$, $I_{3,1_S}$, $I_{5,8_S}$, $I_a$, $I_c$, $I_d$
	\item $J_1$, $J_{2,8_S}$, $J_{2,1_S}$, $J_{3,8_S}$, $J_{3,1_S}$, $J_{4,1_A}$ $J_{a,1_S}$, $J_e$, $I_{5,1_S}$, $I_{b,1_A}$, $I_e$
	\item $I_{3,27_S}$
	\item $J_{3,27_S}$, $J_{a,27_S}$, $J_{b,8_A}$, $I_{2,27_S}$, $I_{2,8_S}$, $I_{5,27_S}$
	\item $J_{4,8_A}$, $J_{a,8_S}$, $J_{b,1_A}$, $I_{2,1_S}$, $I_{b,8_A}$
\end{enumerate}	
Similar to the renormalized currents, the division into these categories is not strict, and some of them may be vague. Each of the first six categories has the same properties as the corresponding categories listed for the renormalized currents. As in the cases of renormalized currents, for category 1, the masses are derived from $\sqrt{\mathcal{R}^1}$; for categories 4 and 7, the masses are derived by the constraint $s_0=(m+\Lambda_\text{QCD})^2$. Some four-quark states in category 6 here yield masses $\gtrsim3\text{GeV}$ based on stability criteria, but it is difficult to find reasonable Borel Windows for them. Category 7 is the same as category 4 with the additional property that $\text{Im}\Pi(s)$ has a minimum which is relatively close to zero. This may help in fixing $s_0$ in $\mathcal{M}^n(\tau,s_0)$, since the integration \ref{moment} is not sensitive to $s_0$ when $\text{Im}\Pi(s_0)\sim 0$. However, it does not provide a better mass estimation. As shown in figure~\ref{fig_s_min}, the $\sqrt{\mathcal{R}^n}$ has worse $\tau$-stability for such choices of $s_0$. The failure to fix $s_0$ by the minimum of $\text{Im}\Pi(s)$ simply reflects the fact that $\text{Im}\Pi(s)$ does not directly provide valid spectral information. Additionally, for $J_{4,[8_S]}$, $I_{2,1_s}$, and $I_{b,[8_A]}$ in category 7, when $\tau$ and $s_0$ are chosen around the Borel windows, the constraint $s_0=(m+\Lambda_\text{QCD})^2$ cannot be applied, and the derived masses may not reliable, as shown at the end of appendices~\ref{sec_mass_b_t} and \ref{sec_mass_b_m}.


\renewcommand{\arraystretch}{1.3}
\begin{table}[t!]
	\centering
	\caption{The mass estimations for bare four-quark currents with $\eta_n=1$, following the same conventions as in table~\ref{mass}. Note that the currents here have no correspondence to the currents in tables~\ref{mass} and \ref{mass_eta}, despite the same subscripts. In some cases, the values are hard to ascertain based on the curves of $\sqrt{\mathcal{R}^n}(\tau)$; refer to the figures in Appendix~\ref{sec_mass_b_t} and \ref{sec_mass_b_m} for details. \label{mass_bare}}
	\stackengine{0pt}{
		\begin{tabular}{|w{c}{0.7cm}!{\vrule width 1pt}w{c}{2.2cm}|w{c}{2.2cm}|w{c}{2.2cm}|w{c}{2.2cm}|w{c}{2.2cm}|}
			\noalign{\hrule height 1pt}
			$J_1$&&&&$0.7-0.9$&$0.8-1$\\
			\noalign{\hrule height 0.2pt}
			$J_2$&$1.7-1.8\,(1.2)$&$0.7-0.9$&$0.9-1.1$&&\\
			\noalign{\hrule height 0.2pt}
			$J_3$&$\times$&$0.7-0.8$&$0.9-1.1$&&\\
			\noalign{\hrule height 0.2pt}
			$J_4$&&&&$0.9-1.9$&$0.9-1.1$\\
			\noalign{\hrule height 1pt}
			$J_a$&$\times$&$0.9-1.1$&$0.8-1$&&\\
			\noalign{\hrule height 0.2pt}
			$J_b$&&&&$\times$&$0.6-1.1$\\
			\noalign{\hrule height 0.2pt}
			$J_c$&$2\,(1.6-1.8)$&$1.6\,(1.2-1.5)$&$1.6\,(1.2-1.4)$&&\\
			\noalign{\hrule height 0.2pt}
			$J_d$&&&&$1.7-1.8\,(1.3)$&$1.5-1.6\,(0.9)$\\
			\noalign{\hrule height 0.2pt}
			$J_e$&&&&$0.7-0.9$&$0.8-1$\\
			\noalign{\hrule height 0.2pt}
			$J_f$&{\small2.7-2.8\,(2.5-2.7)}&{\small1.1-1.2\,(0.9-1)}&$1.2-1.3\,(0.9)$&&\\
			\noalign{\hrule height 1pt}
			$I_1$&$2.2\,(1.8)$&$1.8\,(1.2-1.6)$&$1.6\,(1.2)$&&\\
			\noalign{\hrule height 0.2pt}
			$I_2$&$\times$&$\times$&$1.2-2.1$&&\\
			\noalign{\hrule height 0.2pt}
			$I_3$&$2.4-2.6$&$1.6-1.8\,(1)$&$1.6-1.8\,(1)$&&\\
			\noalign{\hrule height 0.2pt}
			$I_4$&$2.6\,(2.1)$&$2.4\,(1.8)$&$2.4\,(1.8)$&&\\
			\noalign{\hrule height 0.2pt}
			$I_5$&$\times$&$0.9-1\,(0.8)$&$1-1.1$&&\\
			\noalign{\hrule height 1pt}
			$I_a$&&&&$1.4\,(1-1.1)$&$1.2-1.3\,(0.8)$\\
			\noalign{\hrule height 0.2pt}
			$I_b$&&&&$0.7-1.5$&$0.9-1.2$\\
			\noalign{\hrule height 0.2pt}
			$I_c$&&&&$1.9-2\,(1.1)$&$1.8-1.9\,(1.1)$\\
			\noalign{\hrule height 0.2pt}
			$I_d$&&&&$1.8\,(1.2)$&$1.7-1.8\,(1.1)$\\
			\noalign{\hrule height 0.2pt}
			$I_e$&&&&$0.8-1$&$0.6-0.8$\\
			\noalign{\hrule height 1pt}
		\end{tabular}
	}{
		\begin{tabular}{!{\vrule width 1pt}w{c}{2.2cm}|w{c}{2.2cm}|w{c}{2.2cm}|w{c}{2.2cm}|w{c}{2.2cm}|}
			\noalign{\hrule height 1pt}
			$[27_S]$&$[8_S]$&$1_S$&$[8_A]$&$1_A$
		\end{tabular}
	}{O}{r}{F}{F}{S}
\end{table}

\renewcommand{\arraystretch}{1.3}
\begin{table}[t!]
	\centering
	\caption{The mass estimations for bare four-quark currents with $\eta_6=3$ and $\eta_8=\eta_{10}=5$, following the same conventions as in table~\ref{mass}. Note that the currents here have no correspondence to the currents in tables~\ref{mass} and \ref{mass_eta}, despite the same subscripts. In some cases, the values are hard to ascertain based on the curves of $\sqrt{\mathcal{R}^n}(\tau)$; refer to the figures in Appendix~\ref{sec_mass_b_t} and \ref{sec_mass_b_m} for details. \label{mass_bare_eta}}
	\stackengine{0pt}{
		\begin{tabular}{|w{c}{0.7cm}!{\vrule width 1pt}w{c}{2.2cm}|w{c}{2.2cm}|w{c}{2.2cm}|w{c}{2.2cm}|w{c}{2.2cm}|}
			\noalign{\hrule height 1pt}
			$J_1$&&&&$0.8-0.9$&$0.9-1.1$\\
			\noalign{\hrule height 0.2pt}
			$J_2$&$2-2.2\,(1.4)$&$0.9-1.1$&$1-1.2$&&\\
			\noalign{\hrule height 0.2pt}
			$J_3$&$\times$&$0.8-0.9$&$1.1-1.2$&&\\
			\noalign{\hrule height 0.2pt}
			$J_4$&&&&$0.9-1.4$&$1-1.1$\\
			\noalign{\hrule height 1pt}
			$J_a$&$\times$&$0.9-1.2$&$0.9-1.1$&&\\
			\noalign{\hrule height 0.2pt}
			$J_b$&&&&$\times$&$0.7-1.2$\\
			\noalign{\hrule height 0.2pt}
			$J_c$&$2.8\,(2-2.4)$&{\small2-2.1\,(1.7-2)}&$1.9\,(1.7-1.9)$&&\\
			\noalign{\hrule height 0.2pt}
			$J_d$&&&&$2.2\,(1.7-1.9)$&$1.6-1.8\,(1.1)$\\
			\noalign{\hrule height 0.2pt}
			$J_e$&&&&$0.7-0.9$&$0.8-1$\\
			\noalign{\hrule height 0.2pt}
			$J_f$&$2.8-2.9$&{\small1.4-1.5\,(1.2-1.4)}&$1.4-1.6\,(1.1)$&&\\
			\noalign{\hrule height 1pt}
			$I_1$&$2.7\,(2.2)$&$2.2\,(1.6-2.1)$&$2\,(1.6-1.8)$&&\\
			\noalign{\hrule height 0.2pt}
			$I_2$&$\times$&$\times$&$0.8-1.2$&&\\
			\noalign{\hrule height 0.2pt}
			$I_3$&$2.6-3$&{\small2.1-2.2\,(1.6\,-1.8)}&$2.1\,(1.6\,-1.9)$&&\\
			\noalign{\hrule height 0.2pt}
			$I_4$&$2.9\,(2.5)$&$2.8\,(2.2)$&$2.6\,(2.1)$&&\\
			\noalign{\hrule height 0.2pt}
			$I_5$&$\times$&$1-1.1\,(1)$&$1.1-1.2$&&\\
			\noalign{\hrule height 1pt}
			$I_a$&&&&$1.7\,(1.3-1.6)$&$1.4-1.5\,(1.1)$\\
			\noalign{\hrule height 0.2pt}
			$I_b$&&&&$0.8-1$&$0.9-1$\\
			\noalign{\hrule height 0.2pt}
			$I_c$&&&&$2.3-2.4\,(1.6)$&{\small2.1-2.2\,(1.6-1.9)}\\
			\noalign{\hrule height 0.2pt}
			$I_d$&&&&$2\,(1.4)$&$1.9-2\,(1.4)$\\
			\noalign{\hrule height 0.2pt}
			$I_e$&&&&$0.8-1$&$0.6-0.8$\\
			\noalign{\hrule height 1pt}
		\end{tabular}
	}{
		\begin{tabular}{!{\vrule width 1pt}w{c}{2.2cm}|w{c}{2.2cm}|w{c}{2.2cm}|w{c}{2.2cm}|w{c}{2.2cm}|}
			\noalign{\hrule height 1pt}
			$[27_S]$&$[8_S]$&$1_S$&$[8_A]$&$1_A$
		\end{tabular}
	}{O}{r}{F}{F}{S}
\end{table}

The mass estimations listed in tables~\ref{mass_bare} and \ref{mass_bare_eta} show that most of the bare tetraquark currents provide nonet masses around $1\text{GeV}$, while most of the bare molecule currents provide nonet masses around $1-2\text{GeV}$, although tetraquark operators and four-quark molecule operators are related by Fierz rearrangement. Compared with the results listed in tables~\ref{mass} and \ref{mass_eta}, here the masses of 8-fold states and corresponding singlets differ significantly in some cases. On the other hand, most of the 27-fold states are also heavier than most of the nonets.

Recall that after choosing the eigenvectors of renormalization matrices as the basis of operators, the renormalized tetraquark and molecule operators have a one-to-one correspondence. Here the bare tetraquark and molecule currents give different nonet masses, which indicates that the tetraquark and molecule operators prefer to couple to different states. However, it is unclear what the exact states are, and it is not known how to choose the four-quark operators to provide the best distinguishability.

\begin{figure}[t!]
	\begin{subfigure}{0.45\textwidth}
		\includegraphics[width=\textwidth]{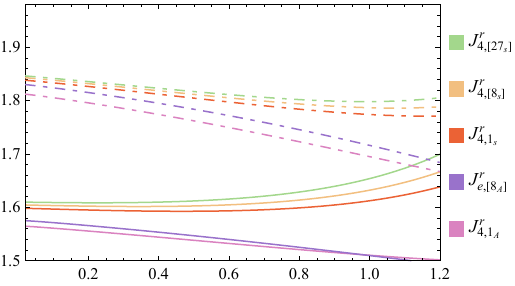}
		\caption{VS-first.}
		\label{r1_t_s0_3_4_vs}
	\end{subfigure}
	\hspace*{\fill}	
	\begin{subfigure}{0.45\textwidth}
		\includegraphics[width=\textwidth]{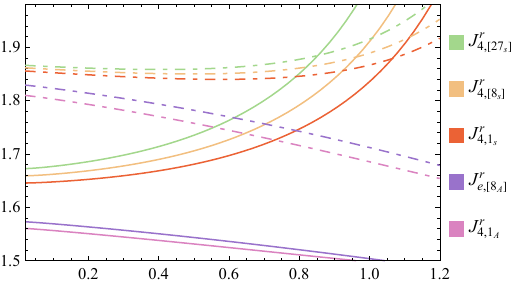}
		\caption{EM-first.}
		\label{r1_t_s0_3_4_em}
	\end{subfigure}
	\caption{Comparison of different factorization procedures for different currents. The figures show $\sqrt{\mathcal{R}^1}(\text{GeV})$ versus $\tau(\text{GeV}^{-2})$, with solid and dot-dashed lines referring to $s_0=3$ and $4(\text{GeV}^2)$, respectively. The left and right figures are obtained by first applying the vacuum saturation hypothesis and equation of motion, respectively.}
	\label{r1_em_vs}
\end{figure}

\subsection{Discussion}

The low precision of QCD sum rules hinders us from obtaining more accurate mass estimations, and the inclusion of factorization deviation factors further reduces the precision. No further conclusions can be derived from tables~\ref{mass} and \ref{mass_eta}, \ref{mass_bare} , \ref{mass_bare_eta}, or the figures in Appendix~\ref{sec_mass}.

To obtain better mass estimations, some other problems need to be solved. The first is the ambiguity of factorization in high-dimensional condensates. As discussed in Appendix~\ref{fac_d8}, applying the equation of motion before or after factorization can cause a significant difference. Comparing tables~\ref{mass} and \ref{mass_eta} with the tables in Appendix~\ref{em-first}, for $\langle\bar{q}q\rangle\langle\bar{q}Gq\rangle$, this ambiguity does not lead to a significant difference in mass estimations. This is because the plateau of $\sqrt{\mathcal{R}^n}(\tau)$ is less affected by this ambiguity, as shown in figure~\ref{r1_em_vs}, and we have already avoided choosing the values from the steep region of $\sqrt{\mathcal{R}^n}(\tau)$. However, for dimension-10 condensates, the effect of factorization ambiguity is not clear. Other condensates like $m\langle\bar{q}Gq\rangle$ or $m\langle\bar{q}q\rangle\langle GG\rangle$ have no such ambiguity. However, the propagator that gives the $mG$ term is infrared divergent \cite{prd_1987_lambda}, which complicates the OPE calculation.

Another problem arises from the naive method adopted to derive the masses. The basic approach involves taking the ratio of moments as in eq.~\ref{ratio}, which only works for the ``pole $+$ continuum'' ansatz. For a meson with a large width or a spectrum in which the peak overlaps with a continuum, such a naive procedure does not perform well.

\begin{figure}[t!]
	\begin{subfigure}{0.45\textwidth}
		\includegraphics[width=\textwidth]{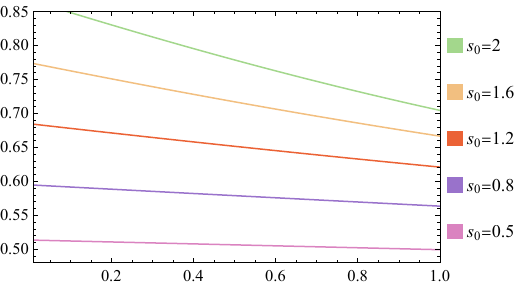}
		\caption{$\sqrt{\mathcal{R}^0}(\text{GeV})$ over $\tau(\text{GeV}^{-2})$.}
		\label{f0(500)_r0_t}
	\end{subfigure}
	\hspace*{\fill}	
	\begin{subfigure}{0.45\textwidth}
		\includegraphics[width=\textwidth]{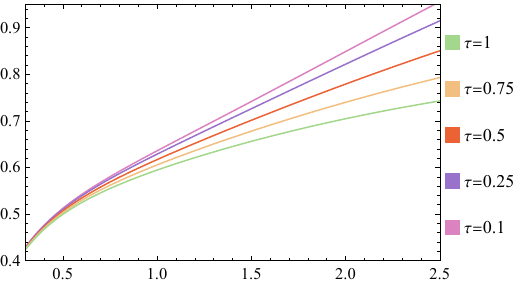}
		\caption{$\sqrt{\mathcal{R}^0}(\text{GeV})$ over $s_0(\text{GeV}^2)$.}
		\label{f0(500)_r0_s}
	\end{subfigure}\\[0.2cm]
	\begin{subfigure}{0.48\textwidth}
		\includegraphics[width=\textwidth]{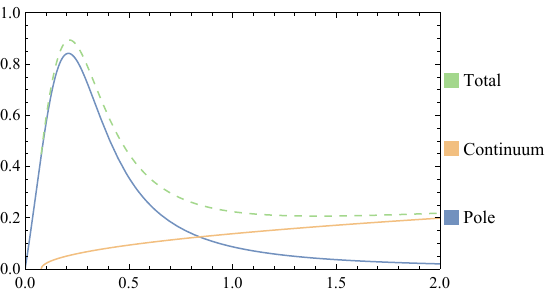}
		\caption{The factitious spectrum.}
		\label{f0(500)_rho}
	\end{subfigure}
	\hspace*{0.8cm}	
	\parbox[b][4.2cm][t]{6.5cm}{
		\caption{Mass estimation based on a manually constructed  spectrum related to $f_0(500)$. The strong $s_0$ dependence makes it difficult to derive the mass. The lower figure shows the $\text{Im}\Pi$ versus $s(\text{GeV}^2)$, and the ``pole $+$ continuum'' ansatz is not feasible for such a spectrum.}\label{f0(500)_borel}}
\end{figure}

To demonstrate that the wide peak causes the failure of ``pole $+$ continuum'' ansatz, consider such a manually constructed spectrum with a Breit-Wigner peak and a $\sqrt{s}$-type continuum starting at the $2\pi$ threshold:
\begin{equation}
	\text{Im}\Pi(s)=\text{tanh}\Big(\frac{s}{M^2-\Gamma^2/4}\Big)\,\text{Im}\Big[\frac{-M\Gamma}{s-(M-i\Gamma/2)^2}\Big]+\frac{1}{25}\theta(s-4m_\pi^2)\sqrt{\frac{s}{4m_\pi^2}-1}.
\end{equation}
Here, $M-i\Gamma/2=(0.475- 0.275\,i)\text{GeV}$ corresponds to $f_0(500)$ \cite{pdg}. The additional factor $\text{tanh}$ is added to ensure that $\text{Im}\Pi(0)=0$, and the factor $1/25$ is added to ensure that the pole is dominant. Such spectral construction is similar to that in ref.~\cite{wang_2017}. The estimated mass is shown in figure~\ref{f0(500)_borel}, where $\sqrt{\mathcal{R}^0}$ behaves like category 4 four-quark states, as shown in figure~\ref{fig_r0_jb}. The dependence on $s_0$ makes it difficult to determine the precise mass. The Gaussian sum rule \cite{gaussian} may help for spectra with wide peaks, but the large uncertainties of dimension-6, -8, and -10 condensates still hinder precise mass estimation.

As shown in tables~\ref{mass}, \ref{mass_eta}, \ref{mass_bare}, \ref{mass_bare_eta}, and Appendix~\ref{sec_mass}, the category 4 four-quark states have masses $\lesssim1\text{GeV}$, and most of them are nonets. These results agree with refs.~\cite{prd_2007_tetraquark, prd_2007_tetraquark_mixing, prd_2008_tetraquark, prd_2019_tetraquark}, although they are based on different operator configurations. This fact implies that the choice of operators to represent a $0^+$ meson $\lesssim1\text{GeV}$ may not be unique. This agrees with the previous discussion that the correspondence between operator and hadron is subtle, and may explain why different QCD sum rule studies~\cite{prd_2007_tetraquark, prd_2007_tetraquark_mixing, prd_2008_tetraquark, prd_2019_tetraquark} give different four-quark configurations for the same particle group.

An interesting observation is that, as shown in tables~\ref{mass}, \ref{mass_eta}, \ref{mass_bare}, \ref{mass_bare_eta}, and the figures in Appendix~\ref{sec_mass}, all four-quark states in categories 2 and 5 have masses $\gtrsim2\text{GeV}$, with some of them even reaching $\sim3\text{GeV}$. The absence of four-quark condensates ($\langle\bar{q}q\rangle^2$, $\langle\bar{q}q\rangle\langle\bar{q}Gq\rangle$, $\langle\bar{q}Gq\rangle^2$, and $\langle GG\rangle\langle\bar{q}q\rangle^2$) in correlators may be related to this property. Since the nonperturbative contribution is given by condensates, the absence of these condensates implies a weak coupling between these currents and low-energy states.

\section{Conclusion}

By exhausting all $J^P=0^+$ scalar four-quark currents and all $su(3)$-flavor configurations, we obtain several $J^P=0^+$ four-quark states with different masses. Some conclusions can be derived despite the uncertainties of the mass estimations.

The first is the existence of four-quark nonets with masses around $1-2\text{GeV}$. Some of the $0^+$ mesons observed in experiments can be interpreted as four-quark states, and the $0^+$ mesons below $1\text{GeV}$ form nonets. The second is the existence of 27-fold states, with typical masses $\gtrsim2\text{GeV}$. The mixing between hybrid states may be relevant to the mass difference between 27-fold states and nonets.

The involvement of hybrid-like operator in the renormalized four-quark operator also introduces a subtlety regarding the four-quark state, implying a mixing between the four-quark states and hybrid states. In principle, such a mixing is possible as long as the symmetry permits.

As discussed at the end of section~\ref{sec2_2}, once the eigenvectors of the renormalization matrix are selected as the renormalized four-quark operators, the corresponding coupled states can be equally interpreted as tetraquarks or molecules. The physical four-quark states might be mixtures of compact tetraquarks and four-quark molecules. On the other hand, since the four-quark molecule is bound by the exchange of color-neutral states like pions, a special renormalization procedure is required for the operator that represents a pure four-quark molecule. Additionally, many of the four-quark states listed in tables~\ref{mass}, \ref{mass_eta}, \ref{mass_bare}, and \ref{mass_bare_eta} have similar masses. Thus, some different operators may actually represent the same four-quark states, or the masses of the four-quark states are highly degenerate.

The main uncertainties in tables~\ref{mass}, \ref{mass_eta}, \ref{mass_bare}, \ref{mass_bare_eta}, and Appendix~\ref{sec_mass} arise from the factorization deviation factors. Precise values of $\eta_n$ are necessary for mass estimation. An interesting observation from these tables is that the mass estimations given by stability criteria are always lower than the values in the Borel Window, indicating that this property may hold in general.

Some useful techniques have been developed in this work. A replacement procedure to obtain the counterterm for an arbitrary multi-quark operator at one-loop level is introduced, which facilitates the calculation of renormalization of multi-quark operators. For the dimension-8 quark condensate, a method to express $\nabla^{\{\mu}\nabla^{\nu\}} \Psi$ as $\Gamma^{\mu\nu\alpha\beta}\, G_{\alpha\beta}^n T^n\Psi$ is provided in Appendix~\ref{fac_d8}, which shows that the factorization ambiguity related to $\langle\bar{q}q\rangle\langle\bar{q}Gq\rangle$ can introduce a discrepancy larger than $O(1/C_A^2)$~\cite{va_sum}. This raises the question of how to establish a factorization procedure without such ambiguity. More work is needed to address all these problems.

\appendix


\section{Gluon Equation of Motion and Hybrid-Like Currents }\label{A_ward}
Consider the QCD generating functional $Z(K)$; an infinitesimal shift $\delta A^a_\nu$ yields:
\begin{equation}
	\int\mathcal{D}\overline{\Psi}\mathcal{D}\Psi\mathcal{D}A\mathcal{D}c\mathcal{D}c^\dagger\, e^{i(S+K^{n\,\rho} A^n_\rho)}\,\big(D^{ab}_\mu G^{b\,\mu\nu}+J^{a\,\nu}+gf^{abc}c^b\partial^\nu\bar{c}^c+\frac{1}{\xi}\partial^\nu\partial^\mu A^a_\mu+K^{a\,\nu}\big)=0.
	\label{ward}
\end{equation}
Here, we use $K^{a\,\nu}$ to denote the source of $A^a_\nu$; $J^{a\,\nu}=g\sum_f \overline{\Psi}_f T^a\gamma^\nu\Psi_f$; $S$ is QCD action. The third and fourth terms originate from the ghost action and the gauge fixing term. Note that the classical equation of motion $D^{ab}_\mu G^{b\,\mu\nu}+J^{a\,\nu}=0$ is not exact.

Nevertheless, this relation still holds for the two-point function of four-quark currents at $O(\alpha_s)$. To see this, it suffices to show that the Green function
\begin{equation}
	\langle T\,\, D^{nm}_\mu G^{m\,\mu\nu}(x)\ \Psi^a_i(0)\overline{\Psi}^b_j(0)\rangle = - \langle T\,\, J^{n\,\nu}(x)\ \Psi^a_i(0)\overline{\Psi}^b_j(0)\rangle 
	\label{eqa2}
\end{equation}
holds at $O(g)$, which corresponds to a subdiagram in the four-quark correlator. Only the $\frac{1}{\xi}\partial^\nu\partial^\mu A^a_\mu$ term in eq.~\ref{ward} requires investigation. The fermion loop integral
\begin{equation}
	\int \frac{d^d k}{(2\pi)^d} S(q+k).\gamma^\nu.S(k)= \, C_1\,\gamma_\mu(\frac{q^\mu q^\nu}{q^2}-g^{\mu\nu})+C_2\, q_\mu\, \gamma^{\mu\nu}
\end{equation}
is transverse, where $S(k)=(\slashed{k}+m)/(k^2-m^2)$, and $C_1$ and $C_2$ are scalar functions of $q^2$, $m$, and $d$. Meanwhile, the $\frac{1}{\xi}\partial^\nu\partial^\mu A^n_\mu$ in eq.~\ref{ward} is longitudinal, thus eq.~\ref{eqa2} holds. Specifically, in momentum space, $D^{nm}_\mu G^{m\,\mu\nu}(x)$ can be written as $(\frac{q^\mu q^\nu}{q^2}- g^{\mu\nu})\,q^2A^n_\mu(q)$ at leading order, thus we have
\begin{equation}
	\begin{split}
	\!\!(\frac{q^\alpha q^\beta}{q^2}\!-\!g^{\alpha\beta})q^2 i\Delta^{nm}_{\beta\nu}(q^2)ig\!\!\int\!\!\frac{d^d k}{(2\pi)^d} S(q\!+\!k).T^m\!\!.\gamma^\nu\!.S(k)\!=&\!-\! gC_1T^n\gamma_\mu(\frac{q^\mu q^\alpha}{q^2}\!-\!g^{\mu\alpha})\!-\!gC_2T^n\! q_\mu\gamma^{\mu\alpha}\\
	=&\!-\!g\!\!\int\!\!\frac{d^d k}{(2\pi)^d} S(q\!+\!k).T^n\!.\gamma^\alpha\!.S(k)
	\end{split}
\end{equation}
at $O(g)$ due to the transverseness, where $\Delta^{nm}_{\beta\nu}(q^2)$ is gluon propagator. Diagrammatically, for the four-quark correlators, we have
\begin{equation}
			\begin{tikzpicture}[baseline=-\the\dimexpr\fontdimen22\textfont2\relax]
				\begin{feynman}
					\vertex (a0);
					\vertex [above=0.1cm of a0](au);
					\vertex [below=0.1cm of a0](ad);
					\vertex [right=2.0cm of a0](b0);
					\vertex [above=0.1cm of b0](bu);
					\vertex [below=0.1cm of b0](bd);
					\vertex [right=1.2cm of a0](g2);
					\diagram*[small]{
						(bu)--[ bend right=40](a0)--[ bend right=40](bd)--[ bend left=20](g2)--[ bend left=20](bu),
						(a0)--[photon, insertion={[style=semithick,size=1.5pt]0}](g2),
					};
					\draw[semithick](a0) circle(0.075cm);
				\end{feynman}
			\end{tikzpicture}\ =
			\begin{tikzpicture}[baseline=-\the\dimexpr\fontdimen22\textfont2\relax]
				\begin{feynman}
					\vertex (a0);
					\vertex [above=0.1cm of a0](au);
					\vertex [left=0.0cm of au](g1);
					\vertex [below=0.1cm of a0](ad);
					\vertex [left=0.0cm of ad](g3);
					\vertex [right=2.0cm of a0](b0);
					\vertex [right=1.5cm of a0](bb);
					\vertex [left=0.05cm of bb](bl);
					\vertex [above=0.03cm of bl](blu);
					\vertex [right=0.05cm of bb](br);
					\vertex [below=0.03cm of br](brd);
					\vertex [above=0.1cm of b0](bu);
					\vertex [right=0.0cm of bu](g2);
					\vertex [below=0.1cm of b0](bd);
					\vertex [right=0.0cm of bd](g4);
					\diagram*[small]{
						(bu)--[ bend right=40](au)--[ bend left=14](blu),
						(brd)--[bend right=9](bd)--[ bend left=40](ad)--[ bend right=15](bb)--[bend left=10](bu)
					};
				\end{feynman}
			\end{tikzpicture}
	\label{dg-4q}\quad\text{and}\quad
		\begin{tikzpicture}[baseline=-\the\dimexpr\fontdimen22\textfont2\relax]
			\begin{feynman}
				\vertex (a0);
				\vertex [above=0.1cm of a0](au);
				\vertex [left=0.0cm of au](g1);
				\vertex [below=0.1cm of a0](ad);
				\vertex [left=0.0cm of ad](g3);
				\vertex [right=2.0cm of a0](b0);
				\vertex [above=0.1cm of b0](bu);
				\vertex [right=0.0cm of bu](g2);
				\vertex [below=0.1cm of b0](bd);
				\vertex [right=0.0cm of bd](g4);
				\vertex [right=1.05cm of a0](g2);
				\vertex [above=0.15cm of g2](g2u);
				\diagram*[small]{
					(bu)--[ bend right=20](g2u)--[ bend right=20](bu),
					(a0)--[ bend right=40](bd)--[ bend left = 15 ](a0),
					(a0)--[photon, bend left=20, insertion={[style=semithick,size=1.5pt]0}](g2u),
				};
				\draw[semithick](a0) circle(0.075cm);
			\end{feynman}
		\end{tikzpicture} \ =
		\begin{tikzpicture}[baseline=-\the\dimexpr\fontdimen22\textfont2\relax]
			\begin{feynman}
				\vertex (a0);
				\vertex [above=0.1cm of a0](au);
				\vertex [left=0.0cm of au](g1);
				\vertex [below=0.1cm of a0](ad);
				\vertex [left=0.0cm of ad](g3);
				\vertex [right=2.0cm of a0](b0);
				\vertex [above=0.1cm of b0](bu);
				\vertex [right=0.0cm of bu](g2);
				\vertex [below=0.1cm of b0](bd);
				\vertex [right=0.0cm of bd](g4);
				\diagram*[small]{
					(bu)--[ bend right=40](au)--[ bend left=10](bu),
					(bd)--[ bend left=10](ad)--[ bend right=40](bd)
				};
			\end{feynman}
		\end{tikzpicture} \, ,
\end{equation}
where the cross-dot denotes the hybrid-like counterterm, and the four-quark vertices are slightly separated to make it clear how the propagators are connected; the relevant factors are omitted.

In contrast, when a gluon is involved in the current, the $\frac{1}{\xi}\partial^\nu\partial^\mu A^n_\mu$ contribution may not vanish. The
\begin{equation}
	\langle T \ \frac{1}{\xi}\partial^\nu\partial^\mu A^n_\mu(q)\ A^{m\,\rho}(q) \rangle
\end{equation}
gives
\begin{equation}
	i\,\frac{1}{\xi} q^\nu q_\mu \,\frac{g^{\mu\rho}-(1-\xi)\frac{q^\mu q^\rho}{q^2}}{q^2}\,\delta^{nm}=\ i \delta^{nm}\,\frac{q^\nu q^\rho}{q^2}\, .
	\label{longitudinal}
\end{equation}

A question still remains about the ghosts and longitudinal gluon, namely, whether $gf^{abc}c^b\partial^\nu\bar{c}^c$ and $\frac{1}{\xi}\partial^\nu\partial^\mu A^a_\mu$ have a physical contribution. Since the gluon inside a hadron can be highly off-shell, the longitudinal gluon here may not be a problem. A direct calculation shows that for
\begin{equation}
	\begin{split}
		\mathcal{O}_1=\ &\,\overline{\Psi}\gamma^\mu T^a ( gf^{abc}c^b\partial^\nu\bar{c}^c) \Psi,\\
		\mathcal{O}_2=\ &\,\overline{\Psi}\gamma^\mu T^a (\frac{1}{\xi}\partial^\nu\partial^\alpha A_\alpha^a) \Psi,\\
		\mathcal{O}_H=\ &\overline{\Psi} T^n G^{n\,\alpha\rho}\gamma^\beta\Psi,
	\end{split}	
\end{equation}
the $\langle T\, \mathcal{O}_1(x) \mathcal{O}_H(0)\rangle+\langle T\, \mathcal{O}_2(x) \mathcal{O}_H(0)\rangle$ is nonvanishing at $O(\alpha_s)$, and the source term in eq.~\ref{ward} has no contribution here. Therefore, the classical equation of motion does not hold in general.


\section{Factorization Ambiguity for Dimension-8 Four-Quark Condensate}\label{fac_d8}
In Fock-Schwinger gauge ($x^\mu A_\mu = 0$)~\cite{condensates_high_order}, expanding the $\Psi_i(x)$ in $\langle \bar{\Psi}_i(x)\Psi_j(x)\bar{\Psi}_k(0)\Psi_l(0)\rangle$ yields terms that contain dimension-8 condensates:
\begin{equation}
		x^\mu x^\nu \Big[\ \frac{1}{2}\langle\bar{\Psi}_i\overleftarrow{\nabla}_{\{\mu}\overleftarrow{\nabla}_{\nu\}}\Psi_j\bar{\Psi}_k\Psi_l\rangle+ \langle\bar{\Psi}_i\overleftarrow{\nabla}_{\{\mu}\nabla_{\nu\}}\Psi_j\bar{\Psi}_k\Psi_l\rangle
		 + \frac{1}{2}\langle\bar{\Psi}_i\nabla_{\{\mu}\nabla_{\nu\}}\Psi_j\bar{\Psi}_k\Psi_l\rangle\ \Big],	
	\label{d8_exp}
\end{equation}
where the derivatives act only on the adjacent fermions, and $\nabla^{\{\mu}\nabla^{\nu\}}=\frac{1}{2}(\nabla^\mu\nabla^\nu+\nabla^\nu\nabla^\mu)$. Consider $|\nabla^\mu\nabla^\nu\Psi(0)\rangle$, which can be written as a linear combination of state vectors. Only $|ig\,\Gamma^{\mu\nu\alpha\beta}\, G_{\alpha\beta}^n T^n\Psi(0)\rangle$ involves gluon field strength and transforms in the same way as $|\nabla^\mu\nabla^\nu\Psi(0)\rangle$. Here $\Gamma^{\mu\nu\alpha\beta}$ represents unknown $\gamma$-matrices. Keeping only the terms that contain gluon field strength, Lorentz symmetry gives
\begin{equation}
	(\nabla^{\{\mu}\nabla^{\nu\}} \Psi_j)^b = \big[A\, g^{\mu\nu}\gamma^{\alpha\beta}+B\,((g^{\mu\alpha}\gamma^{\nu\beta}-g^{\mu\beta}\gamma^{\nu\alpha})+\{\mu\leftrightarrow\nu\})\big]_{jr}\, G^n_{\alpha\beta}T^{n\,be}\Psi^e_r
	\label{linear_1}
\end{equation}
where $\gamma^{\alpha\beta}=\,\frac{1}{2}[\gamma^\alpha,\gamma^\beta]$. By contracting eq.~\ref{linear_1} with $g_{\mu\nu}$ and $\gamma_\nu$ respectively, the linear equations and the equation of motion in the massless limit yield
\begin{equation}
	A=\frac{ig}{2(d+2)}\text{;}\ B=\frac{ig}{4(d+2)},
	\label{linear_3}
\end{equation}
and we have
\begin{equation}
	\nabla^{\{\mu}\nabla^{\nu\}}\Psi=\Gamma^{\mu\nu\alpha\beta}\, G_{\alpha\beta}^n T^n\Psi,
\end{equation}
where
\begin{equation}
	\Gamma^{\mu\nu\alpha\beta}=\frac{i}{4(d+2)}\big[2g^{\mu\nu}\gamma^{\alpha\beta}+((g^{\mu\alpha}\gamma^{\nu\beta}-g^{\mu\beta}\gamma^{\nu\alpha})+\{\mu\leftrightarrow\nu\})\big],
	\label{ggamma}
\end{equation}
and $d$ is the spacetime dimension. This identity has already been used in ref~\cite{1-+}, but it has not been symmetrized there.

Applying the above identities to the condensate
\begin{equation}
	\langle\bar{\Psi}_i^a\, (\nabla^{\{\mu}\nabla^{\nu\}}\Psi_j)^b\, \bar{\Psi}_k^c\,\Psi_l^d\rangle,
	\label{d8_c1}
\end{equation}
after factorization and simplification, it becomes
\begin{equation}
	\begin{split}
		&\frac{\langle \bar{\Psi}G\Psi\rangle\langle \bar{\Psi}\Psi\rangle}{2^5 C_A^2\, d}\,g^{\mu\nu}\big(\delta^{ba}\delta_{ji}\delta^{dc}\delta_{lk}-\delta^{bc}\delta_{jk}\delta^{da}\delta_{li}\big)\\
		&\ +\frac{\langle \bar{\Psi}G\Psi\rangle\langle \bar{\Psi}\Psi\rangle}{2^4C_A^2 C_F\, d(d-1)}\,T^{n\,be}\,\Gamma^{\mu\nu\alpha\beta}_{jr}\Big[T^{n\,dc}(\sigma_{\alpha\beta})_{lk}\delta^{ea}\delta_{ri}-T^{n\,da}(\sigma_{\alpha\beta})_{li}\delta^{ec}\delta_{rk}\Big]
	\end{split}	
	\label{d8_fac_2}
\end{equation}

Thus, applying the equation of motion before or after factorization (EM-first or VS-first) leads to discrepancies. It is easy to verify that the first line in eq.~\ref{d8_fac_2} is identical to the result obtained by the VS-first procedure. This is because when factorizing eq.\ref{d8_c1}, the gluon is paired with $\Psi^e_r$ to give the first line in eq.~\ref{d8_fac_2}, whereas for the second line, the gluon is paired with $\Psi^d_l$. 

Contracting eq.~\ref{d8_fac_2} with $T^{n\,ab}\gamma^\mu_{ij}\ T^{n\,cd}\gamma^\nu_{kl}$ reproduces the conclusion in ref.~\cite{va_sum} (the equation of motion is needed to convert $\nabla_{\{\mu}\nabla_{\nu\}}\gamma^\nu\Psi$ to $\nabla_{[\mu}\nabla_{\nu]}\gamma^\nu\Psi$). In this case, the EM-first and VS-first procedures yield discrepancies of order $\sim1/C_A^2$.

However, the difference can be much larger. For example, if $x_\mu x_\nu T^{m\, ab}T^{m\, cd} \gamma^\rho_{il}\gamma^\sigma_{kj}$ contracts with eq.~\ref{d8_fac_2}, the VS-first procedure yields zero, while the EM-first procedure gives
\begin{equation}
	\frac{(d-1)(d-2)x^2 g^{\rho\sigma}-4(d-2)x^\rho x^\sigma}{2^4 C_A (d-1)d(d+2)}\langle \bar{\Psi}G\Psi\rangle\langle \bar{\Psi}\Psi\rangle.
\end{equation}

Unlike eq.~\ref{d8_c1}, the factorization of
\begin{equation}
	\langle \bar{\Psi}_i^a\overleftarrow{\nabla}^\mu \, (\nabla^\nu \Psi_j)^b \, \bar{\Psi}_k^c\, \Psi_l^d\rangle
	\label{d8_c2}
\end{equation}
is more subtle. In eq.~\ref{d8_c1}, we extract $G_{\mu\nu}$ via a simple linear algebra manipulation, which is feasible because the gluon field strength $G_{\mu\nu}$ is the curvature of the connection of the principal bundle $A^n_\mu(x)$, and the curvature is encoded in the second order covariant derivative. However, the first order covariant derivatives in this case cannot be completely eliminated; the same procedure is not applicable here. Therefore, we factorize eq.~\ref{d8_c2} by the VS-first procedure, which gives the dimension-8 contribution:
\begin{equation}
	\langle \bar{\Psi}_i^a\overleftarrow{\nabla}^\mu\,(\nabla^\nu \Psi_j)^b\,\bar{\Psi}_k^c\, \Psi_l^d\rangle=-\frac{\langle \bar{\Psi}G\Psi\rangle\langle \bar{\Psi}\Psi\rangle}{2^5C_A^2d}\,g^{\mu\nu}\,\delta^{ba}\delta_{ji}\delta^{dc}\delta_{lk}.
	\label{d8_fac_d}
\end{equation}

This result differs  by a sign from the first term in the expansion of the first line in eq.~\ref{d8_fac_2}. Because for $\langle \bar{\Psi}_i^a(x) \Psi_j^b(x)\bar{\Psi}^c_k \Psi_l^d\rangle$, the direct factorization gives a term $\propto\!\langle\bar{\Psi}(x)\Psi(x)\rangle\langle\bar{\Psi}\Psi\rangle=\langle \bar{\Psi}\Psi\rangle^2$ in the result. Since $\langle\bar{\Psi}(x)\Psi(x)\rangle=\langle \bar{\Psi}\Psi\rangle$, this equivalently gives
\begin{equation}
	x_\mu x_\nu \langle\, \bar{\Psi}\overleftarrow{\nabla}^{\{\mu}\overleftarrow{\nabla}^{\nu\}} \,\Psi\, +\, 2\bar{\Psi}\overleftarrow{\nabla}^{\{\mu} \nabla^{\nu\}} \Psi \,+\,\bar{\Psi}\nabla^{\{\mu}\nabla^{\nu\}} \Psi\, \rangle = 0,
\end{equation}
which gives a constraint between eq.~\ref{d8_fac_d} and the corresponding term in eq.~\ref{d8_fac_2}.

A more direct way to demonstrate the factorization ambiguity is to consider the dimension-8 condensate involving different quarks, e.g., $\langle\bar{u}_i\nabla_\mu\nabla_\nu u_j\, \bar{d}_k d_l\rangle$. The VS-first procedure yields a result $\propto \langle\bar{u}Gu\rangle \langle\bar{d}d\rangle$, while the EM-first procedure gives an additional term $\propto \langle\bar{u}u\rangle \langle\bar{d}Gd\rangle$ since the gluon field strength exists before factorization. For light quarks accompanied by $c$ or $b$ quarks, such ambiguity may lead a large difference. In general, the factorization ambiguity arises for all condensates with $\geq4$ quarks and $\geq2$ covariant derivatives.

\section{The OPE Diagrams for the Four-Quark Correlators}\label{dias}

Here, the diagrams differing by permutations, as well as the directions of fermion propagators are omitted. For tetraquark correlators, not all diagrams are involved. The four-quark vertices are slightly separated to makes it clear how the propagators are connected.

\renewcommand{\arraystretch}{1.2}
\begin{table}[h!]
	\centering
	\caption{The corresponding diagrams for each type of condensate.}
	\begin{tabular}{m{0.125\textwidth}m{0.8\textwidth}}
		\noalign{\hrule height 1pt}
		$m\langle \bar{q}q\rangle$ and $\langle GG\rangle$&{\raggedright\includegraphics[width=0.72\textwidth]{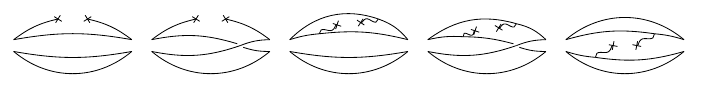}} \vspace{-2pt}\\
		\noalign{\hrule height 0.2pt}
		$\langle \bar{q}q\rangle^2$&{\raggedright\includegraphics[width=0.64\textwidth]{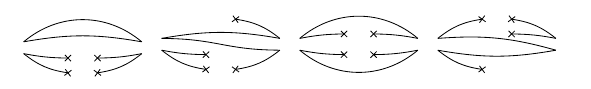}}\vspace{-2pt}\\
		\noalign{\hrule height 0.2pt}
		$\langle \bar{q}q\rangle\langle \bar{q}Gq\rangle$&{\raggedright\includegraphics[width=0.64\textwidth]{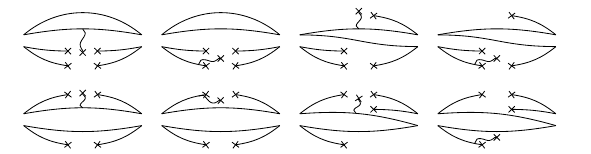}}\vspace{-0pt}\\
		\noalign{\hrule height 0.2pt}
		$\langle \bar{q}Gq\rangle^2$ and $\langle GG\rangle\langle \bar{q}q\rangle^2$&{\raggedright\includegraphics[width=0.82\textwidth]{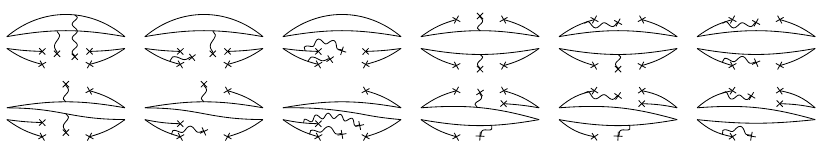}}\vspace{-0pt}\\
		\noalign{\hrule height 0.2pt}
		$m\langle \bar{q}q\rangle^3$&{\raggedright\includegraphics[width=0.32\textwidth]{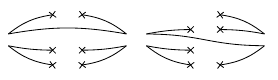}}\vspace{-1pt}\\
		\noalign{\hrule height 1pt}
	\end{tabular}
	\label{ope_dia_ds}
\end{table}

\begin{table}[h!]
	\centering
	\caption{Diagrams and propagator notations related to $\langle G^3\rangle$. The infrared divergence originating from the quark propagators in the background gluon field is canceled, making each diagram here infrared free. Refer to eq.~\ref{q_propagator} for the detailed quark propagator structure.}
	\begin{tabular}{m{0.18\textwidth}m{0.75\textwidth}}
		\noalign{\hrule height 1pt}
		$\langle G^3\rangle$ diagrams&{\raggedright\includegraphics[width=0.75\textwidth]{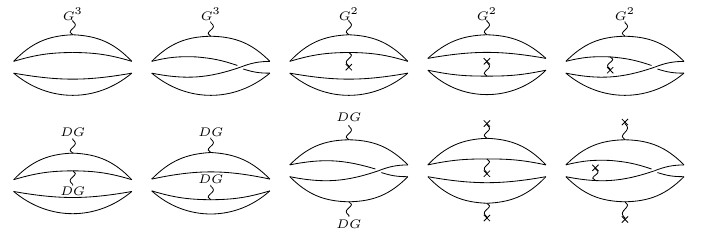}}\vspace{-1pt}\\
		\noalign{\hrule height 0.2pt}
		The propagator notations&{\raggedright\includegraphics[width=0.78\textwidth]{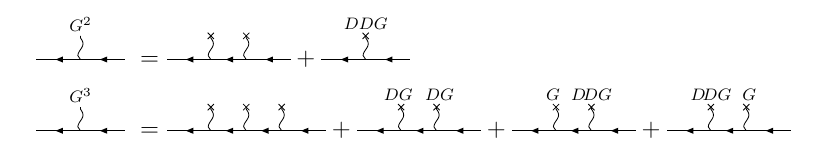}}\vspace{-1pt}\\
		\noalign{\hrule height 1pt}
	\end{tabular}
	\label{ope_dia_d6}
\end{table}


\begin{figure}[h!]
	\centering
	\includegraphics[width=0.89\textwidth]{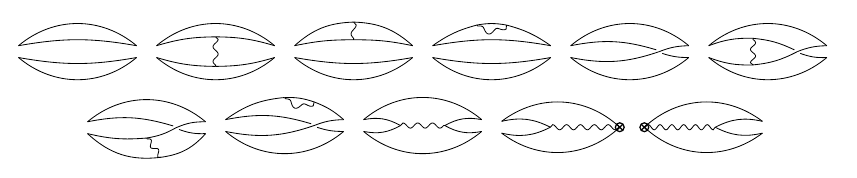}
	\label{ope_d0}
	\caption{Perturbative diagrams. The last two diagrams are related to hybrid-like counterterm.}
\end{figure}


\clearpage
\section{Mass Estimations Referring to Different Factorization Procedures}\label{em-first}


\renewcommand{\arraystretch}{1.35}
\begin{table}[ht!]
	\centering
	\caption{The four-quark mass estimations correspond to $\eta_n=1$, following the same conventions as in table~\ref{mass}. The $\langle\bar{q}q\rangle\langle\bar{q}Gq\rangle$ is obtained by the EM-first procedure (see Appendix~\ref{fac_d8}) here. In some cases, the values are hard to ascertain based on the curves of $\sqrt{\mathcal{R}^n}(\tau)$; the figures are given in ref.~\cite{results}. \label{mass_em-first}}
	\vspace*{-0.3cm}
	\stackengine{0pt}{
		\begin{tabular}{|w{c}{0.7cm}!{\vrule width 1pt}w{c}{2.2cm}|w{c}{2.2cm}|w{c}{2.2cm}|w{c}{2.2cm}|w{c}{2.2cm}|}
			\noalign{\hrule height 1pt}
			$J_1^r$&&&&$0.7-0.9$&$0.8-1$\\
			\noalign{\hrule height 0.2pt}
			$J_2^r$&$2.4\,(2.2)$&$0.6-0.7$&$0.9-1$&&\\
			\noalign{\hrule height 0.2pt}
			$J_3^r$&&&&$2.1-2.2$&$2.2-2.3$\\
			\noalign{\hrule height 0.2pt}
			$J_4^r$&$2.4\,(1.9)$&$2.4\,(2)$&$2.3\,(1.8)$&&\\
			\noalign{\hrule height 1pt}
			$J_a^r$&{\small2.4-2.6\,(2.2-2.5)}&$1.2\,(1-1.2)$&$1.3-1.4\,(1)$&&\\
			\noalign{\hrule height 0.2pt}
			$J_b^r$&&&&$0.6-0.8$&$0.6-0.9$\\
			\noalign{\hrule height 0.2pt}
			$J_c^r$&&&&$1.6-1.8$&$1.6-1.8$\\
			\noalign{\hrule height 0.2pt}
			$J_d^r$&{\small2.8-2.9\,(2.6-2.8)}&{\small2.9-3\,(2.7-2.8)}&{\small2.8-3\,(2.7-2.9)}&&\\
			\noalign{\hrule height 0.2pt}
			$J_e^r$&&&&$2-2.1\,(1.1)$&$1.9-2.1\,(1.3)$\\
			\noalign{\hrule height 0.2pt}
			$J_f^r$&$\times$&$\times$&$\times$&&\\
			\noalign{\hrule height 1pt}
			$I_1^r$&$2.3\,(2.2)$&$0.6-0.9$&$0.9-1$&&\\
			\noalign{\hrule height 0.2pt}
			$I_2^r$&$2.3\,(1.8)$&$2.3\,(1.8)$&$2.3\,(1.8)$&&\\
			\noalign{\hrule height 0.2pt}
			$I_3^r$&$2.7\,(2.4-2.5)$&{\small2.6-2.7\,(2.4-2.6)}&{\small2.5-2.6\,(2.3-2.6)}&&\\
			\noalign{\hrule height 0.2pt}
			$I_4^r$&{\small2.4-2.5\,(2.2-2.4)}&$1.2\,(0.9-1.2)$&$1.3-1.4\,(1)$&&\\
			\noalign{\hrule height 0.2pt}
			$I_5^r$&$\times$&$\times$&$\times$&&\\
			\noalign{\hrule height 1pt}
			$I_a^r$&&&&$0.7-0.8$&$0.8-1$\\
			\noalign{\hrule height 0.2pt}
			$I_b^r$&&&&$2.1-2.2$&$2.3-2.4$\\
			\noalign{\hrule height 0.2pt}
			$I_c^r$&&&&$1.6-1.8$&$1.6-1.8$\\
			\noalign{\hrule height 0.2pt}
			$I_d^r$&&&&$0.6-0.8$&$0.7-0.9$\\
			\noalign{\hrule height 0.2pt}
			$I_e^r$&&&&$1.9-2.1\,(1.1)$&$1.9-2\,(1.4)$\\
			\noalign{\hrule height 1pt}
		\end{tabular}
	}{
		\begin{tabular}{!{\vrule width 1pt}w{c}{2.2cm}|w{c}{2.2cm}|w{c}{2.2cm}|w{c}{2.2cm}|w{c}{2.2cm}|}
			\noalign{\hrule height 1pt}
			$[27_S]$&$[8_S]$&$1_S$&$[8_A]$&$1_A$
		\end{tabular}
	}{O}{r}{F}{F}{S}
\end{table}

\renewcommand{\arraystretch}{1.35}
\begin{table}[ht!]
	\centering
	\caption{The four-quark mass estimations correspond to $\eta_6=3$ and $\eta_8=\eta_{10}=5$, following the same conventions as in table~\ref{mass}. The $\langle\bar{q}q\rangle\langle\bar{q}Gq\rangle$ is obtained by the EM-first procedure (see Appendix~\ref{fac_d8}) here. In some cases, the values are hard to ascertain based on the curves of $\sqrt{\mathcal{R}^n}(\tau)$; the figures are given in ref.~\cite{results}. \label{mass_em-first_eta}}
	\vspace*{-0.3cm}
	\stackengine{0pt}{
		\begin{tabular}{|w{c}{0.7cm}!{\vrule width 1pt}w{c}{2.2cm}|w{c}{2.2cm}|w{c}{2.2cm}|w{c}{2.2cm}|w{c}{2.2cm}|}
			\noalign{\hrule height 1pt}
			$J_1^r$&&&&$0.7-0.9$&$0.9-1.1$\\
			\noalign{\hrule height 0.2pt}
			$J_2^r$&$2.8\,(2.6)$&$0.7-0.8$&$1.1-1.2$&&\\
			\noalign{\hrule height 0.2pt}
			$J_3^r$&&&&$2.3$&$2.3-2.4$\\
			\noalign{\hrule height 0.2pt}
			$J_4^r$&$2.8\,(2.4)$&$2.8\,(2.4)$&$2.7\,(2.4)$&&\\
			\noalign{\hrule height 1pt}
			$J_a^r$&$2.6$&{\small1.4-1.5\,(1.2\,-1.5)}&$1.6\,(1.2-1.3)$&&\\
			\noalign{\hrule height 0.2pt}
			$J_b^r$&&&&$0.6-0.8$&$0.7-0.9$\\
			\noalign{\hrule height 0.2pt}
			$J_c^r$&&&&$2-2.2$&$2-2.2$\\
			\noalign{\hrule height 0.2pt}
			$J_d^r$&$2.9-3$&$2.9-3$&$2.9-3$&&\\
			\noalign{\hrule height 0.2pt}
			$J_e^r$&&&&$2.4-2.5\,(1.7)$&{\small2.3-2.4\,(1.9-2.2)}\\
			\noalign{\hrule height 0.2pt}
			$J_f^r$&$\times$&$\times$&$\times$&&\\
			\noalign{\hrule height 1pt}
			$I_1^r$&$2.7\,(2.5)$&$0.7-0.9$&$1-1.1$&&\\
			\noalign{\hrule height 0.2pt}
			$I_2^r$&$2.7\,(2.4)$&$2.7\,(2.4)$&$2.7\,(2.4)$&&\\
			\noalign{\hrule height 0.2pt}
			$I_3^r$&$2.7-2.8$&$2.7-2.9$&$2.7-2.8$&&\\
			\noalign{\hrule height 0.2pt}
			$I_4^r$&$2.4-2.5$&{\small1.4-1.5\,(1.2\,-1.5)}&{\small1.5-1.6\,(1.2-1.4)}&&\\
			\noalign{\hrule height 0.2pt}
			$I_5^r$&$\times$&$\times$&$\times$&&\\
			\noalign{\hrule height 1pt}
			$I_a^r$&&&&$0.7-0.8$&$0.9-1.1$\\
			\noalign{\hrule height 0.2pt}
			$I_b^r$&&&&$2.2$&$2.4$\\
			\noalign{\hrule height 0.2pt}
			$I_c^r$&&&&$2-2.2$&$2-2.2$\\
			\noalign{\hrule height 0.2pt}
			$I_d^r$&&&&$0.6-0.8$&$0.7-0.9$\\
			\noalign{\hrule height 0.2pt}
			$I_e^r$&&&&$2.4\,(1.7)$&{\small2.3-2.4\,(1.9-2.2)}\\
			\noalign{\hrule height 1pt}
		\end{tabular}
	}{
		\begin{tabular}{!{\vrule width 1pt}w{c}{2.2cm}|w{c}{2.2cm}|w{c}{2.2cm}|w{c}{2.2cm}|w{c}{2.2cm}|}
			\noalign{\hrule height 1pt}
			$[27_S]$&$[8_S]$&$1_S$&$[8_A]$&$1_A$
		\end{tabular}
	}{O}{r}{F}{F}{S}
\end{table}

\vspace*{-0.2cm}


\section{Figures for Mass Estimations}\label{sec_mass}

The conventions adopted in the following figures are the same as those in figures~\ref{fig_r_tau} and \ref{fig_r0_jb}; we present them here for convenience. The dot-dashed and solid lines represent the cases with and without including the factorization deviation factors $\eta_n$, respectively. The vertical dashed lines indicate the Borel Window, with gray and blue referring to including and not including $\eta_n$, respectively; the $s_0$ for Borel windows are shown beside their corresponding lines in matching colors. If two $s_0$ overlap, it means that they are identical. For category 1, we derive the masses using only $\sqrt{\mathcal{R}^1}$ because the pole exists near the Borel Window in $\sqrt{\mathcal{R}^0}$. For categories 2 and 5, as well as some currents with flavors specified to $27_S$, the four-quark condensates $\langle\bar{q}q\rangle^2$, $\langle\bar{q}q\rangle\langle\bar{q}Gq\rangle$, $\langle\bar{q}Gq\rangle^2$, and $\langle GG\rangle\langle\bar{q}q\rangle^2$ are absent. Since the $m\langle\bar{q}q\rangle^3$ only affects $\sqrt{\mathcal{R}^0}$, for these cases, only the cases with $\eta_n=1$ are shown for $\sqrt{\mathcal{R}^1}$.

The corresponding values in tables~\ref{mass} and \ref{mass_eta}, \ref{mass_bare}, and \ref{mass_em-first} are obtained as follows: We select the values where $\sqrt{\mathcal{R}^n}(\tau)$ is most stable to determine the range of masses, which are enclosed in parentheses. Additionally, we select $\sqrt{\mathcal{R}^n}(\tau)$ within the Borel Window, to obtain another range of masses for the same states. 

In cases without plateaus , the markers on the curves show the results where $\tau$ is fixed by the constraint $s_0=(\sqrt{\mathcal{R}^n}+\Lambda)^2$, and $s_0$ is selected to be the same as the $s_0$ specified for the Borel Window. The filled and empty markers refer to including $\eta_n$ and not including $\eta_n$, respectively. The triangle, circle, and square markers refer to choosing $\Lambda=0.3$, $0.35$, and $0.4\text{GeV}$ respectively. The markers outside the visible range are omitted.

\begin{figure}[h!]
	\centering
	\includegraphics[width=10cm,height=4cm,]{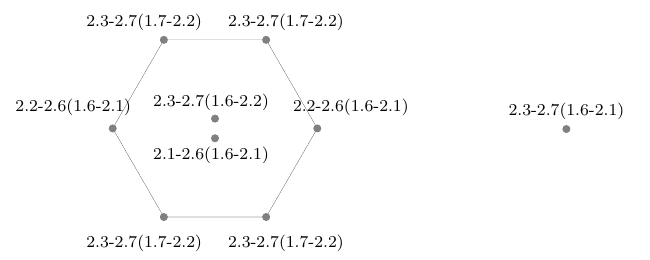}\\[0cm]
	\caption{Nonet masses ($\sqrt{\mathcal{R}^1}$) for $J^r_{4,8_S}$ and $J^r_{4,1_S}$. The values in parentheses are obtained by stability criteria; other values are derived from the values in the Borel Window. The values are not decisive because it is hard to determine them solely based on the curves of $\sqrt{\mathcal{R}^n}(\tau)$. Refer to figures~\ref{fig_r_ca_1_1}, \ref{fig_r_j4_nonet}, and ref.~\cite{results} for details.}
	\label{su(3)_j4_nonet}
\end{figure}

As previously discussed in Section~\ref{sec_3_mass}, within each current, states that belong to the same $su(3)$-flavor representation exhibit similar masses. Here we show the masses of a nonet in figure~\ref{su(3)_j4_nonet}, corresponding to a current in category 1. The stability criteria are applicable here, which reduce the bias introduced when selecting $s_0$.

\subsection{Mass Estimations for the Renormalized Four-Quark Currents}\label{sec_mass_r}

\begin{figure}[ht!]
	\begin{subfigure}{0.45\textwidth}
		\includegraphics[width=\textwidth]{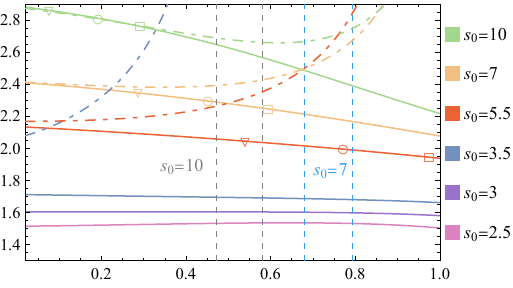}
		\caption{$\sqrt{\mathcal{R}^0}$ for $J^r_{4,[8_S]}$.}
		\label{r0_t_j4_8_s}
	\end{subfigure}
	\hspace*{\fill}	
	\begin{subfigure}{0.45\textwidth}
		\includegraphics[width=\textwidth]{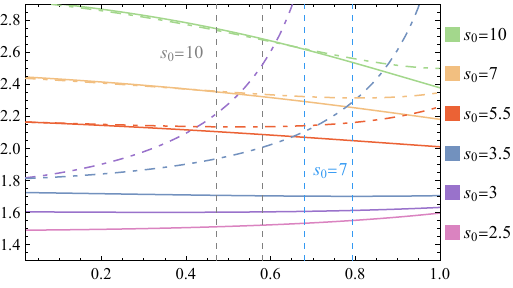}
		\caption{$\sqrt{\mathcal{R}^1}$ for $J^r_{4,[8_S]}$.}
		\label{r1_t_j4_8_s}
	\end{subfigure}\\[0cm]
	\begin{subfigure}{0.45\textwidth}
		\includegraphics[width=\textwidth]{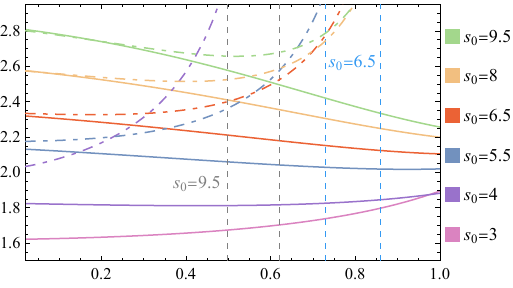}
		\caption{$\sqrt{\mathcal{R}^0}$ for $J^r_{4,1_S}$.}
		\label{r0_t_j4_1_s_apdix}
	\end{subfigure}
	\hspace*{\fill}	
	\begin{subfigure}{0.45\textwidth}
		\includegraphics[width=\textwidth]{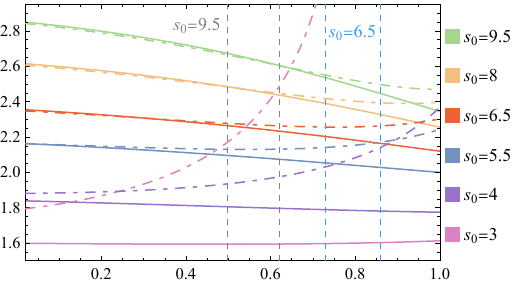}
		\caption{$\sqrt{\mathcal{R}^1}$ for $J^r_{4,1_S}$.}
		\label{r1_t_j4_1_s}
	\end{subfigure}
	\caption{$\sqrt{\mathcal{R}^n}(\text{GeV})$ versus $\tau(\text{GeV}^{-2})$ for different values of $s_0(\text{GeV}^2)$, corresponding to the renormalized four-quark currents in category 1.}
	\label{fig_r_ca_1_1}
\end{figure}

\begin{figure}[h!]
	\begin{subfigure}{0.45\textwidth}
		\includegraphics[width=\textwidth]{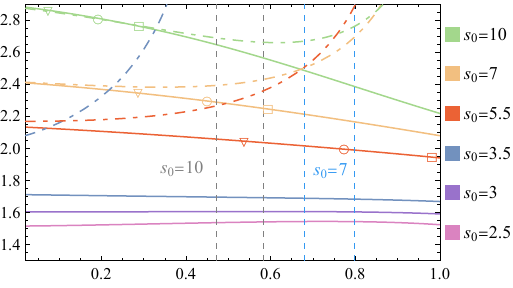}
		\caption{$\sqrt{\mathcal{R}^0}$ for $J^r_{4,8_S}$ with $i_3=\frac{1}{2}$ and $Y=\frac{3}{2}$.}
		\label{r0_t_j4_8_s_2}
	\end{subfigure}
	\hspace*{\fill}	
	\begin{subfigure}{0.45\textwidth}
		\includegraphics[width=\textwidth]{./j4_8_s/r1_t_1}
		\caption{$\sqrt{\mathcal{R}^1}$ for $J^r_{4,8_S}$ with $i_3=\frac{1}{2}$ and $Y=\frac{3}{2}$.}
		\label{r1_t_j4_8_s_2}
	\end{subfigure}\\[0.3cm]
	\begin{subfigure}{0.45\textwidth}
		\includegraphics[width=\textwidth]{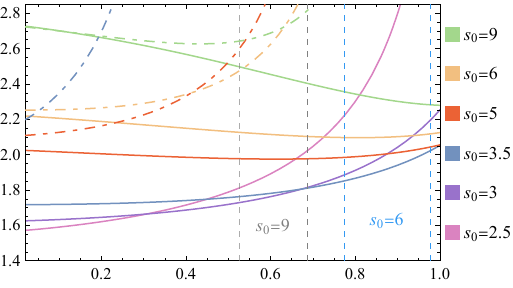}
		\caption{$\sqrt{\mathcal{R}^0}$ for $J^r_{4,8_S}$ with $i_3=1$ and $Y=0$.}
		\label{r0_t_j4_8_s_3}
	\end{subfigure}
	\hspace*{\fill}	
	\begin{subfigure}{0.45\textwidth}
		\includegraphics[width=\textwidth]{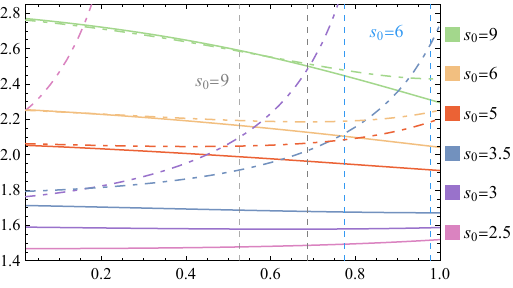}
		\caption{$\sqrt{\mathcal{R}^1}$ for $J^r_{4,8_S}$ with $i_3=1$ and $Y=0$.}
		\label{r1_t_j4_8_s_3}
	\end{subfigure}\\[0.3cm]
		\begin{subfigure}{0.45\textwidth}
		\includegraphics[width=\textwidth]{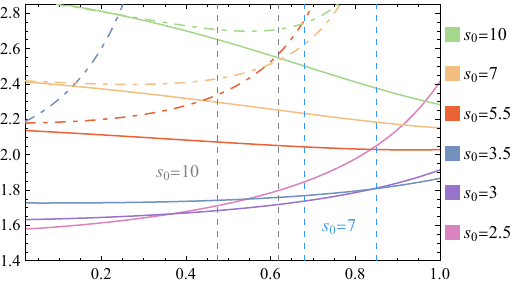}
		\caption{$\sqrt{\mathcal{R}^0}$ for $J^r_{4,8_S}$ with $i_3=Y=0$.}
		\label{r0_t_j4_8_s_7}
	\end{subfigure}
	\hspace*{\fill}	
	\begin{subfigure}{0.45\textwidth}
		\includegraphics[width=\textwidth]{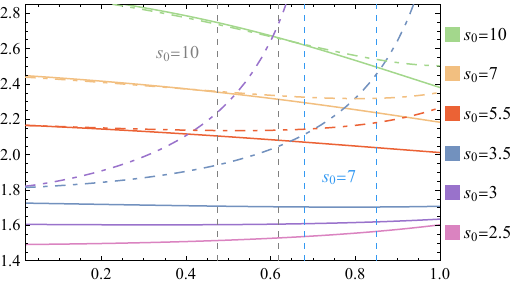}
		\caption{$\sqrt{\mathcal{R}^1}$ for $J^r_{4,8_S}$ with $i_3=Y=0$.}
		\label{r1_t_j4_8_s_7}
	\end{subfigure}\\[0.3cm]
	\begin{subfigure}{0.45\textwidth}
		\includegraphics[width=\textwidth]{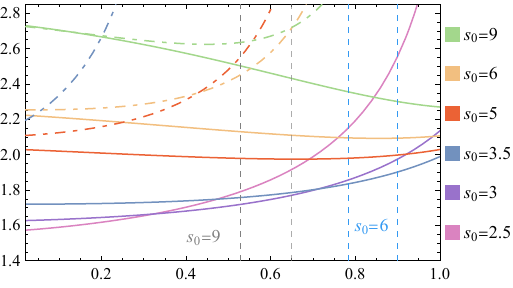}
		\caption{$\sqrt{\mathcal{R}^0}$ for $J^r_{4,8_S}$ with $i_3=Y=0$.}
		\label{r0_t_j4_8_s_8}
	\end{subfigure}
	\hspace*{\fill}	
	\begin{subfigure}{0.45\textwidth}
		\includegraphics[width=\textwidth]{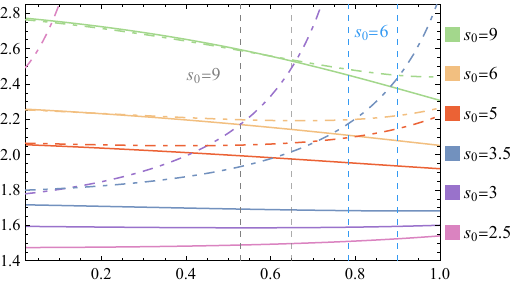}
		\caption{$\sqrt{\mathcal{R}^1}$ for $J^r_{4,8_S}$ with $i_3=Y=0$.}
		\label{r1_t_j4_8_s_8}
	\end{subfigure}
	\caption{$\sqrt{\mathcal{R}^n}(\text{GeV})$ versus $\tau(\text{GeV}^{-2})$ for different values of $s_0(\text{GeV}^2)$, corresponding to $J^r_{4,8_S}$. Here $i_3$ denotes the third component of isospin. Figures (e) and (f) correspond to the upper $i_3=0$ state in figure~\ref{su(3)_j4_nonet}; figures (g) and (h) correspond to the lower $i_3=0$ state in figure~\ref{su(3)_j4_nonet}. }
	\label{fig_r_j4_nonet}
\end{figure}

\begin{figure}[h!]
	\begin{subfigure}{0.45\textwidth}
		\includegraphics[width=\textwidth]{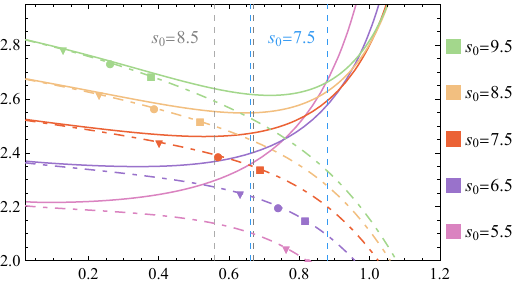}
		\caption{$\sqrt{\mathcal{R}^0}$ for $J^r_{a,[27_S]}$.}
		\label{r0_t_ja_27_s}
	\end{subfigure}
	\hspace*{\fill}	
	\begin{subfigure}{0.45\textwidth}
		\includegraphics[width=\textwidth]{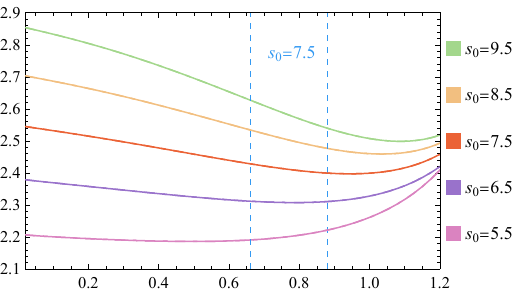}
		\caption{$\sqrt{\mathcal{R}^1}$ for $J^r_{a,[27_S]}$.}
		\label{r1_t_ja_27_s}
	\end{subfigure}\\[0.3cm]
	\begin{subfigure}{0.45\textwidth}
		\includegraphics[width=\textwidth]{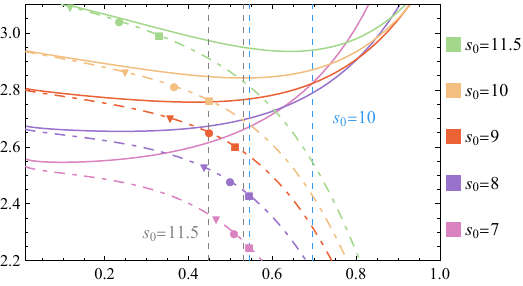}
		\caption{$\sqrt{\mathcal{R}^0}$ for $J^r_{d,[27_S]}$.}
		\label{r0_t_jd_27_s}
	\end{subfigure}
	\hspace*{\fill}	
	\begin{subfigure}{0.45\textwidth}
		\includegraphics[width=\textwidth]{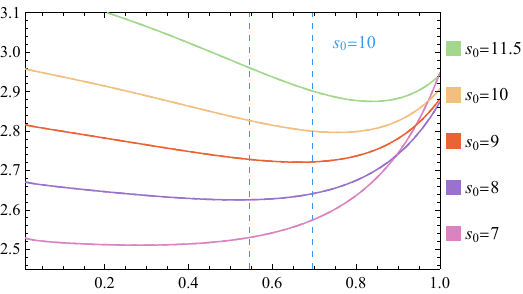}
		\caption{$\sqrt{\mathcal{R}^1}$ for $J^r_{d,[27_S]}$.}
		\label{r1_t_jd_27_s}
	\end{subfigure}\\[0.3cm]
	\begin{subfigure}{0.45\textwidth}
		\includegraphics[width=\textwidth]{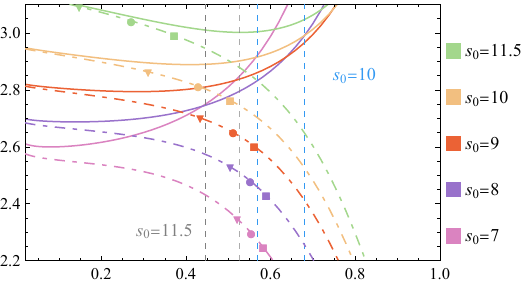}
		\caption{$\sqrt{\mathcal{R}^0}$ for $J^r_{d,[8_S]}$.}
		\label{r0_t_jd_8_s}
	\end{subfigure}
	\hspace*{\fill}	
	\begin{subfigure}{0.45\textwidth}
		\includegraphics[width=\textwidth]{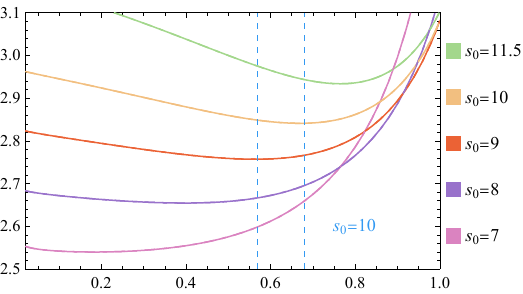}
		\caption{$\sqrt{\mathcal{R}^1}$ for $J^r_{d,[8_S]}$.}
		\label{r1_t_jd_8_s}
	\end{subfigure}\\[0.3cm]
	\begin{subfigure}{0.45\textwidth}
		\includegraphics[width=\textwidth]{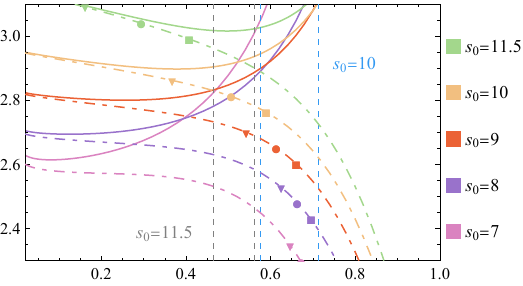}
		\caption{$\sqrt{\mathcal{R}^0}$ for $J^r_{d,1_S}$.}
		\label{r0_t_jd_1_s}
	\end{subfigure}
	\hspace*{\fill}	
	\begin{subfigure}{0.45\textwidth}
		\includegraphics[width=\textwidth]{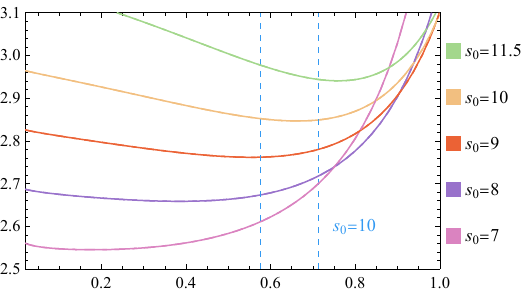}
		\caption{$\sqrt{\mathcal{R}^1}$ for $J^r_{d,1_S}$.}
		\label{r1_t_jd_1_s}
	\end{subfigure}\\[0.3cm]
	\caption{$\sqrt{\mathcal{R}^n}(\text{GeV})$ versus $\tau(\text{GeV}^{-2})$ for different values of $s_0(\text{GeV}^2)$, corresponding to the renormalized four-quark currents in category 2, with masses close to $3\text{GeV}$.}
	\label{fig_r_ca_2_3}
\end{figure}

\begin{figure}[h!]
	\begin{subfigure}{0.45\textwidth}
		\includegraphics[width=\textwidth]{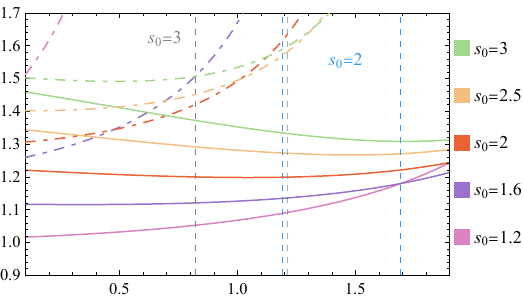}
		\caption{$\sqrt{\mathcal{R}^0}$ for $J^r_{a,[8_S]}$.}
		\label{r0_t_ja_8_s}
	\end{subfigure}
	\hspace*{\fill}	
	\begin{subfigure}{0.45\textwidth}
		\includegraphics[width=\textwidth]{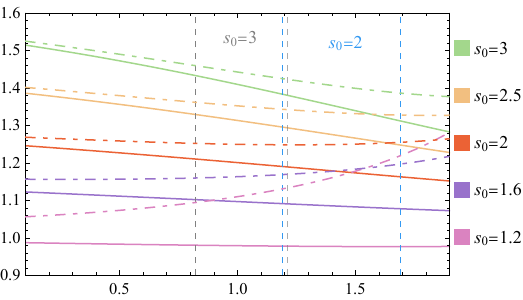}
		\caption{$\sqrt{\mathcal{R}^1}$ for $J^r_{a,[8_S]}$.}
		\label{r1_t_ja_8_s}
	\end{subfigure}\\[0.3cm]
	\begin{subfigure}{0.45\textwidth}
		\includegraphics[width=\textwidth]{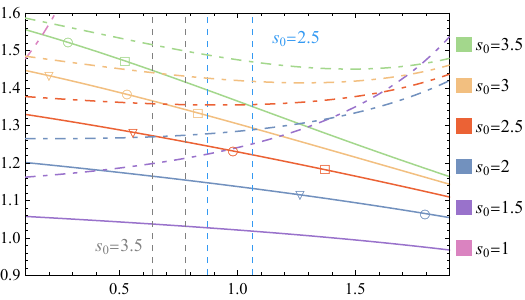}
		\caption{$\sqrt{\mathcal{R}^0}$ for $J^r_{a,1_S}$.}
		\label{r0_t_ja_1_s}
	\end{subfigure}
	\hspace*{\fill}	
	\begin{subfigure}{0.45\textwidth}
		\includegraphics[width=\textwidth]{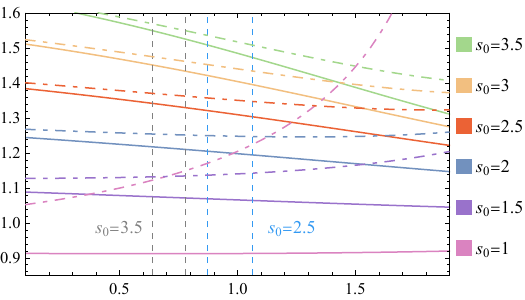}
		\caption{$\sqrt{\mathcal{R}^1}$ for $J^r_{a,1_S}$.}
		\label{r1_t_ja_1_s}
	\end{subfigure}\\[0.3cm]
	\begin{subfigure}{0.45\textwidth}
		\includegraphics[width=\textwidth]{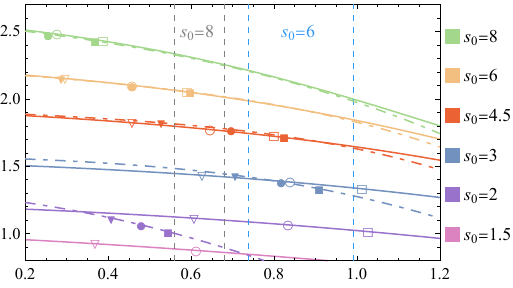}
		\caption{$\sqrt{\mathcal{R}^0}$ for $J^r_{e,[8_A]}$.}
		\label{r0_t_je_8_a}
	\end{subfigure}
	\hspace*{\fill}	
	\begin{subfigure}{0.45\textwidth}
		\includegraphics[width=\textwidth]{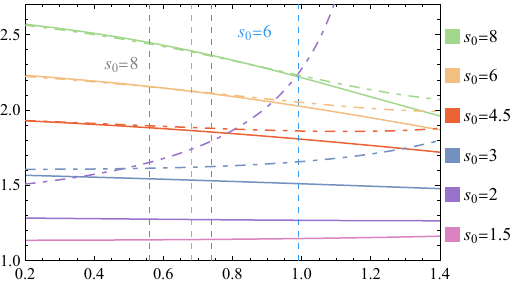}
		\caption{$\sqrt{\mathcal{R}^1}$ for $J^r_{e,[8_A]}$.}
		\label{r1_t_je_8_a}
	\end{subfigure}\\[0.3cm]
	\begin{subfigure}{0.45\textwidth}
		\includegraphics[width=\textwidth]{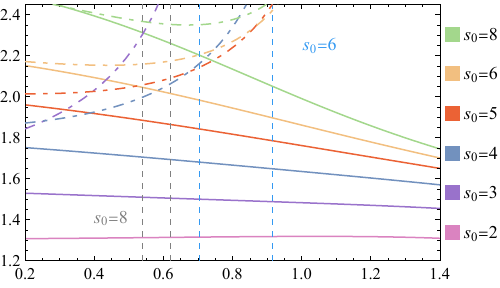}
		\caption{$\sqrt{\mathcal{R}^0}$ for $J^r_{e,1_A}$.}
		\label{r0_t_je_1_a}
	\end{subfigure}
	\hspace*{\fill}	
	\begin{subfigure}{0.45\textwidth}
		\includegraphics[width=\textwidth]{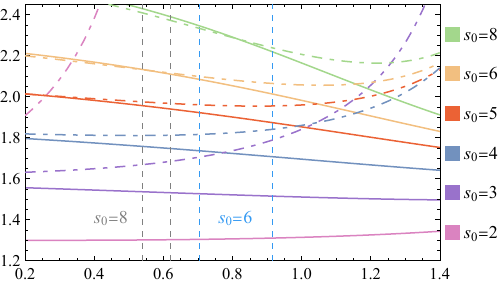}
		\caption{$\sqrt{\mathcal{R}^1}$ for $J^r_{e,1_A}$.}
		\label{r1_t_je_1_a}
	\end{subfigure}\\[0.3cm]
	\caption{$\sqrt{\mathcal{R}^n}(\text{GeV})$ versus $\tau(\text{GeV}^{-2})$ for different values of $s_0(\text{GeV}^2)$, corresponding to the renormalized four-quark currents in category 3.}
	\label{fig_r_ca_3}
\end{figure}

\begin{figure}[h!]
	\begin{subfigure}{0.45\textwidth}
		\includegraphics[width=\textwidth]{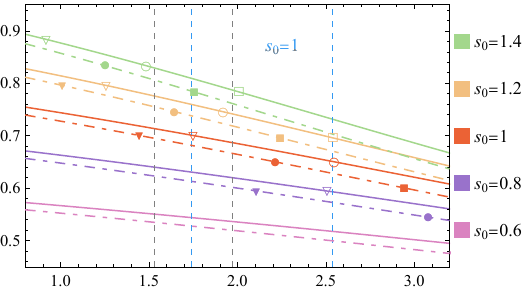}
		\caption{$\sqrt{\mathcal{R}^0}$ for $J^r_{1,[8_A]}$.}
		\label{r0_t_j1_8_a}
	\end{subfigure}
	\hspace*{\fill}	
	\begin{subfigure}{0.45\textwidth}
		\includegraphics[width=\textwidth]{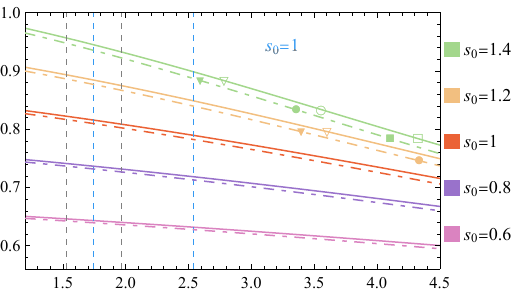}
		\caption{$\sqrt{\mathcal{R}^1}$ for $J^r_{1,[8_A]}$.}
		\label{r1_t_j1_8_a}
	\end{subfigure}\\[0.3cm]
	\begin{subfigure}{0.45\textwidth}
		\includegraphics[width=\textwidth]{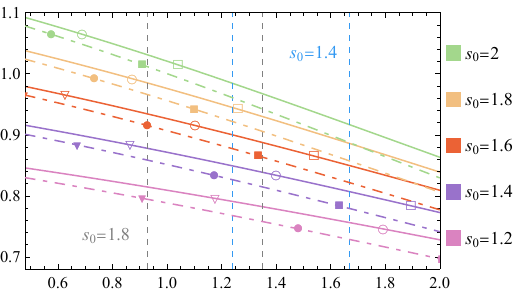}
		\caption{$\sqrt{\mathcal{R}^0}$ for $J^r_{1,1_A}$.}
		\label{r0_t_j1_1_a}
	\end{subfigure}
	\hspace*{\fill}	
	\begin{subfigure}{0.45\textwidth}
		\includegraphics[width=\textwidth]{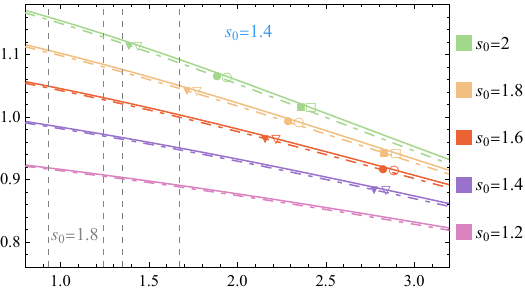}
		\caption{$\sqrt{\mathcal{R}^1}$ for $J^r_{1,1_A}$.}
		\label{r1_t_j1_1_a}
	\end{subfigure}\\[0.3cm]
	\begin{subfigure}{0.45\textwidth}
		\includegraphics[width=\textwidth]{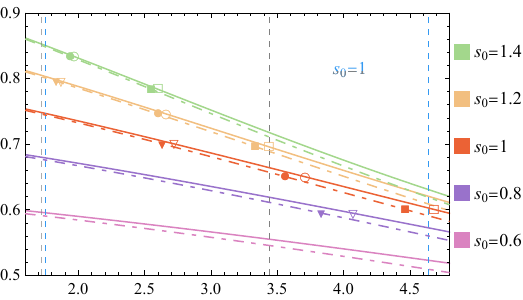}
		\caption{$\sqrt{\mathcal{R}^0}$ for $J^r_{2,[8_S]}$.}
		\label{r0_t_j2_8_s}
	\end{subfigure}
	\hspace*{\fill}	
	\begin{subfigure}{0.45\textwidth}
		\includegraphics[width=\textwidth]{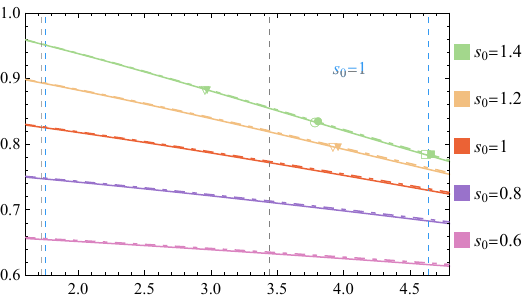}
		\caption{$\sqrt{\mathcal{R}^1}$ for $J^r_{2,[8_S]}$.}
		\label{r1_t_j2_8_s}
	\end{subfigure}\\[0.3cm]
	\begin{subfigure}{0.45\textwidth}
		\includegraphics[width=\textwidth]{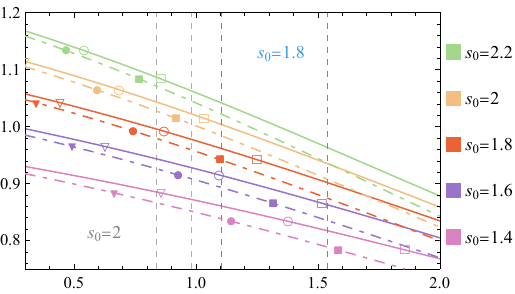}
		\caption{$\sqrt{\mathcal{R}^0}$ for $J^r_{2,1_S}$.}
		\label{r0_t_j2_1_s}
	\end{subfigure}
	\hspace*{\fill}	
	\begin{subfigure}{0.45\textwidth}
		\includegraphics[width=\textwidth]{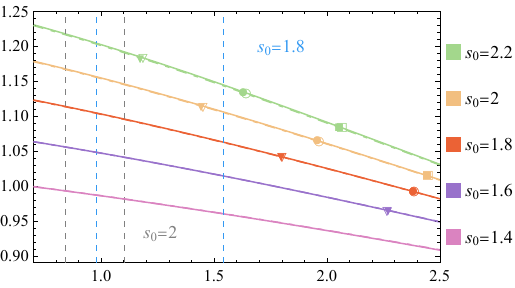}
		\caption{$\sqrt{\mathcal{R}^1}$ for $J^r_{2,1_S}$.}
		\label{r1_t_j2_1_s}
	\end{subfigure}\\[0.3cm]
	\caption{$\sqrt{\mathcal{R}^n}(\text{GeV})$ versus $\tau(\text{GeV}^{-2})$ for different values of $s_0(\text{GeV}^2)$, corresponding to the renormalized four-quark currents in category 4.}
	\label{fig_r_ca_4}
\end{figure}

\begin{figure}[h!]
	\begin{subfigure}{0.45\textwidth}
		\includegraphics[width=\textwidth]{./plot/r0_t_jb_8_a}
		\caption{$\sqrt{\mathcal{R}^0}$ for $J^r_{b,[8_A]}$.}
	\end{subfigure}
	\hspace*{\fill}	
	\begin{subfigure}{0.45\textwidth}
		\includegraphics[width=\textwidth]{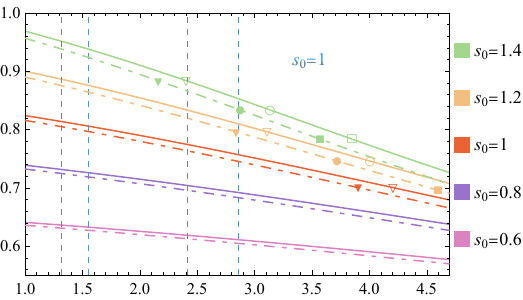}
		\caption{$\sqrt{\mathcal{R}^1}$ for $J^r_{b,[8_A]}$.}
	\end{subfigure}\\[0.1cm]
	\begin{subfigure}{0.45\textwidth}
		\includegraphics[width=\textwidth]{./plot/r0_t_jb_1_a}
		\caption{$\sqrt{\mathcal{R}^0}$ for $J^r_{b,1_A}$.}
	\end{subfigure}
	\hspace*{\fill}	
	\begin{subfigure}{0.45\textwidth}
		\includegraphics[width=\textwidth]{./plot/r1_t_jb_1_a}
		\caption{$\sqrt{\mathcal{R}^1}$ for $J^r_{b,1_A}$.}
	\end{subfigure}
	\caption{$\sqrt{\mathcal{R}^n}(\text{GeV})$ versus $\tau(\text{GeV}^{-2})$ for different values of $s_0(\text{GeV}^2)$, corresponding to the renormalized four-quark currents in category 4, with masses lower than $1\text{GeV}$.}
	\label{fig_r_ca_4_1}
\end{figure}

\begin{figure}[h!]
	\begin{subfigure}{0.45\textwidth}
		\includegraphics[width=\textwidth]{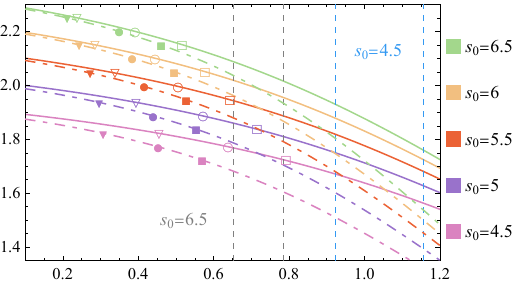}
		\caption{$\sqrt{\mathcal{R}^0}$ for $J^r_{c,[8_A]}$.}
	\end{subfigure}
	\hspace*{\fill}	
	\begin{subfigure}{0.45\textwidth}
		\includegraphics[width=\textwidth]{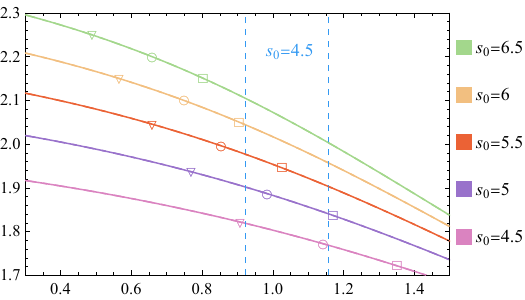}
		\caption{$\sqrt{\mathcal{R}^1}$ for $J^r_{c,[8_A]}$.}
	\end{subfigure}\\[0.1cm]
	\begin{subfigure}{0.45\textwidth}
		\includegraphics[width=\textwidth]{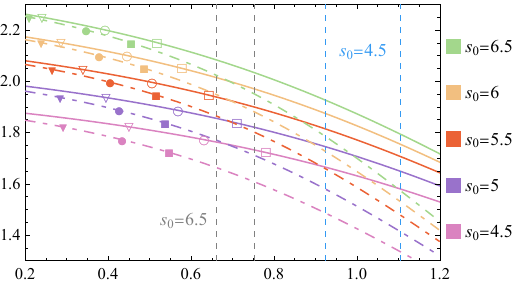}
		\caption{$\sqrt{\mathcal{R}^0}$ for $J^r_{c,1_A}$.}
	\end{subfigure}
	\hspace*{\fill}	
	\begin{subfigure}{0.45\textwidth}
		\includegraphics[width=\textwidth]{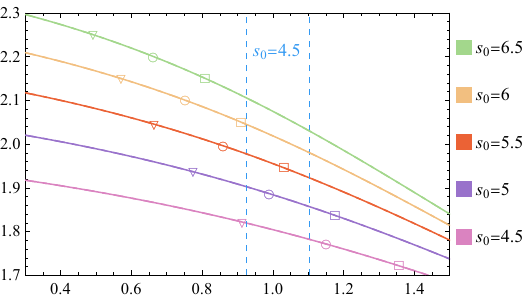}
		\caption{$\sqrt{\mathcal{R}^1}$ for $J^r_{c,1_A}$.}
	\end{subfigure}
	\caption{$\sqrt{\mathcal{R}^n}(\text{GeV})$ versus $\tau(\text{GeV}^{-2})$ for different values of $s_0(\text{GeV}^2)$, corresponding to the renormalized four-quark currents in category 5, with masses around $2\text{GeV}$.}
	\label{fig_r_ca_5}
\end{figure}


\begin{figure}[ht!]
	\begin{subfigure}{0.45\textwidth}
		\includegraphics[width=\textwidth]{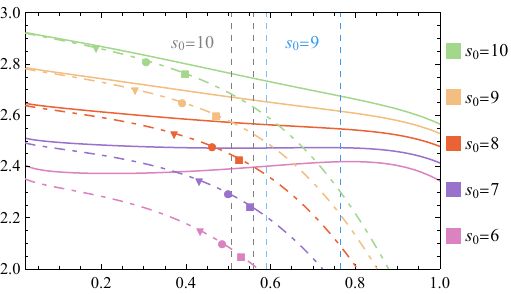}
		\caption{$\sqrt{\mathcal{R}^0}$ for $I^r_{3,[27_S]}$.}
		\label{r0_t_i3_27_s}
	\end{subfigure}
	\hspace*{\fill}	
	\begin{subfigure}{0.45\textwidth}
		\includegraphics[width=\textwidth]{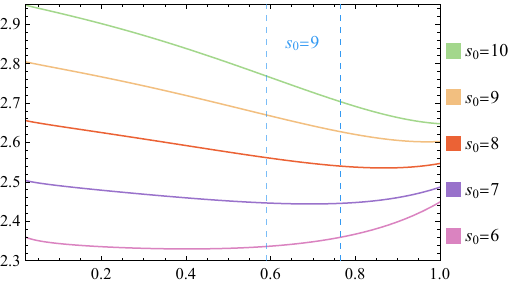}
		\caption{$\sqrt{\mathcal{R}^1}$ for $I^r_{3,[27_S]}$.}
		\label{r1_t_i3_27_s}
	\end{subfigure}\\[0.3cm]
	\begin{subfigure}{0.45\textwidth}
		\includegraphics[width=\textwidth]{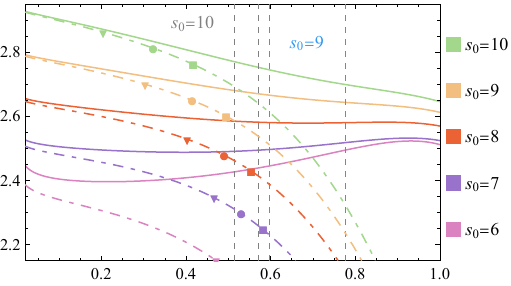}
		\caption{$\sqrt{\mathcal{R}^0}$ for $I^r_{3,[8_S]}$.}
		\label{r0_t_i3_8_s}
	\end{subfigure}
	\hspace*{\fill}	
	\begin{subfigure}{0.45\textwidth}
		\includegraphics[width=\textwidth]{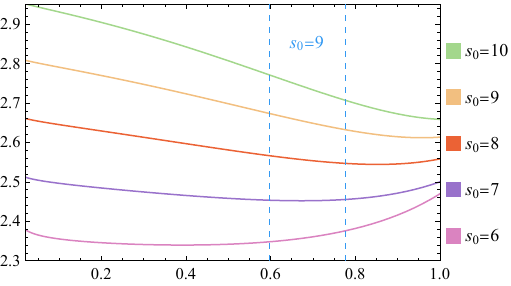}
		\caption{$\sqrt{\mathcal{R}^1}$ for $I^r_{3,[8_S]}$.}
		\label{r1_t_i3_8_s}
	\end{subfigure}\\[0.3cm]
	\begin{subfigure}{0.45\textwidth}
		\includegraphics[width=\textwidth]{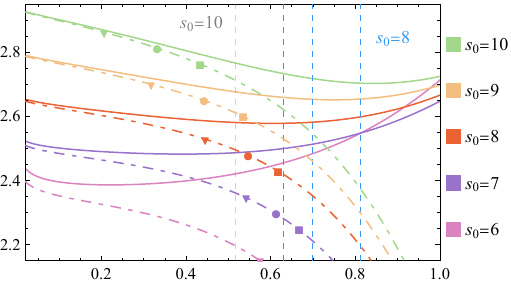}
		\caption{$\sqrt{\mathcal{R}^0}$ for $I^r_{3,1_S}$.}
		\label{r0_t_i3_1_s}
	\end{subfigure}
	\hspace*{\fill}	
	\begin{subfigure}{0.45\textwidth}
		\includegraphics[width=\textwidth]{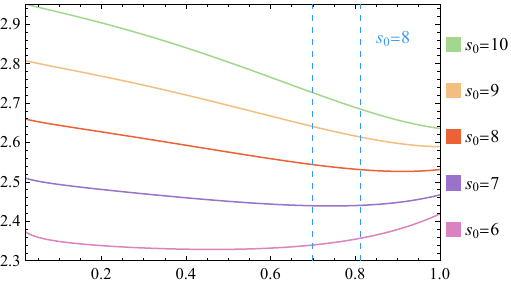}
		\caption{$\sqrt{\mathcal{R}^1}$ for $I^r_{3,1_S}$.}
		\label{r1_t_i3_1_s}
	\end{subfigure}
	\caption{$\sqrt{\mathcal{R}^n}(\text{GeV})$ versus $\tau(\text{GeV}^{-2})$ for different values of $s_0(\text{GeV}^2)$, corresponding to the renormalized four-quark currents in category 2. Compared to figure~\ref{fig_r_ca_2_3}, the masses here are lower by $\sim0.3\text{GeV}$. This difference originates from the $\gamma^5$ involved in Fierz rearrangement, which breaks the correspondence listed in table~\ref{bases}.}
	\label{fig_r_ca_2_3_m}
\end{figure}

\begin{figure}[ht!]
	\begin{subfigure}{0.45\textwidth}
		\includegraphics[width=\textwidth]{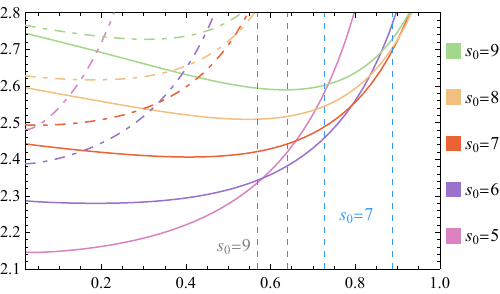}
		\caption{$\sqrt{\mathcal{R}^0}$ for $J^r_{2,[27_S]}$.}
		\label{r0_t_j2_27_s}
	\end{subfigure}
	\hspace*{\fill}	
	\begin{subfigure}{0.45\textwidth}
		\includegraphics[width=\textwidth]{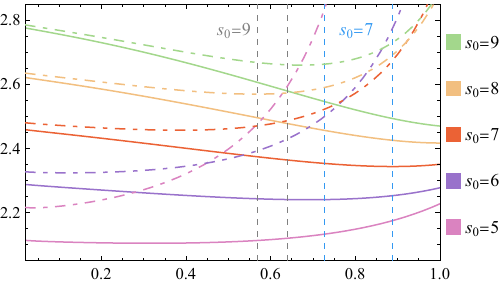}
		\caption{$\sqrt{\mathcal{R}^1}$ for $J^r_{2,[27_S]}$.}
		\label{r1_t_j2_27_s}
	\end{subfigure}\\[0.1cm]
	\begin{subfigure}{0.45\textwidth}
		\includegraphics[width=\textwidth]{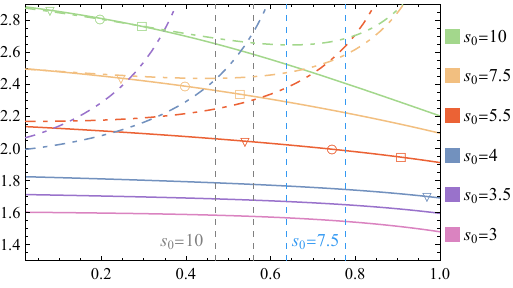}
		\caption{$\sqrt{\mathcal{R}^0}$ for $J^r_{4,[27_S]}$.}
		\label{r0_t_j4_27_s}
	\end{subfigure}
	\hspace*{\fill}	
	\begin{subfigure}{0.45\textwidth}
		\includegraphics[width=\textwidth]{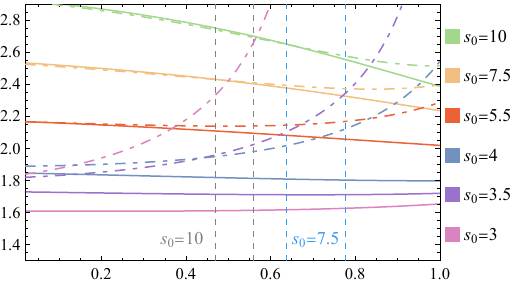}
		\caption{$\sqrt{\mathcal{R}^1}$ for $J^r_{4,[27_S]}$.}
		\label{r1_t_j4_27_s}
	\end{subfigure}
	\caption{$\sqrt{\mathcal{R}^n}(\text{GeV})$ versus $\tau(\text{GeV}^{-2})$ for different values of $s_0(\text{GeV}^2)$, corresponding to the renormalized four-quark currents in category 1.}
	\label{fig_r_ca_1}
\end{figure}

\begin{figure}[h!]
	\begin{subfigure}{0.45\textwidth}
		\includegraphics[width=\textwidth]{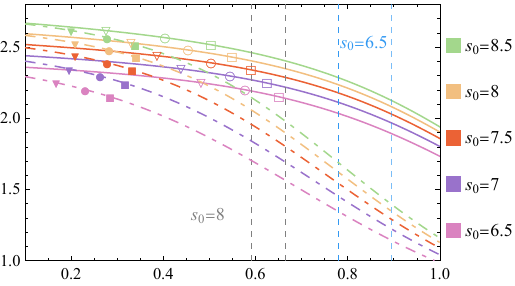}
		\caption{$\sqrt{\mathcal{R}^0}$ for $J^r_{3,[8_A]}$.}
		\label{r0_t_j3_8_a}
	\end{subfigure}
	\hspace*{\fill}	
	\begin{subfigure}{0.45\textwidth}
		\includegraphics[width=\textwidth]{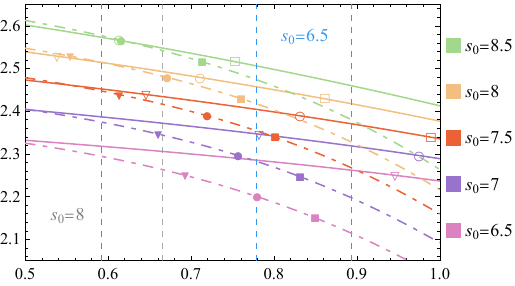}
		\caption{$\sqrt{\mathcal{R}^1}$ for $J^r_{3,[8_A]}$.}
		\label{r1_t_j3_8_a}
	\end{subfigure}\\[0.1cm]
	\begin{subfigure}{0.45\textwidth}
		\includegraphics[width=\textwidth]{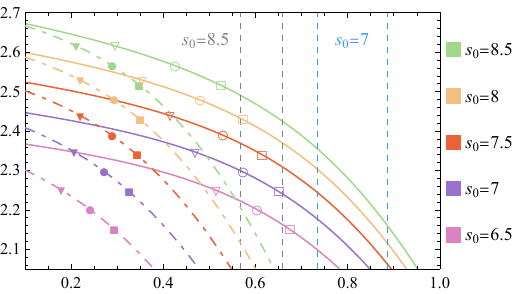}
		\caption{$\sqrt{\mathcal{R}^0}$ for $J^r_{3,1_A}$.}
		\label{r0_t_j3_1_a}
	\end{subfigure}
	\hspace*{\fill}	
	\begin{subfigure}{0.45\textwidth}
		\includegraphics[width=\textwidth]{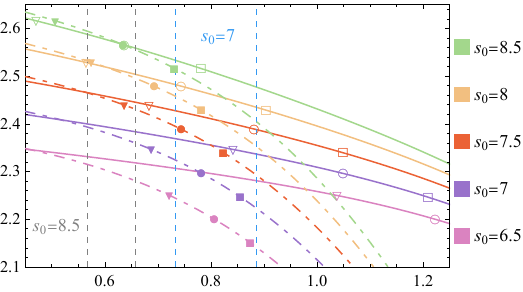}
		\caption{$\sqrt{\mathcal{R}^1}$ for $J^r_{3,1_A}$.}
		\label{r1_t_j3_1_a}
	\end{subfigure}
	\caption{$\sqrt{\mathcal{R}^n}(\text{GeV})$ versus $\tau(\text{GeV}^{-2})$ for different values of $s_0(\text{GeV}^2)$, corresponding to the renormalized four-quark currents in category 1; apply the constraint $s_0=(m+\Lambda)^2$.}
	\label{fig_r_ca_1_c}
\end{figure}

\clearpage
\subsection{Mass Estimations for the Bare Tetraquark Currents}\label{sec_mass_b_t}

\begin{figure}[h]
	\begin{subfigure}{0.45\textwidth}
		\includegraphics[width=\textwidth]{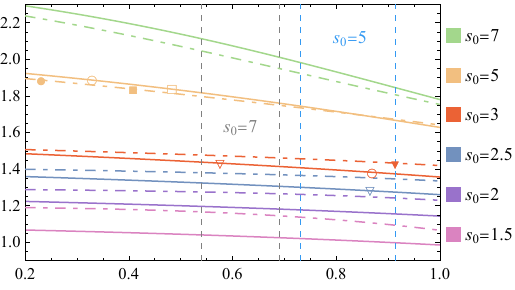}
		\caption{$\sqrt{\mathcal{R}^0}$ for $J_{2,[27_S]}$.}
		\label{r0_t_j2_27_s_bare}
	\end{subfigure}
	\hspace*{\fill}	
	\begin{subfigure}{0.45\textwidth}
		\includegraphics[width=\textwidth]{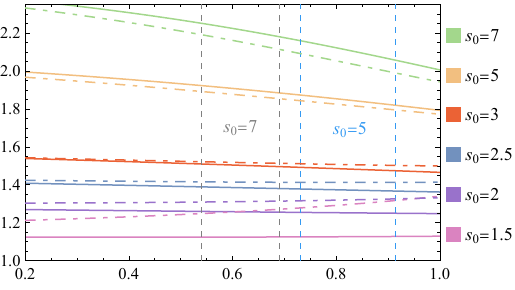}
		\caption{$\sqrt{\mathcal{R}^1}$ for $J_{2,[27_S]}$.}
		\label{r1_t_j2_27_s_bare}
	\end{subfigure}\\[0.3cm]
	\begin{subfigure}{0.45\textwidth}
		\includegraphics[width=\textwidth]{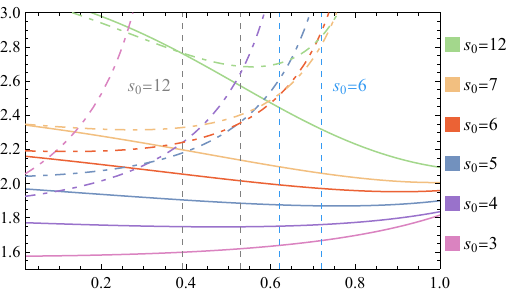}
		\caption{$\sqrt{\mathcal{R}^0}$ for $J_{c,[27_S]}$.}
		\label{r0_t_jc_27_s_bare}
	\end{subfigure}
	\hspace*{\fill}	
	\begin{subfigure}{0.45\textwidth}
		\includegraphics[width=\textwidth]{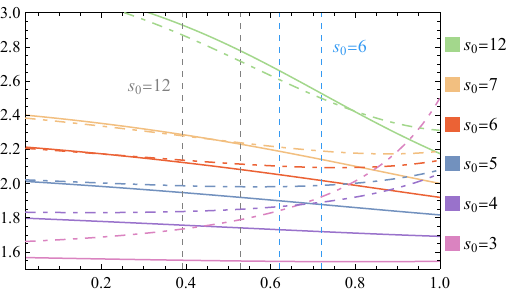}
		\caption{$\sqrt{\mathcal{R}^1}$ for $J_{c,[27_S]}$.}
		\label{r1_t_jc_27_s_bare}
	\end{subfigure}\\[0.3cm]
	\begin{subfigure}{0.45\textwidth}
		\includegraphics[width=\textwidth]{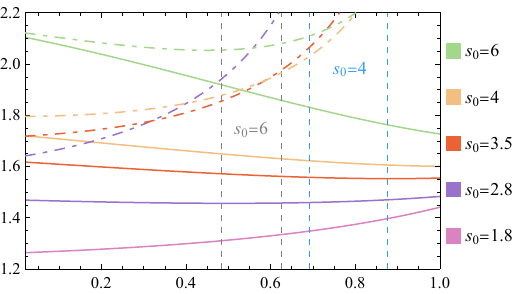}
		\caption{$\sqrt{\mathcal{R}^0}$ for $J_{c,[8_S]}$.}
		\label{r0_t_jc_8_s_bare}
	\end{subfigure}
	\hspace*{\fill}	
	\begin{subfigure}{0.45\textwidth}
		\includegraphics[width=\textwidth]{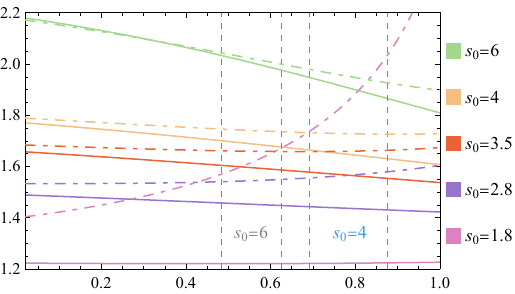}
		\caption{$\sqrt{\mathcal{R}^1}$ for $J_{c,[8_S]}$.}
		\label{r1_t_jc_8_s_bare}
	\end{subfigure}\\[0.3cm]
	\begin{subfigure}{0.45\textwidth}
		\includegraphics[width=\textwidth]{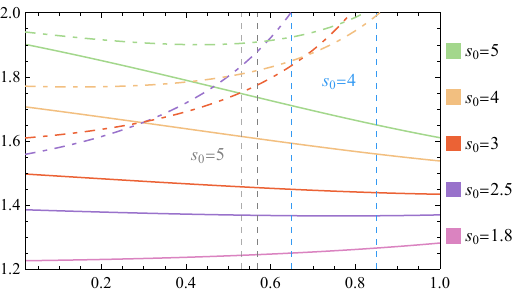}
		\caption{$\sqrt{\mathcal{R}^0}$ for $J_{c,1_S}$.}
		\label{r0_t_jc_1_s_bare}
	\end{subfigure}
	\hspace*{\fill}	
	\begin{subfigure}{0.45\textwidth}
		\includegraphics[width=\textwidth]{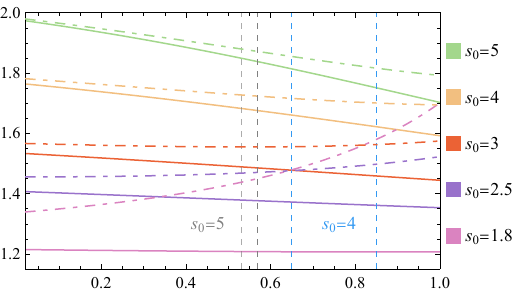}
		\caption{$\sqrt{\mathcal{R}^1}$ for $J_{c,1_S}$.}
		\label{r1_t_jc_1_s_bare}
	\end{subfigure}
	\caption{$\sqrt{\mathcal{R}^n}(\text{GeV})$ versus $\tau(\text{GeV}^{-2})$ for different values of $s_0(\text{GeV}^2)$, corresponding to the bare tetraquark currents in category 3.}
	\label{fig_t_ca_3_1}
\end{figure}

\begin{figure}[h]
	\begin{subfigure}{0.45\textwidth}
		\includegraphics[width=\textwidth]{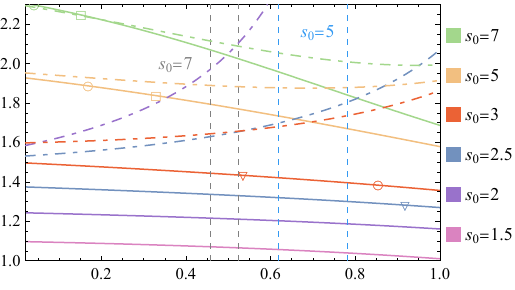}
		\caption{$\sqrt{\mathcal{R}^0}$ for $J_{d,[8_A]}$.}
		\label{r0_t_jd_8_a_bare}
	\end{subfigure}
	\hspace*{\fill}	
	\begin{subfigure}{0.45\textwidth}
		\includegraphics[width=\textwidth]{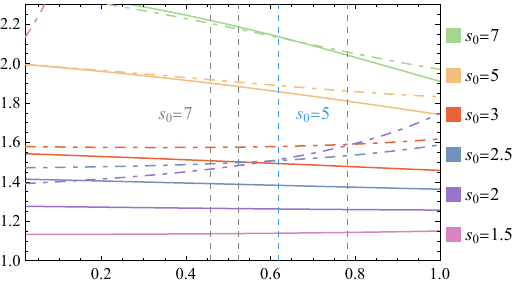}
		\caption{$\sqrt{\mathcal{R}^1}$ for $J_{d,[8_A]}$.}
		\label{r1_t_jd_8_a_bare}
	\end{subfigure}\\[0.3cm]
	\begin{subfigure}{0.45\textwidth}
		\includegraphics[width=\textwidth]{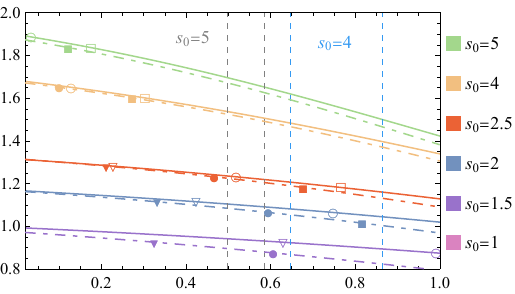}
		\caption{$\sqrt{\mathcal{R}^0}$ for $J_{d,1_A}$.}
		\label{r0_t_jd_1_a_bare}
	\end{subfigure}
	\hspace*{\fill}	
	\begin{subfigure}{0.45\textwidth}
		\includegraphics[width=\textwidth]{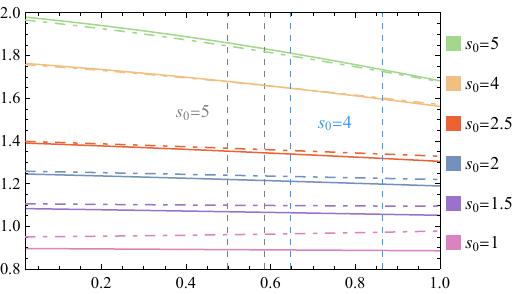}
		\caption{$\sqrt{\mathcal{R}^1}$ for $J_{d,1_A}$.}
		\label{r1_t_jd_1_a_bare}
	\end{subfigure}\\[0.3cm]
	\begin{subfigure}{0.45\textwidth}
		\includegraphics[width=\textwidth]{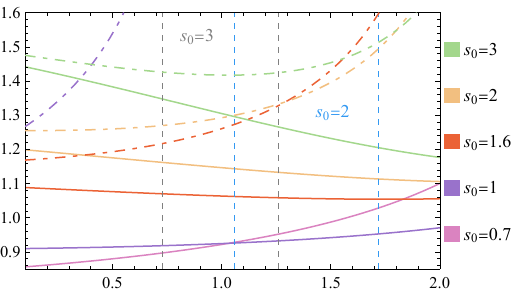}
		\caption{$\sqrt{\mathcal{R}^0}$ for $J_{f,[8_S]}$.}
		\label{r0_t_jf_8_s_bare}
	\end{subfigure}
	\hspace*{\fill}	
	\begin{subfigure}{0.45\textwidth}
		\includegraphics[width=\textwidth]{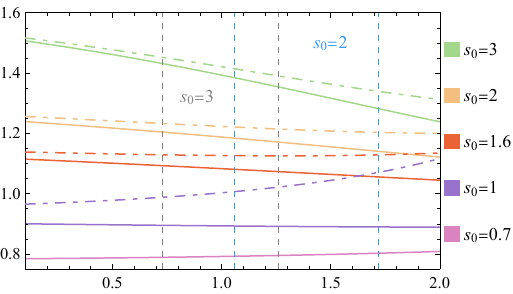}
		\caption{$\sqrt{\mathcal{R}^1}$ for $J_{f,[8_S]}$.}
		\label{r1_t_jf_8_s_bare}
	\end{subfigure}\\[0.3cm]
	\begin{subfigure}{0.45\textwidth}
		\includegraphics[width=\textwidth]{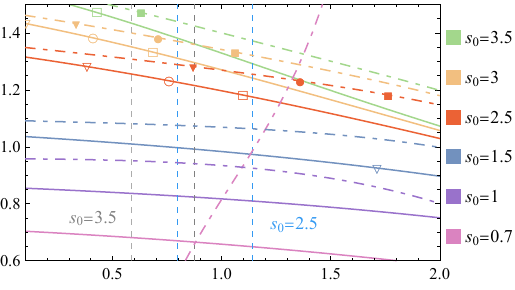}
		\caption{$\sqrt{\mathcal{R}^0}$ for $J_{f,1_S}$.}
		\label{r0_t_jf_1_s_bare}
	\end{subfigure}
	\hspace*{\fill}	
	\begin{subfigure}{0.45\textwidth}
		\includegraphics[width=\textwidth]{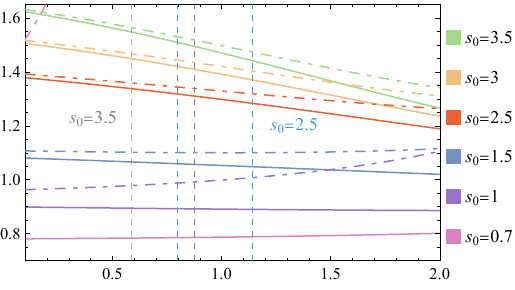}
		\caption{$\sqrt{\mathcal{R}^1}$ for $J_{f,1_S}$.}
		\label{r1_t_jf_1_s_bare}
	\end{subfigure}
	\caption{$\sqrt{\mathcal{R}^n}(\text{GeV})$ versus $\tau(\text{GeV}^{-2})$ for different values of $s_0(\text{GeV}^2)$, corresponding to the bare tetraquark currents in category 3.}
	\label{fig_t_ca_3_2}
\end{figure}

\begin{figure}[h]
	\begin{subfigure}{0.45\textwidth}
		\includegraphics[width=\textwidth]{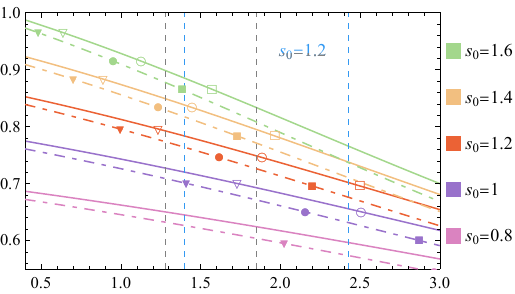}
		\caption{$\sqrt{\mathcal{R}^0}$ for $J_{1,[8_A]}$.}
		\label{r0_t_j1_8_a_bare}
	\end{subfigure}
	\hspace*{\fill}	
	\begin{subfigure}{0.45\textwidth}
		\includegraphics[width=\textwidth]{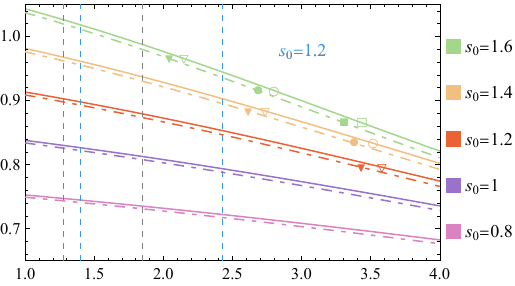}
		\caption{$\sqrt{\mathcal{R}^1}$ for $J_{1,[8_A]}$.}
		\label{r1_t_j1_8_a_bare}
	\end{subfigure}\\[0.3cm]
	\begin{subfigure}{0.45\textwidth}
		\includegraphics[width=\textwidth]{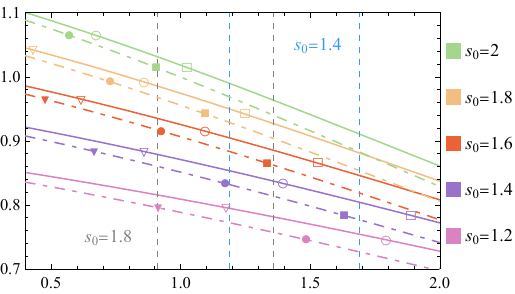}
		\caption{$\sqrt{\mathcal{R}^0}$ for $J_{1,1_A}$.}
		\label{r0_t_j1_1_a_bare}
	\end{subfigure}
	\hspace*{\fill}	
	\begin{subfigure}{0.45\textwidth}
		\includegraphics[width=\textwidth]{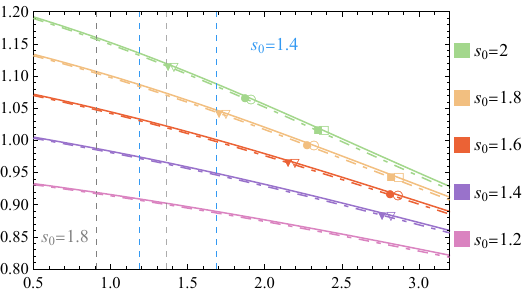}
		\caption{$\sqrt{\mathcal{R}^1}$ for $J_{1,1_A}$.}
		\label{r1_t_j1_1_a_bare}
	\end{subfigure}\\[0.3cm]
	\begin{subfigure}{0.45\textwidth}
		\includegraphics[width=\textwidth]{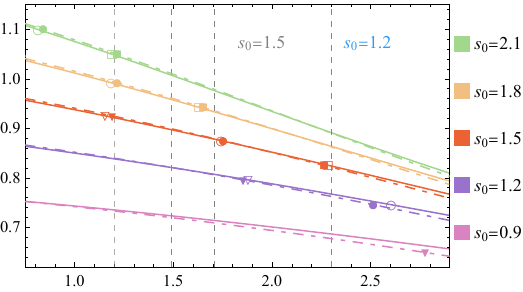}
		\caption{$\sqrt{\mathcal{R}^0}$ for $J_{2,[8_S]}$.}
		\label{r0_t_j2_8_s_bare}
	\end{subfigure}
	\hspace*{\fill}	
	\begin{subfigure}{0.45\textwidth}
		\includegraphics[width=\textwidth]{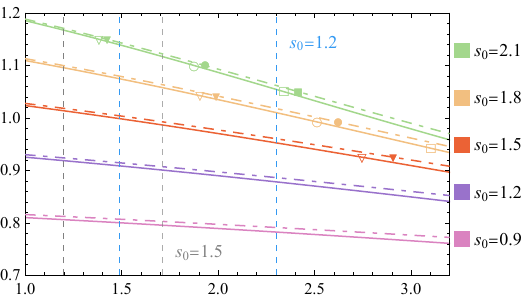}
		\caption{$\sqrt{\mathcal{R}^1}$ for $J_{2,[8_S]}$.}
		\label{r1_t_j2_8_s_bare}
	\end{subfigure}\\[0.3cm]
	\begin{subfigure}{0.45\textwidth}
		\includegraphics[width=\textwidth]{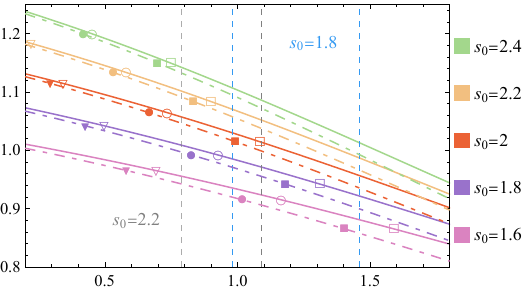}
		\caption{$\sqrt{\mathcal{R}^0}$ for $J_{2,1_S}$.}
		\label{r0_t_j2_1_s_bare}
	\end{subfigure}
	\hspace*{\fill}	
	\begin{subfigure}{0.45\textwidth}
		\includegraphics[width=\textwidth]{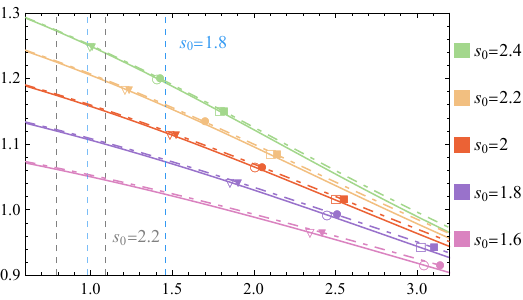}
		\caption{$\sqrt{\mathcal{R}^1}$ for $J_{2,1_S}$.}
		\label{r1_t_j2_1_s_bare}
	\end{subfigure}
	\caption{$\sqrt{\mathcal{R}^n}(\text{GeV})$ versus $\tau(\text{GeV}^{-2})$ for different values of $s_0(\text{GeV}^2)$, corresponding to the bare tetraquark currents in category 4.}
	\label{fig_t_ca_4}
\end{figure}

\begin{figure}[h]
	\begin{subfigure}{0.45\textwidth}
		\includegraphics[width=\textwidth]{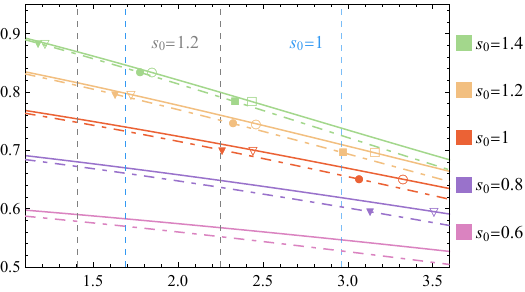}
		\caption{$\sqrt{\mathcal{R}^0}$ for $J_{3,[8_S]}$.}
		\label{r0_t_j3_8_s_bare}
	\end{subfigure}
	\hspace*{\fill}	
	\begin{subfigure}{0.45\textwidth}
		\includegraphics[width=\textwidth]{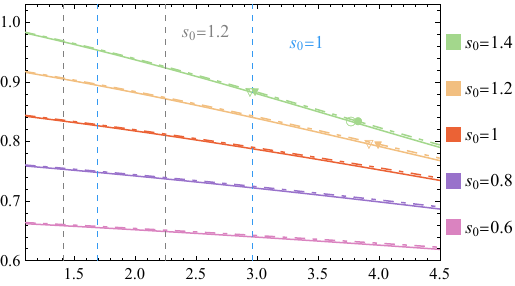}
		\caption{$\sqrt{\mathcal{R}^1}$ for $J_{3,[8_S]}$.}
		\label{r1_t_j3_8_s_bare}
	\end{subfigure}\\[0.3cm]
	\begin{subfigure}{0.45\textwidth}
		\includegraphics[width=\textwidth]{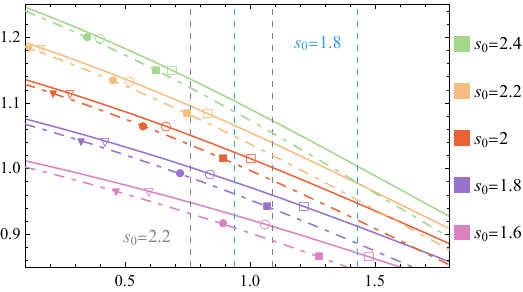}
		\caption{$\sqrt{\mathcal{R}^0}$ for $J_{3,1_S}$.}
		\label{r0_t_j3_1_s_bare}
	\end{subfigure}
	\hspace*{\fill}	
	\begin{subfigure}{0.45\textwidth}
		\includegraphics[width=\textwidth]{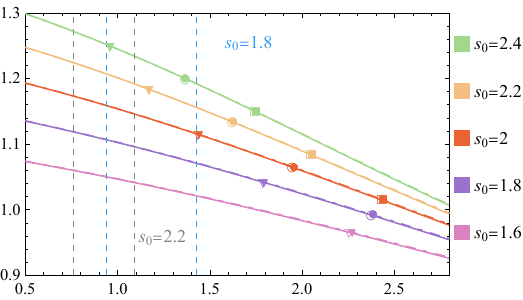}
		\caption{$\sqrt{\mathcal{R}^1}$ for $J_{3,1_S}$.}
		\label{r1_t_j3_1_s_bare}
	\end{subfigure}\\[0.3cm]
		\begin{subfigure}{0.45\textwidth}
		\includegraphics[width=\textwidth]{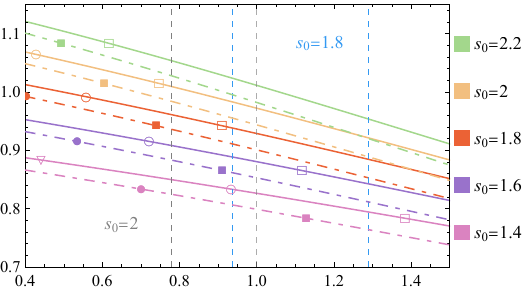}
		\caption{$\sqrt{\mathcal{R}^0}$ for $J_{4,1_A}$.}
		\label{r0_t_j4_1_a_bare}
	\end{subfigure}
	\hspace*{\fill}	
	\begin{subfigure}{0.45\textwidth}
		\includegraphics[width=\textwidth]{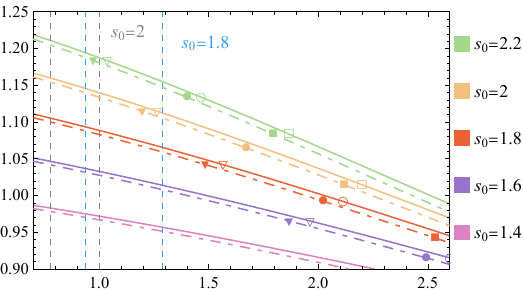}
		\caption{$\sqrt{\mathcal{R}^1}$ for $J_{4,1_A}$.}
		\label{r1_t_j4_1_a_bare}
	\end{subfigure}\\[0.3cm]
		\begin{subfigure}{0.45\textwidth}
		\includegraphics[width=\textwidth]{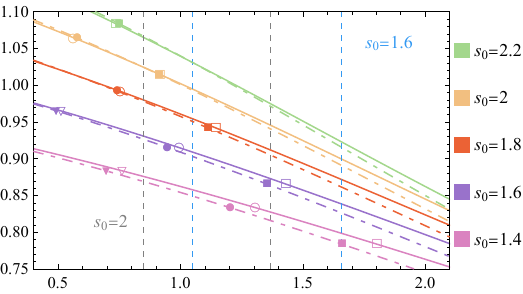}
		\caption{$\sqrt{\mathcal{R}^0}$ for $J_{a,1_S}$.}
		\label{r0_t_ja_1_s_bare}
	\end{subfigure}
	\hspace*{\fill}	
	\begin{subfigure}{0.45\textwidth}
		\includegraphics[width=\textwidth]{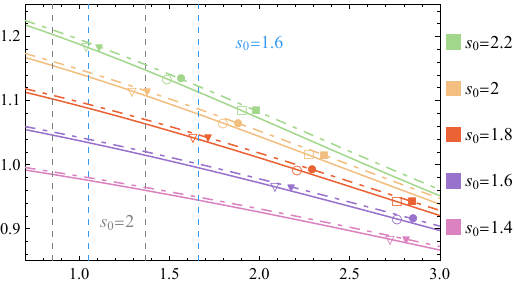}
		\caption{$\sqrt{\mathcal{R}^1}$ for $J_{a,1_S}$.}
		\label{r1_t_ja_1_s_bare}
	\end{subfigure}
	\caption{$\sqrt{\mathcal{R}^n}(\text{GeV})$ versus $\tau(\text{GeV}^{-2})$ for different values of $s_0(\text{GeV}^2)$, corresponding to the bare tetraquark currents in category 4.}
	\label{fig_t_ca_4_2}
\end{figure}

\begin{figure}[h]
	\begin{subfigure}{0.45\textwidth}
		\includegraphics[width=\textwidth]{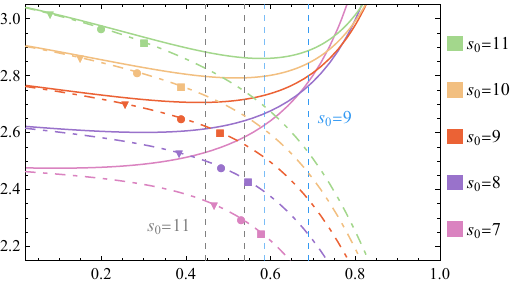}
		\caption{$\sqrt{\mathcal{R}^0}$ for $J_{f,[27_S]}$.}
		\label{r0_t_jf_27_s_bare}
	\end{subfigure}
	\hspace*{\fill}	
	\begin{subfigure}{0.45\textwidth}
		\includegraphics[width=\textwidth]{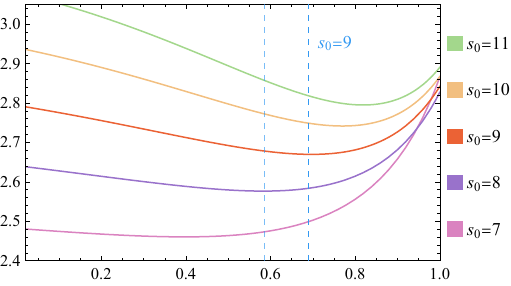}
		\caption{$\sqrt{\mathcal{R}^1}$ for $J_{f,[27_S]}$.}
		\label{r1_t_jf_27_s_bare}
	\end{subfigure}
	\caption{$\sqrt{\mathcal{R}^n}(\text{GeV})$ versus $\tau(\text{GeV}^{-2})$ for different values of $s_0(\text{GeV}^2)$, corresponding to the bare tetraquark current in category 2.}
	\label{fig_t_ca_2}
\end{figure}

\begin{figure}[h]
	\begin{subfigure}{0.45\textwidth}
		\includegraphics[width=\textwidth]{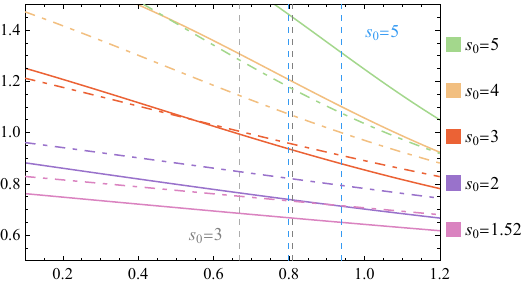}
		\caption{$\sqrt{\mathcal{R}^0}$ for $J_{4,[8_A]}$.}
		\label{r0_t_j4_8_a_bare}
	\end{subfigure}
	\hspace*{\fill}	
	\begin{subfigure}{0.45\textwidth}
		\includegraphics[width=\textwidth]{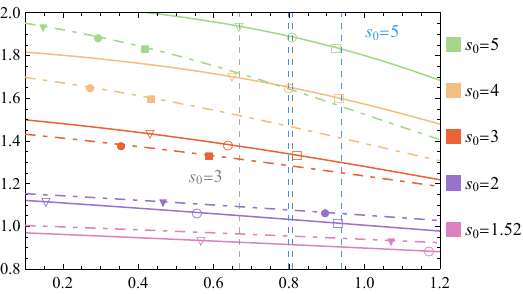}
		\caption{$\sqrt{\mathcal{R}^1}$ for $J_{4,[8_A]}$.}
		\label{r1_t_j4_8_a_bare}
	\end{subfigure}\\[0.1cm]
	\begin{subfigure}{0.45\textwidth}
		\includegraphics[width=\textwidth]{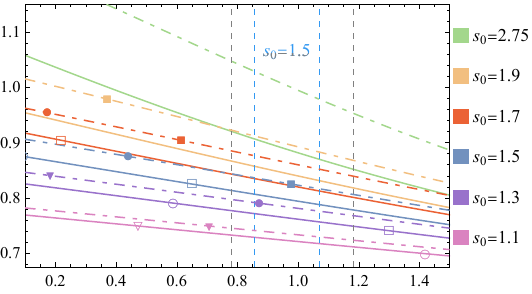}
		\caption{$\sqrt{\mathcal{R}^0}$ for $J_{a,[8_S]}$.}
		\label{r0_t_ja_8_s_bare}
	\end{subfigure}
	\hspace*{\fill}	
	\begin{subfigure}{0.45\textwidth}
		\includegraphics[width=\textwidth]{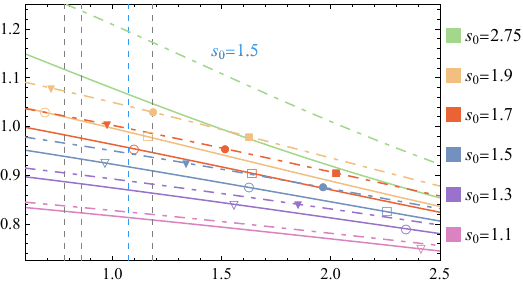}
		\caption{$\sqrt{\mathcal{R}^1}$ for $J_{a,[8_S]}$.}
		\label{r1_t_js_8_s_bare}
	\end{subfigure}\\[0.1cm]
	\begin{subfigure}{0.45\textwidth}
		\includegraphics[width=\textwidth]{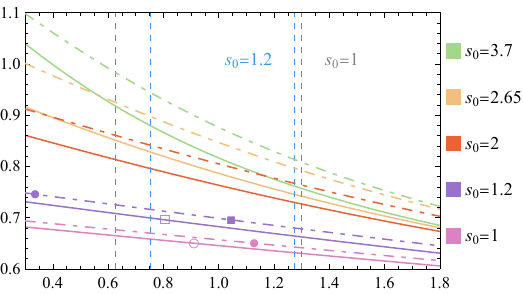}
		\caption{$\sqrt{\mathcal{R}^0}$ for $J_{b,1_A}$.}
		\label{r0_t_jb_1_a_bare}
	\end{subfigure}
	\hspace*{\fill}	
	\begin{subfigure}{0.45\textwidth}
		\includegraphics[width=\textwidth]{./other/r1_t_jb_1_a}
		\caption{$\sqrt{\mathcal{R}^1}$ for $J_{b,1_A}$.}
		\label{r1_t_jb_1_a_bare}
	\end{subfigure}
	\caption{$\sqrt{\mathcal{R}^n}(\text{GeV})$ versus $\tau(\text{GeV}^{-2})$ for different values of $s_0(\text{GeV}^2)$, corresponding to the bare tetraquark currents in category 7. For $J_{4,[8_A]}$, when $\tau$ and $s_0$ are chosen around the Borel windows, the constraint $s_0=(m+\Lambda_\text{QCD})^2$ cannot be applied, and the derived masses may not reliable.}
	\label{fig_t_ca_7}
\end{figure}

\begin{figure}[h]
	\begin{subfigure}{0.4\textwidth}
		\includegraphics[width=\textwidth]{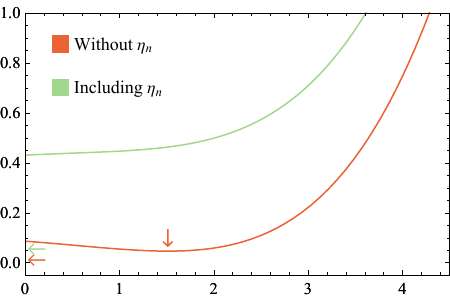}
		\caption{$J_{4,[8_A]}$.}
		\label{rho_s_j4_8_a}
	\end{subfigure}
	\hspace*{0.135\textwidth}	
	\begin{subfigure}{0.4\textwidth}
		\includegraphics[width=\textwidth]{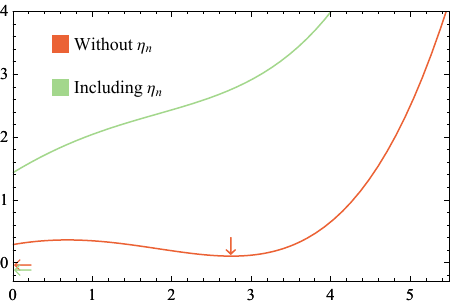}
		\caption{$J_{a,[8_S]}$.}
		\label{rho_s_ja_8_s}
	\end{subfigure}\\[0.2cm]
	\begin{subfigure}{0.4\textwidth}
		\includegraphics[width=\textwidth]{./other/rho_s_jb_1_a_10xx3}
		\caption{$J_{b,1_A}$.}
		\label{rho_s_jb_1_a}
	\end{subfigure}
	\hspace*{\fill}	
	\parbox[b][4.4cm][t]{7cm}{
		\caption{$\text{Im}\Pi(s)(10^3\text{GeV}^8)$ versus $s(\text{GeV}^2)$ when $\mu=1\text{GeV}$. The term {\small$\propto\delta(s)$} cannot be plotted; the left-pointing arrows indicate the coefficient $c$ in {\small$c\,\delta(s)$}. The downward arrows indicate the positions of the minima of $\text{Im}\Pi(s)$.}\label{rho_s_tetra}}
\end{figure}

\begin{figure}[h]
	\begin{subfigure}{0.45\textwidth}
		\includegraphics[width=\textwidth]{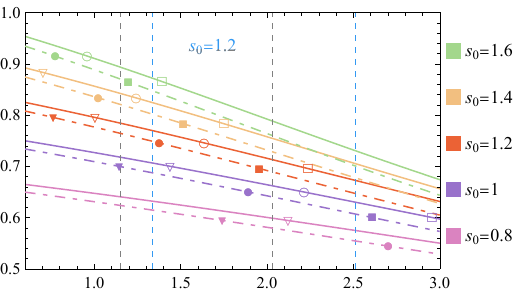}
		\caption{$\sqrt{\mathcal{R}^0}$ for $J_{e,[8_A]}$.}
		\label{r0_t_je_8_a_bare}
	\end{subfigure}
	\hspace*{\fill}	
	\begin{subfigure}{0.45\textwidth}
		\includegraphics[width=\textwidth]{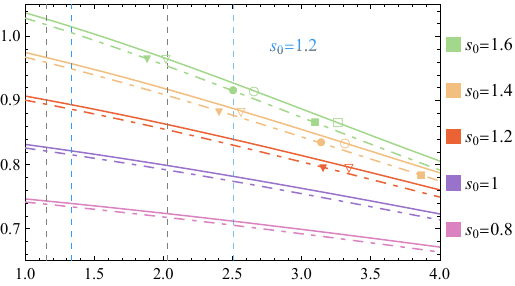}
		\caption{$\sqrt{\mathcal{R}^1}$ for $J_{e,[8_A]}$.}
		\label{r1_t_je_8_a_bare}
	\end{subfigure}\\[0.2cm]
	\begin{subfigure}{0.45\textwidth}
		\includegraphics[width=\textwidth]{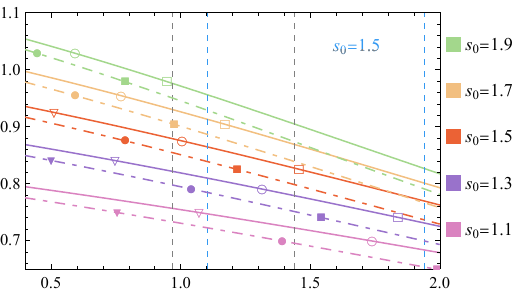}
		\caption{$\sqrt{\mathcal{R}^0}$ for $J_{e,1_A}$.}
		\label{r0_t_je_1_a_bare}
	\end{subfigure}
	\hspace*{\fill}	
	\begin{subfigure}{0.45\textwidth}
		\includegraphics[width=\textwidth]{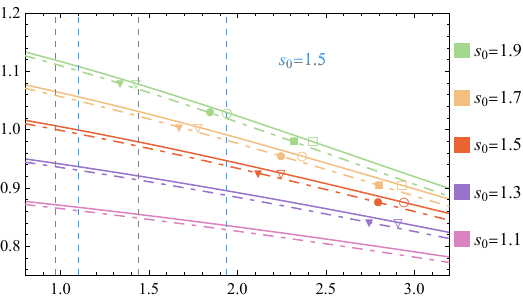}
		\caption{$\sqrt{\mathcal{R}^1}$ for $J_{e,1_A}$.}
		\label{r1_t_je_1_a_bare}
	\end{subfigure}
	\caption{$\sqrt{\mathcal{R}^n}(\text{GeV})$ versus $\tau(\text{GeV}^{-2})$ for different values of $s_0(\text{GeV}^2)$, corresponding to the bare tetraquark currents in category 4.}
	\label{fig_t_ca_4_3}
\end{figure}

\clearpage
\subsection{Mass Estimations for the Bare Molecule Currents}\label{sec_mass_b_m}

\begin{figure}[h]
	\begin{subfigure}{0.45\textwidth}
		\includegraphics[width=\textwidth]{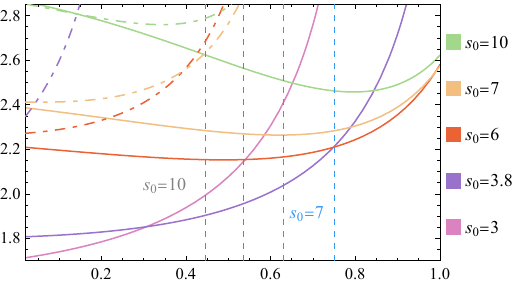}
		\caption{$\sqrt{\mathcal{R}^0}$ for $I_{1,[27_S]}$.}
		\label{r0_t_i1_27_s_bare}
	\end{subfigure}
	\hspace*{\fill}	
	\begin{subfigure}{0.45\textwidth}
		\includegraphics[width=\textwidth]{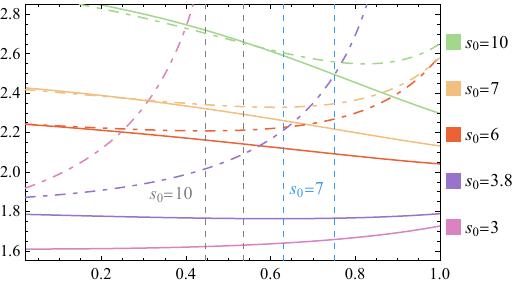}
		\caption{$\sqrt{\mathcal{R}^1}$ for $I_{1,[27_S]}$.}
		\label{r1_t_i1_27_s_bare}
	\end{subfigure}\\[0.3cm]
	\begin{subfigure}{0.45\textwidth}
		\includegraphics[width=\textwidth]{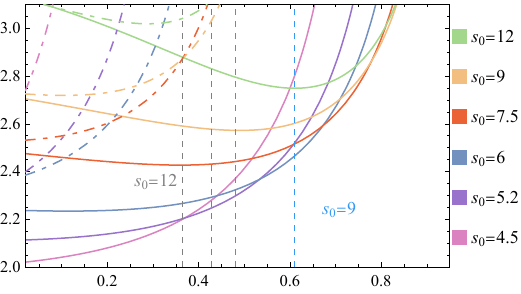}
		\caption{$\sqrt{\mathcal{R}^0}$ for $I_{4,[27_S]}$.}
		\label{r0_t_i4_27_s_bare}
	\end{subfigure}
	\hspace*{\fill}	
	\begin{subfigure}{0.45\textwidth}
		\includegraphics[width=\textwidth]{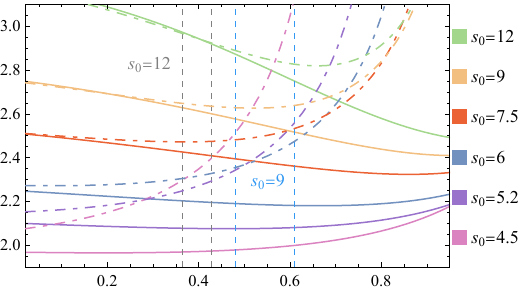}
		\caption{$\sqrt{\mathcal{R}^1}$ for $I_{4,[27_S]}$.}
		\label{r1_t_i4_27_s_bare}
	\end{subfigure}\\[0.3cm]
	\begin{subfigure}{0.45\textwidth}
		\includegraphics[width=\textwidth]{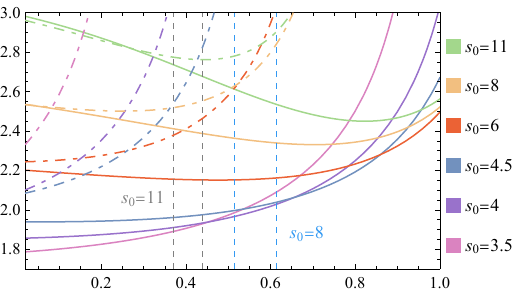}
		\caption{$\sqrt{\mathcal{R}^0}$ for $I_{4,[8_S]}$.}
		\label{r0_t_i4_8_s_bare}
	\end{subfigure}
	\hspace*{\fill}	
	\begin{subfigure}{0.45\textwidth}
		\includegraphics[width=\textwidth]{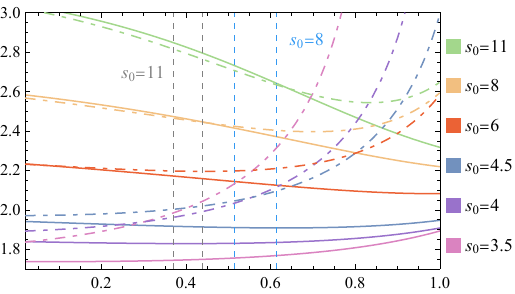}
		\caption{$\sqrt{\mathcal{R}^1}$ for $I_{4,[8_S]}$.}
		\label{r1_t_i4_8_s_bare}
	\end{subfigure}\\[0.3cm]
	\begin{subfigure}{0.45\textwidth}
		\includegraphics[width=\textwidth]{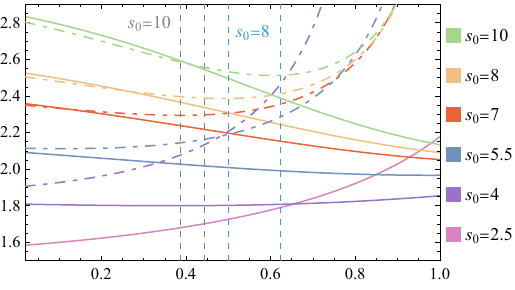}
		\caption{$\sqrt{\mathcal{R}^0}$ for $I_{4,1_S}$.}
		\label{r0_t_i4_1_s_bare}
	\end{subfigure}
	\hspace*{\fill}	
	\begin{subfigure}{0.45\textwidth}
		\includegraphics[width=\textwidth]{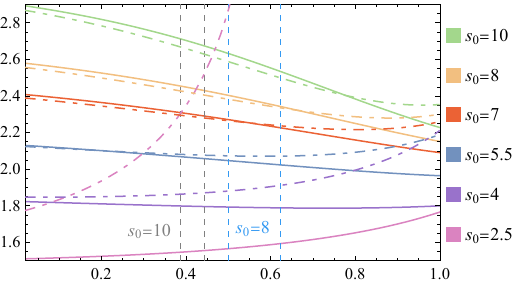}
		\caption{$\sqrt{\mathcal{R}^1}$ for $I_{4,1_S}$.}
		\label{r1_t_i4_1_s_bare}
	\end{subfigure}
	\caption{$\sqrt{\mathcal{R}^n}(\text{GeV})$ versus $\tau(\text{GeV}^{-2})$ for different values of $s_0(\text{GeV}^2)$, corresponding to the bare molecule currents in category 1.}
	\label{fig_m_ca_1}
\end{figure}

\begin{figure}[h]
	\begin{subfigure}{0.45\textwidth}
		\includegraphics[width=\textwidth]{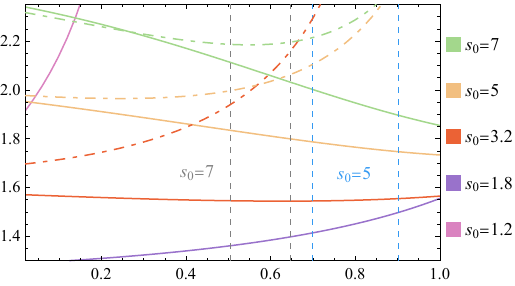}
		\caption{$\sqrt{\mathcal{R}^0}$ for $I_{1,[8_S]}$.}
		\label{r0_t_i1_8_s_bare}
	\end{subfigure}
	\hspace*{\fill}	
	\begin{subfigure}{0.45\textwidth}
		\includegraphics[width=\textwidth]{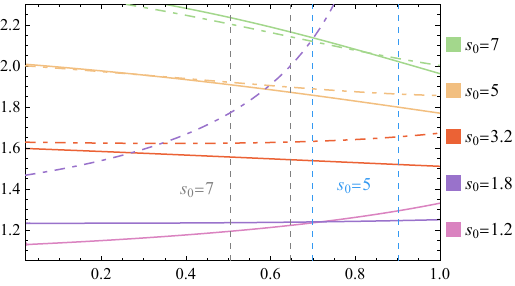}
		\caption{$\sqrt{\mathcal{R}^1}$ for $I_{1,[8_S]}$.}
		\label{r1_t_i1_8_s_bare}
	\end{subfigure}\\[0.3cm]
	\begin{subfigure}{0.45\textwidth}
		\includegraphics[width=\textwidth]{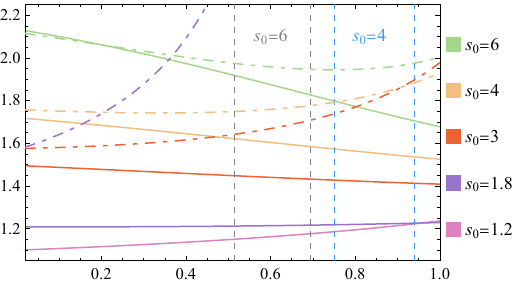}
		\caption{$\sqrt{\mathcal{R}^0}$ for $I_{1,1_S}$.}
		\label{r0_t_i1_1_s_bare}
	\end{subfigure}
	\hspace*{\fill}	
	\begin{subfigure}{0.45\textwidth}
		\includegraphics[width=\textwidth]{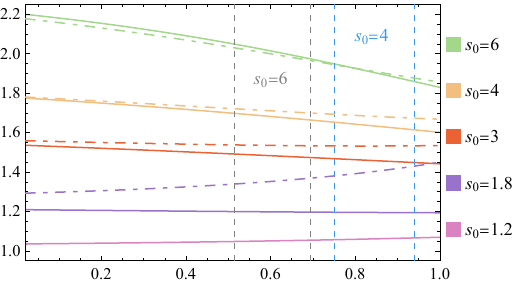}
		\caption{$\sqrt{\mathcal{R}^1}$ for $I_{1,1_S}$.}
		\label{r1_t_i1_1_s_bare}
	\end{subfigure}\\[0.3cm]
	\begin{subfigure}{0.45\textwidth}
		\includegraphics[width=\textwidth]{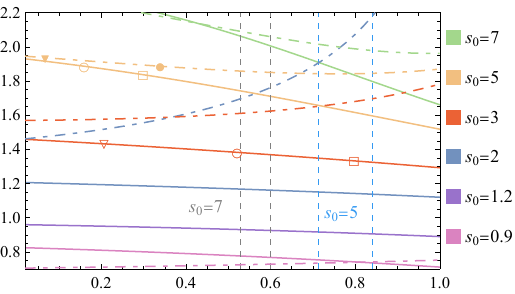}
		\caption{$\sqrt{\mathcal{R}^0}$ for $I_{3,[8_S]}$.}
		\label{r0_t_i3_8_s_bare}
	\end{subfigure}
	\hspace*{\fill}	
	\begin{subfigure}{0.45\textwidth}
		\includegraphics[width=\textwidth]{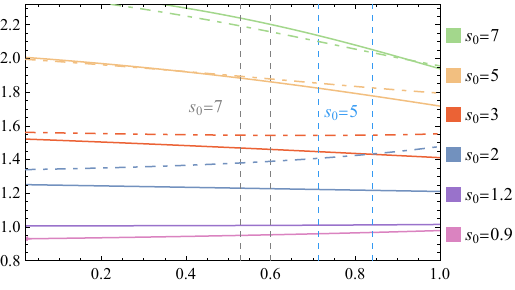}
		\caption{$\sqrt{\mathcal{R}^1}$ for $I_{3,[8_S]}$.}
		\label{r1_t_i3_8_s_bare}
	\end{subfigure}\\[0.3cm]
	\begin{subfigure}{0.45\textwidth}
		\includegraphics[width=\textwidth]{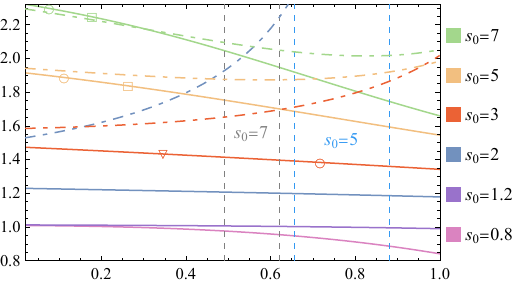}
		\caption{$\sqrt{\mathcal{R}^0}$ for $I_{3,1_S}$.}
		\label{r0_t_i3_1_s_bare}
	\end{subfigure}
	\hspace*{\fill}	
	\begin{subfigure}{0.45\textwidth}
		\includegraphics[width=\textwidth]{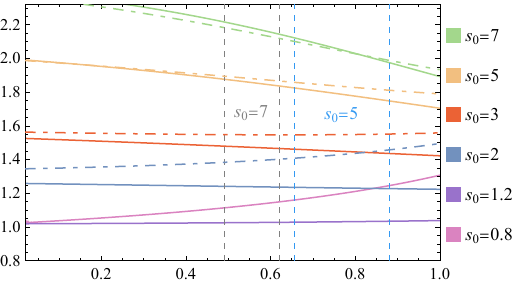}
		\caption{$\sqrt{\mathcal{R}^1}$ for $I_{3,1_S}$.}
		\label{r1_t_i3_1_s_bare}
	\end{subfigure}
	\caption{$\sqrt{\mathcal{R}^n}(\text{GeV})$ versus $\tau(\text{GeV}^{-2})$ for different values of $s_0(\text{GeV}^2)$, corresponding to the bare molecule currents in category 3.}
	\label{fig_m_ca_3}
\end{figure}

\begin{figure}[h]
	\begin{subfigure}{0.45\textwidth}
		\includegraphics[width=\textwidth]{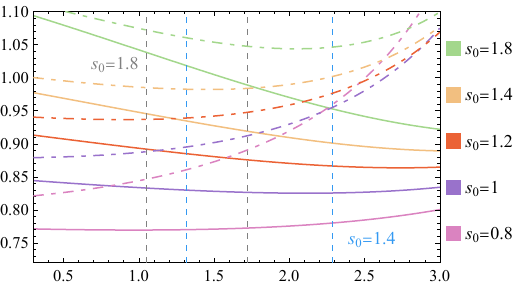}
		\caption{$\sqrt{\mathcal{R}^0}$ for $I_{5,[8_S]}$.}
		\label{r0_t_i5_8_s_bare}
	\end{subfigure}
	\hspace*{\fill}	
	\begin{subfigure}{0.45\textwidth}
		\includegraphics[width=\textwidth]{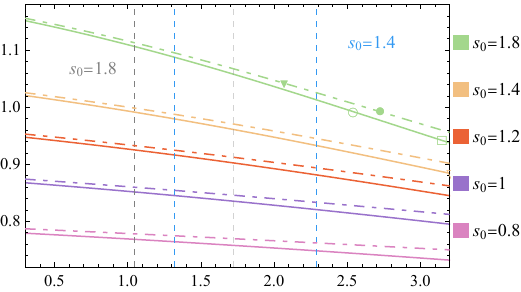}
		\caption{$\sqrt{\mathcal{R}^1}$ for $I_{5,[8_S]}$.}
		\label{r1_t_i5_8_s_bare}
	\end{subfigure}\\[0.3cm]
	\begin{subfigure}{0.45\textwidth}
		\includegraphics[width=\textwidth]{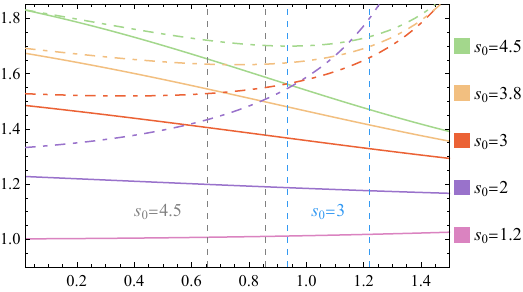}
		\caption{$\sqrt{\mathcal{R}^0}$ for $I_{a,[8_A]}$.}
		\label{r0_t_ia_8_a_bare}
	\end{subfigure}
	\hspace*{\fill}	
	\begin{subfigure}{0.45\textwidth}
		\includegraphics[width=\textwidth]{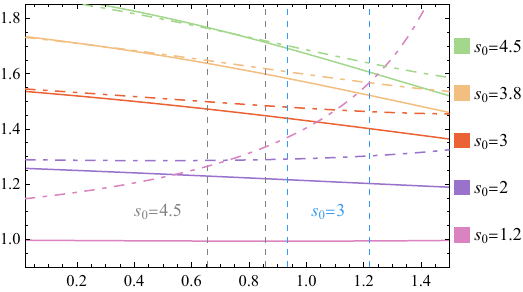}
		\caption{$\sqrt{\mathcal{R}^1}$ for $I_{a,[8_A]}$.}
		\label{r1_t_ia_8_a_bare}
	\end{subfigure}\\[0.3cm]
	\begin{subfigure}{0.45\textwidth}
		\includegraphics[width=\textwidth]{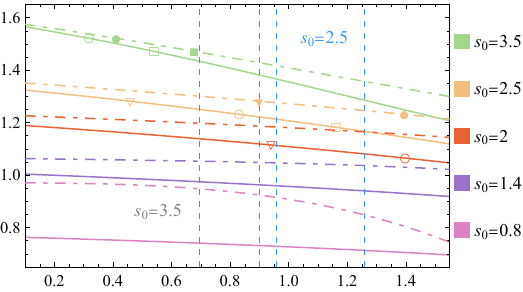}
		\caption{$\sqrt{\mathcal{R}^0}$ for $I_{a,1_A}$.}
		\label{r0_t_ia_1_a_bare}
	\end{subfigure}
	\hspace*{\fill}	
	\begin{subfigure}{0.45\textwidth}
		\includegraphics[width=\textwidth]{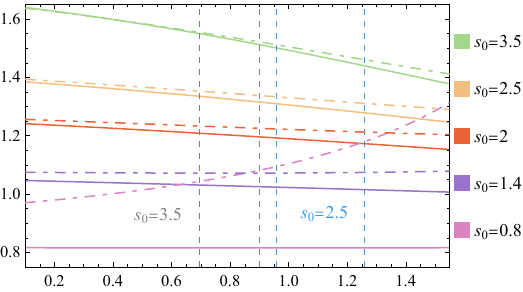}
		\caption{$\sqrt{\mathcal{R}^1}$ for $I_{a,1_A}$.}
		\label{r1_t_ia_1_a_bare}
	\end{subfigure}\\[0.3cm]
	\begin{subfigure}{0.45\textwidth}
		\includegraphics[width=\textwidth]{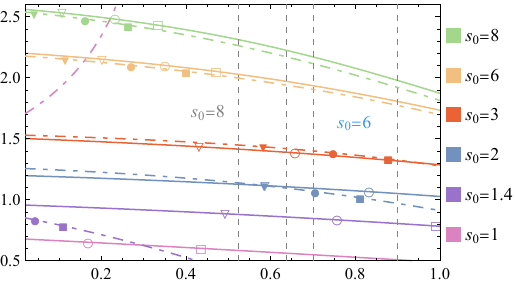}
		\caption{$\sqrt{\mathcal{R}^0}$ for $I_{c,[8_A]}$.}
		\label{r0_t_ic_8_a_bare}
	\end{subfigure}
	\hspace*{\fill}	
	\begin{subfigure}{0.45\textwidth}
		\includegraphics[width=\textwidth]{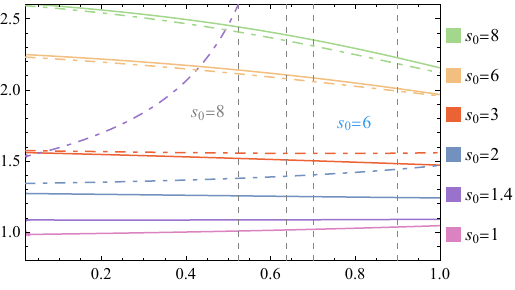}
		\caption{$\sqrt{\mathcal{R}^1}$ for $I_{c,[8_A]}$.}
		\label{r1_t_ic_8_a_bare}
	\end{subfigure}
	\caption{$\sqrt{\mathcal{R}^n}(\text{GeV})$ versus $\tau(\text{GeV}^{-2})$ for different values of $s_0(\text{GeV}^2)$, corresponding to the bare molecule currents in category 3.}
	\label{fig_m_ca_3_2}
\end{figure}

\begin{figure}[h]
	\begin{subfigure}{0.45\textwidth}
		\includegraphics[width=\textwidth]{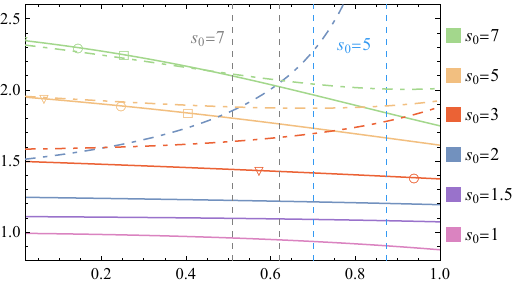}
		\caption{$\sqrt{\mathcal{R}^0}$ for $I_{c,1_A}$.}
		\label{r0_t_ic_1_a_bare}
	\end{subfigure}
	\hspace*{\fill}	
	\begin{subfigure}{0.45\textwidth}
		\includegraphics[width=\textwidth]{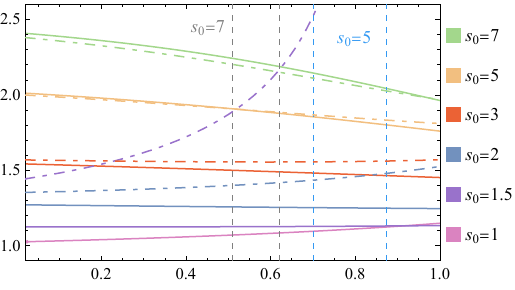}
		\caption{$\sqrt{\mathcal{R}^1}$ for $I_{c,1_A}$.}
		\label{r1_t_ic_1_a_bare}
	\end{subfigure}\\[0.1cm]
	\begin{subfigure}{0.45\textwidth}
		\includegraphics[width=\textwidth]{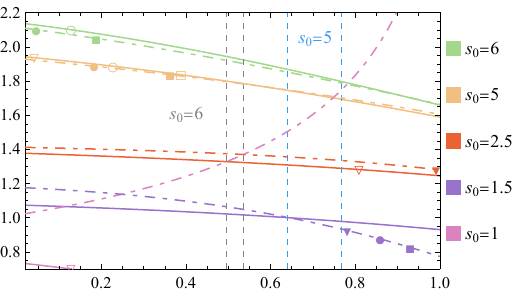}
		\caption{$\sqrt{\mathcal{R}^0}$ for $I_{d,[8_A]}$.}
		\label{r0_t_id_8_a_bare}
	\end{subfigure}
	\hspace*{\fill}	
	\begin{subfigure}{0.45\textwidth}
		\includegraphics[width=\textwidth]{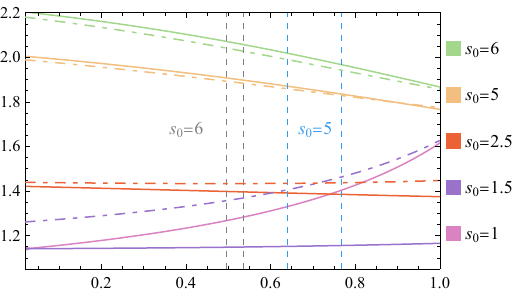}
		\caption{$\sqrt{\mathcal{R}^1}$ for $I_{d,[8_A]}$.}
		\label{r1_t_id_8_a_bare}
	\end{subfigure}\\[0.1cm]
	\begin{subfigure}{0.45\textwidth}
		\includegraphics[width=\textwidth]{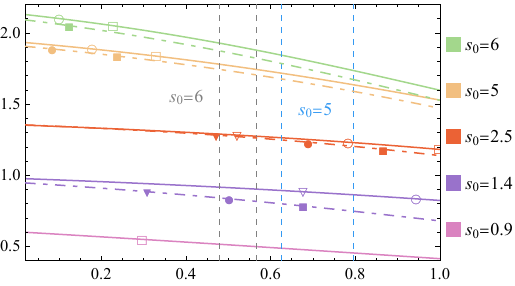}
		\caption{$\sqrt{\mathcal{R}^0}$ for $I_{d,1_A}$.}
		\label{r0_t_id_1_a_bare}
	\end{subfigure}
	\hspace*{\fill}	
	\begin{subfigure}{0.45\textwidth}
		\includegraphics[width=\textwidth]{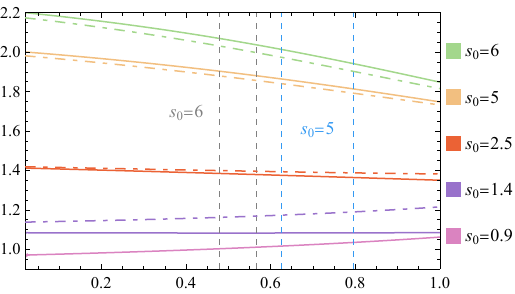}
		\caption{$\sqrt{\mathcal{R}^1}$ for $I_{d,1_A}$.}
		\label{r1_t_id_1_a_bare}
	\end{subfigure}
	\caption{$\sqrt{\mathcal{R}^n}(\text{GeV})$ versus $\tau(\text{GeV}^{-2})$ for different values of $s_0(\text{GeV}^2)$, corresponding to the bare molecule currents in category 3.}
	\label{fig_m_ca_3_3}
\end{figure}

\begin{figure}[h]
	\begin{subfigure}{0.45\textwidth}
		\includegraphics[width=\textwidth]{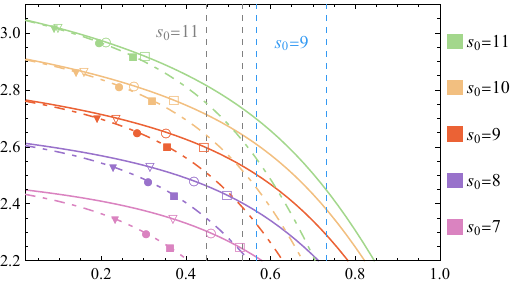}
		\caption{$\sqrt{\mathcal{R}^0}$ for $I_{3,[27_S]}$.}
		\label{r0_t_i3_27_s_bare}
	\end{subfigure}
	\hspace*{\fill}	
	\begin{subfigure}{0.45\textwidth}
		\includegraphics[width=\textwidth]{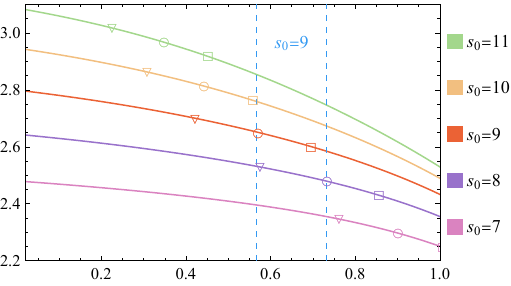}
		\caption{$\sqrt{\mathcal{R}^1}$ for $I_{3,[27_S]}$.}
		\label{r1_t_i3_27_s_bare}
	\end{subfigure}
	\caption{$\sqrt{\mathcal{R}^n}(\text{GeV})$ versus $\tau(\text{GeV}^{-2})$ for different values of $s_0(\text{GeV}^2)$, corresponding to the bare molecule current in category 5.}
	\label{fig_m_ca_5}
\end{figure}

\begin{figure}[h]
	\begin{subfigure}{0.45\textwidth}
		\includegraphics[width=\textwidth]{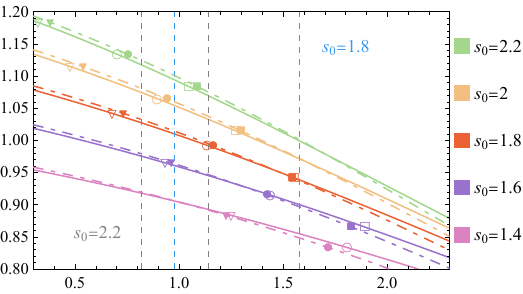}
		\caption{$\sqrt{\mathcal{R}^0}$ for $I_{5,1_S}$.}
		\label{r0_t_i5_1_s_bare}
	\end{subfigure}
	\hspace*{\fill}	
	\begin{subfigure}{0.45\textwidth}
		\includegraphics[width=\textwidth]{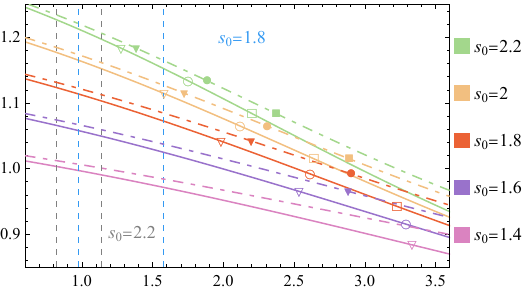}
		\caption{$\sqrt{\mathcal{R}^1}$ for $I_{5,1_S}$.}
		\label{r1_t_i5_1_s_bare}
	\end{subfigure}\\[0.3cm]
		\begin{subfigure}{0.45\textwidth}
		\includegraphics[width=\textwidth]{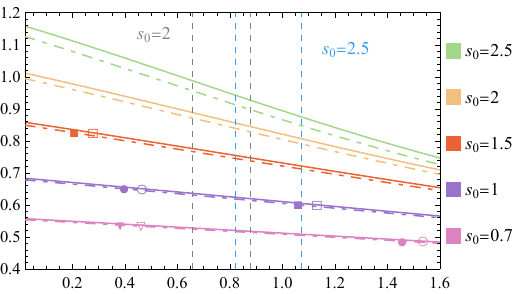}
		\caption{$\sqrt{\mathcal{R}^0}$ for $I_{b,1_A}$.}
		\label{r0_t_ib_1_a_bare}
	\end{subfigure}
	\hspace*{\fill}	
	\begin{subfigure}{0.45\textwidth}
		\includegraphics[width=\textwidth]{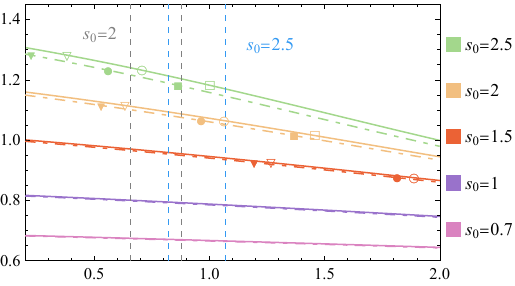}
		\caption{$\sqrt{\mathcal{R}^1}$ for $I_{b,1_A}$.}
		\label{r1_t_ib_1_a_bare}
	\end{subfigure}\\[0.3cm]
	\begin{subfigure}{0.45\textwidth}
		\includegraphics[width=\textwidth]{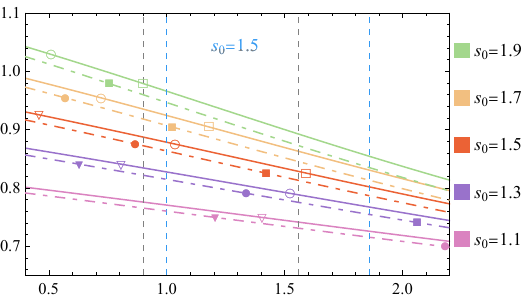}
		\caption{$\sqrt{\mathcal{R}^0}$ for $I_{e,[8_A]}$.}
		\label{r0_t_ie_8_a_bare}
	\end{subfigure}
	\hspace*{\fill}	
	\begin{subfigure}{0.45\textwidth}
		\includegraphics[width=\textwidth]{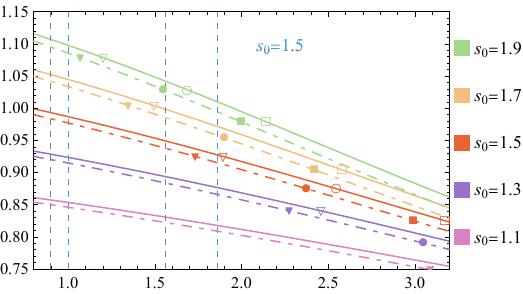}
		\caption{$\sqrt{\mathcal{R}^1}$ for $I_{e,[8_A]}$.}
		\label{r1_t_ie_8_a_bare}
	\end{subfigure}\\[0.3cm]
	\begin{subfigure}{0.45\textwidth}
		\includegraphics[width=\textwidth]{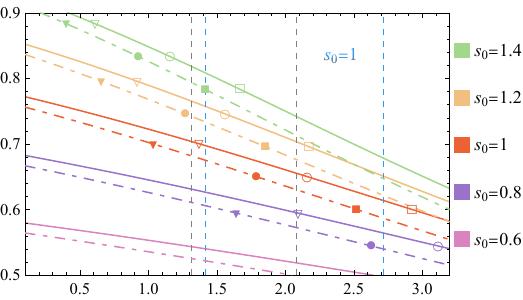}
		\caption{$\sqrt{\mathcal{R}^0}$ for $I_{e,1_A}$.}
		\label{r0_t_ie_1_a_bare}
	\end{subfigure}
	\hspace*{\fill}	
	\begin{subfigure}{0.45\textwidth}
		\includegraphics[width=\textwidth]{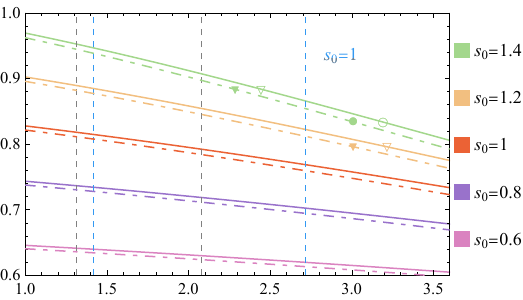}
		\caption{$\sqrt{\mathcal{R}^1}$ for $I_{e,1_A}$.}
		\label{r1_t_ie_1_a_bare}
	\end{subfigure}
	\caption{$\sqrt{\mathcal{R}^n}(\text{GeV})$ versus $\tau(\text{GeV}^{-2})$ for different values of $s_0(\text{GeV}^2)$, corresponding to the bare molecule currents in category 4.}
	\label{fig_m_ca_4}
\end{figure}

\begin{figure}[h]
	\begin{subfigure}{0.45\textwidth}
		\includegraphics[width=\textwidth]{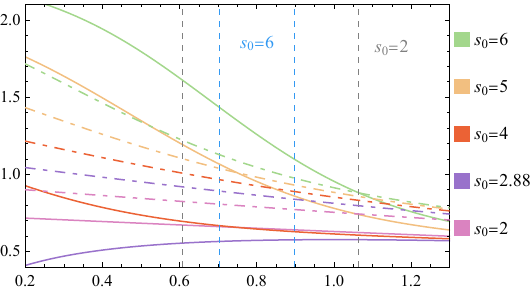}
		\caption{$\sqrt{\mathcal{R}^0}$ for $I_{2,1_S}$.}
		\label{r0_t_i2_1_s_bare}
	\end{subfigure}
	\hspace*{\fill}	
	\begin{subfigure}{0.45\textwidth}
		\includegraphics[width=\textwidth]{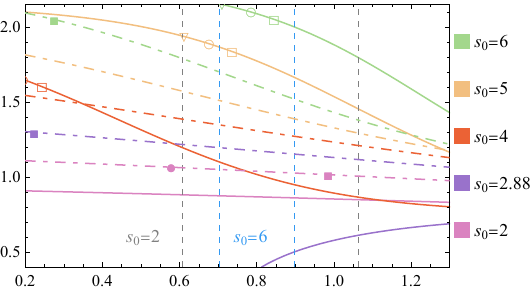}
		\caption{$\sqrt{\mathcal{R}^1}$ for $I_{2,1_S}$.}
		\label{r1_t_i2_1_s_bare}
	\end{subfigure}\\[0.3cm]
	\begin{subfigure}{0.45\textwidth}
		\includegraphics[width=\textwidth]{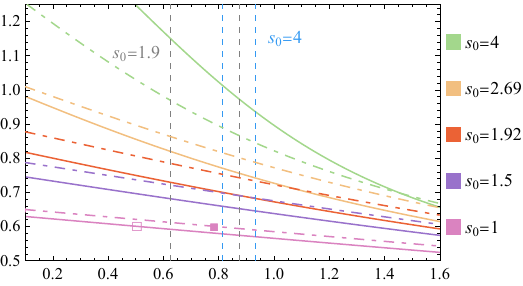}
		\caption{$\sqrt{\mathcal{R}^0}$ for $I_{b,[8_A]}$.}
		\label{r0_t_ib_8_a_bare}
	\end{subfigure}
	\hspace*{\fill}	
	\begin{subfigure}{0.45\textwidth}
		\includegraphics[width=\textwidth]{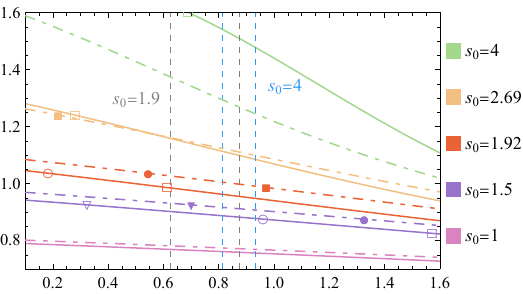}
		\caption{$\sqrt{\mathcal{R}^1}$ for $I_{b,[8_A]}$.}
		\label{r1_t_ib_8_a_bare}
	\end{subfigure}
	\caption{$\sqrt{\mathcal{R}^n}(\text{GeV})$ versus $\tau(\text{GeV}^{-2})$ for different values of $s_0(\text{GeV}^2)$, corresponding to the bare molecule currents in category 7. When $\tau$ and $s_0$ are chosen around the Borel windows, the constraint $s_0=(m+\Lambda_\text{QCD})^2$ cannot be applied, so the derived masses may not reliable for these two currents.}
	\label{fig_m_ca_7}
\end{figure}

\begin{figure}[h]
	\begin{subfigure}{0.4\textwidth}
		\includegraphics[width=\textwidth]{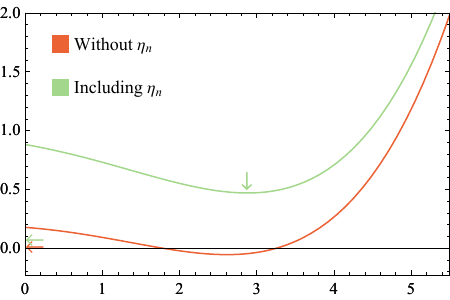}
		\caption{$I_{2,1_S}$, scale up by $10^2$.}
		\label{rho_s_i2_1_s}
	\end{subfigure}
	\hspace*{0.135\textwidth}	
	\begin{subfigure}{0.4\textwidth}
		\includegraphics[width=\textwidth]{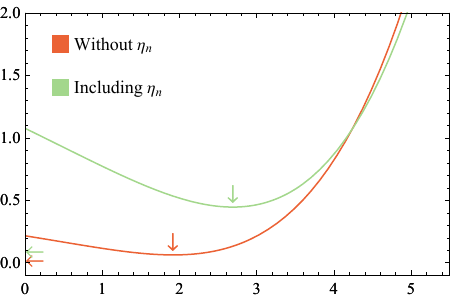}
		\caption{$I_{b,[8_A]}$, scale up by $10^3$.}
		\label{rho_s_i7_8_a}
	\end{subfigure}
		\caption{$\text{Im}\Pi(s)(\text{GeV}^8)$ versus $s(\text{GeV}^2)$ with $\mu=1\text{GeV}$. The term {\small$\propto\delta(s)$} cannot be plotted; the left-pointing arrows indicate the coefficient $c$ in {\small$c\,\delta(s)$}. The downward arrows indicate the positions of the minima of $\text{Im}\Pi(s)$, with the negative minimum excluded.}
		\label{rho_s_mole}
\end{figure}


\clearpage
\pagebreak
\newpage

\section{The OPE Results for the Four-Quark Correlators}\label{ope_results}

\begingroup
\allowdisplaybreaks

The following presents the correlators with unspecified flavors corresponding to the renormalized tetraquark currents in eqs.~\ref{tetra_1_r} and \ref{tetra_2_r}. Some notations are adopted here: $\delta_{f_if_j}$ is abbreviated as $\delta_{ij}$; $\Psi_{f_i}$ is abbreviated as $f_i$; $\Pi_J(q^2)$ denotes $\langle T\,\,{J(q^2)J^\dagger(q^2)}\rangle$ and the flavor indices in $J^\dagger$ are renamed according to the rules: $f_1\rightarrow f_a$, $f_2\rightarrow f_b$, $f_3\rightarrow f_c$, $f_4\rightarrow f_d$. The dimension-8 condensate $\langle \bar{q}q\rangle\langle \bar{q}Gq\rangle$ is obtained by the VS-first procedure here.

For comparison, the $\Pi_{I_3^r}(q^2)$ is also presented at the end of this section. One can verify that after exchanging $f_1\leftrightarrow f_4$ and $f_b\leftrightarrow f_d$ to match their flavor conventions, $\Pi_{J^r_d}(q^2)$ and $\Pi_{I^r_3}(q^2)$ differ only by an overall factor 4, since the currents are not normalized, except for the term $\propto g_s^2\,\text{Log}(-\frac{q^2}{\mu^2})$ due to the $\gamma^5$. This is consistent with table~\ref{bases}.

For other correlators, refer to ref.~\cite{results} for details.

{\setstretch{0.5}

{\tiny
	\begin{align}
		\Pi_{J^r_1}(q^2)\!=&\!\frac{q^8 g_s^2 \!\log\! \!\left(\!-\!\frac{q^2}{\mu ^2}\right)\!}{232243200 \pi ^8}\!\Big[2923 \!\left(\!\left(\!\left(\!\left(\delta _{24} \delta _{1b} \delta _{3d} \delta _{ac}\!-\!\{\!3\!\!\leftrightarrow\!\! 4,a\!\!\leftrightarrow\!\! c\!\}\right)\!\!-\!\{\!c\!\!\leftrightarrow\!\! d\!\}\right)\!\!-\!\{\!1\!\!\leftrightarrow\!\! 2\!\}\right)\!\!-\!\!\left(\!\left(\!\left(\delta _{24} \delta _{1a} \delta _{3d} \delta _{bc}\!-\!\{\!3\!\!\leftrightarrow\!\! 4,b\!\!\leftrightarrow\!\! c\!\}\right)\!\!-\!\{\!c\!\!\leftrightarrow\!\! d\!\}\right)\!\!-\!\{\!1\!\!\leftrightarrow\!\! 2\!\}\right)\!\right)\nonumber \\
		&\qquad\qquad\qquad\qquad-\!263322 \!\left(\!\left(\delta _{2a} \delta _{1b} \delta _{4c} \delta _{3d}\!-\!\{\!c\!\!\leftrightarrow\!\! d\!\}\right)\!\!-\!\{\!1\!\!\leftrightarrow\!\! 2\!\}\right)\!\Big]\!\nonumber \\
		&\!+\!\frac{q^8 g_s^2 \!\log\! ^2\!\left(\!-\!\frac{q^2}{\mu ^2}\right)\!}{552960 \pi ^8}\!\Big[54 \!\left(\!\left(\delta _{2a} \delta _{1b} \delta _{4c} \delta _{3d}\!-\!\{\!c\!\!\leftrightarrow\!\! d\!\}\right)\!\!-\!\{\!1\!\!\leftrightarrow\!\! 2\!\}\right)\nonumber \\
		&\qquad\qquad\qquad-\!\!\left(\!\left(\!\left(\!\left(\delta _{24} \delta _{1b} \delta _{3d} \delta _{ac}\!-\!\{\!3\!\!\leftrightarrow\!\! 4,a\!\!\leftrightarrow\!\! c\!\}\right)\!\!-\!\{\!c\!\!\leftrightarrow\!\! d\!\}\right)\!\!-\!\{\!1\!\!\leftrightarrow\!\! 2\!\}\right)\!\!-\!\!\left(\!\left(\!\left(\delta _{24} \delta _{1a} \delta _{3d} \delta _{bc}\!-\!\{\!3\!\!\leftrightarrow\!\! 4,b\!\!\leftrightarrow\!\! c\!\}\right)\!\!-\!\{\!c\!\!\leftrightarrow\!\! d\!\}\right)\!\!-\!\{\!1\!\!\leftrightarrow\!\! 2\!\}\right)\!\right)\!\Big]\!\nonumber \\
		&\!-\!\frac{q^8 \!\log\! \!\left(\!-\!\frac{q^2}{\mu ^2}\right)\!}{5120 \pi ^6}\!\left(\!\left(\delta _{2a} \delta _{1b} \delta _{4c} \delta _{3d}\!-\!\{\!c\!\!\leftrightarrow\!\! d\!\}\right)\!\!-\!\{\!1\!\!\leftrightarrow\!\! 2\!\}\right)\!-\!\frac{q^4 \!\log\! \!\left(\!-\!\frac{q^2}{\mu ^2}\right)\!}{64 \pi ^5}\langle\! \, GG\, \!\rangle\!  \!\left(\!\left(\delta _{2a} \delta _{1b} \delta _{4c} \delta _{3d}\!-\!\{\!c\!\!\leftrightarrow\!\! d\!\}\right)\!\!-\!\{\!1\!\!\leftrightarrow\!\! 2\!\}\right)\!\nonumber \\
		&\!+\!\frac{q^4 \!\log\! \!\left(\!-\!\frac{q^2}{\mu ^2}\right)\!}{96 \pi ^4}\!\left(\!\!\left(m_1\!\!-\!3 m_2\right)\!\!\langle\!\overline{f_2}f_2\!\rangle\! \!+\!\!\left(m_2\!\!-\!3 m_1\right)\!\!\langle\!\overline{f_1}f_1\!\rangle\! \!+\!\!\left(m_3\!\!-\!3 m_4\right)\!\!\langle\!\overline{f_4}f_4\!\rangle\! \!+\!\!\left(m_4\!\!-\!3 m_3\right)\!\!\langle\!\overline{f_3}f_3\!\rangle\! \right)\! \!\left(\!\left(\delta _{2a} \delta _{1b} \delta _{4c} \delta _{3d}\!-\!\{\!c\!\!\leftrightarrow\!\! d\!\}\right)\!\!-\!\{\!1\!\!\leftrightarrow\!\! 2\!\}\right)\!\nonumber \\
		&\!+\!\frac{11 q^2 \!\log\! \!\left(\!-\!\frac{q^2}{\mu ^2}\right)\!}{576 \pi ^6}\langle\! G^3\!\rangle\!  \!\left(\!\left(\delta _{2a} \delta _{1b} \delta _{4c} \delta _{3d}\!-\!\{\!c\!\!\leftrightarrow\!\! d\!\}\right)\!\!-\!\{\!1\!\!\leftrightarrow\!\! 2\!\}\right)\!-\!\frac{q^2 \!\log\! \!\left(\!-\!\frac{q^2}{\mu ^2}\right)\!}{12 \pi ^2}\!\left(\!\!\langle\!\overline{f_1}f_1\!\rangle\!  \langle\! \, \overline{f_2}f_2\!\rangle\! \!+\!\langle\! \, \overline{f_3}f_3\!\rangle\!  \langle\! \, \overline{f_4}f_4\!\rangle\! \right)\! \!\left(\!\left(\delta _{2a} \delta _{1b} \delta _{4c} \delta _{3d}\!-\!\{\!c\!\!\leftrightarrow\!\! d\!\}\right)\!\!-\!\{\!1\!\!\leftrightarrow\!\! 2\!\}\right)\!\nonumber \\
		&\!-\!\frac{3 q^2 \!\log\! \!\left(\!-\!\frac{q^2}{\mu ^2}\right)\!}{2 \pi ^2}\!\left(\!\left(\!\left(\!\left(\delta _{24} \delta _{1b} \delta _{3d} \delta _{ac} \langle\! \, \overline{f_1}f_1\!\rangle\!  \langle\! \, \overline{f_b}f_b\!\rangle\! \!-\!\{\!1\!\!\leftrightarrow\!\! 2,3\!\!\leftrightarrow\!\! 4\!\}\right)\!\!+\!\{\!1\!\!\leftrightarrow\!\! 2,a\!\!\leftrightarrow\!\! b\!\}\right)\!\!-\!\{\!c\!\!\leftrightarrow\!\! d\!\}\right)\!\!-\!\{\!3\!\!\leftrightarrow\!\! 4\!\}\right)\!\nonumber \\
		&\!-\!\frac{\!\log\! \!\left(\!-\!\frac{q^2}{\mu ^2}\right)\!}{24 \pi ^2}\!\left(\!\!\langle\!\overline{f_2}f_2\!\rangle\!  \langle\! \, \overline{f_1}\text{G}f_1\!\rangle\! \!+\!\langle\! \, \overline{f_1}f_1\!\rangle\!  \langle\! \, \overline{f_2}\text{G}f_2\!\rangle\! \!+\!\langle\! \, \overline{f_4}f_4\!\rangle\!  \langle\! \, \overline{f_3}\text{G}f_3\!\rangle\! \!+\!\langle\! \, \overline{f_3}f_3\!\rangle\!  \langle\! \, \overline{f_4}\text{G}f_4\!\rangle\! \right)\! \!\left(\!\left(\delta _{2a} \delta _{1b} \delta _{4c} \delta _{3d}\!-\!\{\!c\!\!\leftrightarrow\!\! d\!\}\right)\!\!-\!\{\!1\!\!\leftrightarrow\!\! 2\!\}\right)\!\nonumber \\
		&\!-\!\frac{1}{36 \pi  q^2}\langle\! \, GG\, \!\rangle\!  \!\left(\!\!\langle\!\overline{f_1}f_1\!\rangle\!  \langle\! \, \overline{f_2}f_2\!\rangle\! \!+\!\langle\! \, \overline{f_3}f_3\!\rangle\!  \langle\! \, \overline{f_4}f_4\!\rangle\! \right)\! \!\left(\!\left(\delta _{2a} \delta _{1b} \delta _{4c} \delta _{3d}\!-\!\{\!c\!\!\leftrightarrow\!\! d\!\}\right)\!\!-\!\{\!1\!\!\leftrightarrow\!\! 2\!\}\right)\!\nonumber \\
		&\!-\!\frac{1}{2 \pi  q^2}\!\left(\!\left(\!\left(\!\left(\delta _{24} \delta _{1b} \delta _{3d} \delta _{ac} \langle\! \, GG\, \!\rangle\!  \langle\! \, \overline{f_1}f_1\!\rangle\!  \langle\! \, \overline{f_b}f_b\!\rangle\! \!-\!\{\!1\!\!\leftrightarrow\!\! 2,3\!\!\leftrightarrow\!\! 4\!\}\right)\!\!+\!\{\!1\!\!\leftrightarrow\!\! 2,a\!\!\leftrightarrow\!\! b\!\}\right)\!\!-\!\{\!c\!\!\leftrightarrow\!\! d\!\}\right)\!\!-\!\{\!3\!\!\leftrightarrow\!\! 4\!\}\right)\!\nonumber \\
		&\!+\!\frac{1}{96 \pi ^2 q^2}\!\left(\!\!\langle\!\overline{f_1}\text{G}f_1\!\rangle\!  \langle\! \, \overline{f_2}\text{G}f_2\!\rangle\! \!+\!\langle\! \, \overline{f_3}\text{G}f_3\!\rangle\!  \langle\! \, \overline{f_4}\text{G}f_4\!\rangle\! \right)\! \!\left(\!\left(\delta _{2a} \delta _{1b} \delta _{4c} \delta _{3d}\!-\!\{\!c\!\!\leftrightarrow\!\! d\!\}\right)\!\!-\!\{\!1\!\!\leftrightarrow\!\! 2\!\}\right)\!\nonumber \\
		&\!+\!\frac{1}{9 q^2}\!\!\left(\!\left(12 m_1\!-\!m_2\right)\!\!\langle\!\overline{f_2}f_2\!\rangle\!  \langle\! \, \overline{f_3}f_3\!\rangle\!  \langle\! \, \overline{f_4}f_4\!\rangle\! \!-\!\langle\! \, \overline{f_1}f_1\!\rangle\!  \!\left(\!\!\langle\!\overline{f_3}f_3\!\rangle\!  \!\left(\!\!\left(m_1\!\!-\!12 m_2\right)\!\!\langle\!\overline{f_4}f_4\!\rangle\! \!+\!\!\left(m_3\!\!-\!12 m_4\right)\!\!\langle\!\overline{f_2}f_2\!\rangle\! \right)\!\!+\!\!\left(m_4\!\!-\!12 m_3\right)\!\!\langle\!\overline{f_2}f_2\!\rangle\!  \langle\! \, \overline{f_4}f_4\!\rangle\! \right)\!\right)\nonumber \\
		&\qquad\qquad\times\left(\!\left(\delta _{2a} \delta _{1b} \delta _{4c} \delta _{3d}\!-\!\{\!c\!\!\leftrightarrow\!\! d\!\}\right)\!\!-\!\{\!1\!\!\leftrightarrow\!\! 2\!\}\right)\nonumber \\
		&\!-\!\frac{2}{q^2}\!\left(\!\left(\!\left(\!\left(\delta _{24} \delta _{1b} \delta _{3d} \delta _{ac} \langle\! \, \overline{f_2}f_2\!\rangle\!  \langle\! \, \overline{f_a}f_a\!\rangle\!  \!\left(\!\left(2 m_1\!+\!m_3\right)\!\!\langle\!\overline{f_3}f_3\!\rangle\! \!+\!\!\left(m_1\!\!+\!2 m_3\right)\!\!\langle\!\overline{f_1}f_1\!\rangle\! \right)\!\!-\!\{\!3\!\!\leftrightarrow\!\! 4\!\}\right)\!\!-\!\{\!c\!\!\leftrightarrow\!\! d\!\}\right)\!\!-\!\{\!a\!\!\leftrightarrow\!\! b\!\}\right)\!\!-\!\{\!1\!\!\leftrightarrow\!\! 2\!\}\right)
	\end{align}
}


{\tiny
	\begin{align}
		\Pi_{J^r_2}(q^2)\!=&\!\frac{q^8 g_s^2 \!\log\! \!\left(\!-\!\frac{q^2}{\mu ^2}\right)\!}{232243200 \pi ^8}\!\Big[2923 \!\left(\!\left(\!\left(\!\left(\delta _{24} \delta _{1b} \delta _{3d} \delta _{ac}\!+\!\{\!3\!\!\leftrightarrow\!\! 4,a\!\!\leftrightarrow\!\! c\!\}\right)\!\!+\!\{\!c\!\!\leftrightarrow\!\! d\!\}\right)\!\!+\!\{\!1\!\!\leftrightarrow\!\! 2\!\}\right)\!\!+\!\!\left(\!\left(\!\left(\delta _{24} \delta _{1a} \delta _{3d} \delta _{bc}\!+\!\{\!3\!\!\leftrightarrow\!\! 4,b\!\!\leftrightarrow\!\! c\!\}\right)\!\!+\!\{\!c\!\!\leftrightarrow\!\! d\!\}\right)\!\!+\!\{\!1\!\!\leftrightarrow\!\! 2\!\}\right)\!\right)\nonumber \\
		&\qquad\qquad\qquad\qquad-\!248922 \!\left(\!\left(\delta _{2a} \delta _{1b} \delta _{4c} \delta _{3d}\!+\!\{\!c\!\!\leftrightarrow\!\! d\!\}\right)\!\!+\!\{\!1\!\!\leftrightarrow\!\! 2\!\}\right)\!\Big]\!\nonumber \\
		&\!-\!\frac{q^8 g_s^2 \!\log\! ^2\!\left(\!-\!\frac{q^2}{\mu ^2}\right)\!}{552960 \pi ^8}\!\Big[\!\left(\!\left(\!\left(\!\left(\delta _{24} \delta _{1b} \delta _{3d} \delta _{ac}\!+\!\{\!3\!\!\leftrightarrow\!\! 4,a\!\!\leftrightarrow\!\! c\!\}\right)\!\!+\!\{\!c\!\!\leftrightarrow\!\! d\!\}\right)\!\!+\!\{\!1\!\!\leftrightarrow\!\! 2\!\}\right)\!\!+\!\!\left(\!\left(\!\left(\delta _{24} \delta _{1a} \delta _{3d} \delta _{bc}\!+\!\{\!3\!\!\leftrightarrow\!\! 4,b\!\!\leftrightarrow\!\! c\!\}\right)\!\!+\!\{\!c\!\!\leftrightarrow\!\! d\!\}\right)\!\!+\!\{\!1\!\!\leftrightarrow\!\! 2\!\}\right)\!\right)\nonumber \\
		&\qquad\qquad\qquad\qquad-\!54 \!\left(\!\left(\delta _{2a} \delta _{1b} \delta _{4c} \delta _{3d}\!+\!\{\!c\!\!\leftrightarrow\!\! d\!\}\right)\!\!+\!\{\!1\!\!\leftrightarrow\!\! 2\!\}\right)\!\Big]\!\nonumber \\
		&\!-\!\frac{q^8 \!\log\! \!\left(\!-\!\frac{q^2}{\mu ^2}\right)\!}{5120 \pi ^6}\!\left(\!\left(\delta _{2a} \delta _{1b} \delta _{4c} \delta _{3d}\!+\!\{\!c\!\!\leftrightarrow\!\! d\!\}\right)\!\!+\!\{\!1\!\!\leftrightarrow\!\! 2\!\}\right)\!-\!\frac{q^4 \!\log\! \!\left(\!-\!\frac{q^2}{\mu ^2}\right)\!}{64 \pi ^5}\langle\! \, GG\, \!\rangle\!  \!\left(\!\left(\delta _{2a} \delta _{1b} \delta _{4c} \delta _{3d}\!+\!\{\!c\!\!\leftrightarrow\!\! d\!\}\right)\!\!+\!\{\!1\!\!\leftrightarrow\!\! 2\!\}\right)\!\nonumber \\
		&\!-\!\frac{q^4 \!\log\! \!\left(\!-\!\frac{q^2}{\mu ^2}\right)\!}{96 \pi ^4}\!\left(\!\left(3 m_1\!+\!m_2\right)\!\!\langle\!\overline{f_1}f_1\!\rangle\! \!+\!\!\left(m_1\!\!+\!3 m_2\right)\!\!\langle\!\overline{f_2}f_2\!\rangle\! \!+\!\!\left(3 m_3\!+\!m_4\right)\!\!\langle\!\overline{f_3}f_3\!\rangle\! \!+\!\!\left(m_3\!\!+\!3 m_4\right)\!\!\langle\!\overline{f_4}f_4\!\rangle\! \right)\! \!\left(\!\left(\delta _{2a} \delta _{1b} \delta _{4c} \delta _{3d}\!+\!\{\!c\!\!\leftrightarrow\!\! d\!\}\right)\!\!+\!\{\!1\!\!\leftrightarrow\!\! 2\!\}\right)\!\nonumber \\
		&\!+\!\frac{11 q^2 \!\log\! \!\left(\!-\!\frac{q^2}{\mu ^2}\right)\!}{576 \pi ^6}\langle\! G^3\!\rangle\!  \!\left(\!\left(\delta _{2a} \delta _{1b} \delta _{4c} \delta _{3d}\!+\!\{\!c\!\!\leftrightarrow\!\! d\!\}\right)\!\!+\!\{\!1\!\!\leftrightarrow\!\! 2\!\}\right)\!+\!\frac{q^2 \!\log\! \!\left(\!-\!\frac{q^2}{\mu ^2}\right)\!}{12 \pi ^2}\!\left(\!\!\langle\!\overline{f_1}f_1\!\rangle\!  \langle\! \, \overline{f_2}f_2\!\rangle\! \!+\!\langle\! \, \overline{f_3}f_3\!\rangle\!  \langle\! \, \overline{f_4}f_4\!\rangle\! \right)\! \!\left(\!\left(\delta _{2a} \delta _{1b} \delta _{4c} \delta _{3d}\!+\!\{\!c\!\!\leftrightarrow\!\! d\!\}\right)\!\!+\!\{\!1\!\!\leftrightarrow\!\! 2\!\}\right)\!\nonumber \\
		&\!-\!\frac{3 q^2 \!\log\! \!\left(\!-\!\frac{q^2}{\mu ^2}\right)\!}{2 \pi ^2}\!\left(\!\left(\!\left(\!\left(\delta _{14} \delta _{2b} \delta _{3d} \delta _{ac} \langle\! \, \overline{f_1}f_1\!\rangle\!  \langle\! \, \overline{f_b}f_b\!\rangle\! \!+\!\{\!1\!\!\leftrightarrow\!\! 2\!\}\right)\!\!+\!\{\!1\!\!\leftrightarrow\!\! 4,a\!\!\leftrightarrow\!\! b\!\}\right)\!\!+\!\{\!c\!\!\leftrightarrow\!\! d\!\}\right)\!\!+\!\{\!3\!\!\leftrightarrow\!\! 4\!\}\right)\!\nonumber \\
		&\!+\!\frac{\!\log\! \!\left(\!-\!\frac{q^2}{\mu ^2}\right)\!}{24 \pi ^2}\!\left(\!\!\langle\!\overline{f_2}f_2\!\rangle\!  \langle\! \, \overline{f_1}\text{G}f_1\!\rangle\! \!+\!\langle\! \, \overline{f_1}f_1\!\rangle\!  \langle\! \, \overline{f_2}\text{G}f_2\!\rangle\! \!+\!\langle\! \, \overline{f_4}f_4\!\rangle\!  \langle\! \, \overline{f_3}\text{G}f_3\!\rangle\! \!+\!\langle\! \, \overline{f_3}f_3\!\rangle\!  \langle\! \, \overline{f_4}\text{G}f_4\!\rangle\! \right)\! \!\left(\!\left(\delta _{2a} \delta _{1b} \delta _{4c} \delta _{3d}\!+\!\{\!c\!\!\leftrightarrow\!\! d\!\}\right)\!\!+\!\{\!1\!\!\leftrightarrow\!\! 2\!\}\right)\!\nonumber \\
		&\!-\!\frac{1}{96 \pi ^2 q^2}\!\left(\!\left(\delta _{2a} \delta _{1b} \delta _{4c} \delta _{3d}\!+\!\{\!c\!\!\leftrightarrow\!\! d\!\}\right)\!\!+\!\{\!1\!\!\leftrightarrow\!\! 2\!\}\right)\!\left(\!\!\langle\!\overline{f_1}\text{G}f_1\!\rangle\!  \langle\! \, \overline{f_2}\text{G}f_2\!\rangle\! \!+\!\langle\! \, \overline{f_3}\text{G}f_3\!\rangle\!  \langle\! \, \overline{f_4}\text{G}f_4\!\rangle\! \right)\! \!\!\nonumber \\
		&\!+\!\frac{1}{36 \pi  q^2}\langle\! \, GG\, \!\rangle\!  \!\left(\!\!\langle\!\overline{f_1}f_1\!\rangle\!  \langle\! \, \overline{f_2}f_2\!\rangle\! \!+\!\langle\! \, \overline{f_3}f_3\!\rangle\!  \langle\! \, \overline{f_4}f_4\!\rangle\! \right)\! \!\left(\!\left(\delta _{2a} \delta _{1b} \delta _{4c} \delta _{3d}\!+\!\{\!c\!\!\leftrightarrow\!\! d\!\}\right)\!\!+\!\{\!1\!\!\leftrightarrow\!\! 2\!\}\right)\!\nonumber \\
		&\!-\!\frac{1}{2 \pi  q^2}\!\left(\!\left(\!\left(\!\left(\delta _{14} \delta _{2b} \delta _{3d} \delta _{ac} \langle\! \, GG\, \!\rangle\!  \langle\! \, \overline{f_1}f_1\!\rangle\!  \langle\! \, \overline{f_b}f_b\!\rangle\! \!+\!\{\!1\!\!\leftrightarrow\!\! 2\!\}\right)\!\!+\!\{\!1\!\!\leftrightarrow\!\! 4,a\!\!\leftrightarrow\!\! b\!\}\right)\!\!+\!\{\!c\!\!\leftrightarrow\!\! d\!\}\right)\!\!+\!\{\!3\!\!\leftrightarrow\!\! 4\!\}\right)\!\nonumber \\
		&\!+\!\frac{1}{9 q^2}\!\left(\!\left(12 m_1\!+\!m_2\right)\!\!\langle\!\overline{f_2}f_2\!\rangle\!  \langle\! \, \overline{f_3}f_3\!\rangle\!  \langle\! \, \overline{f_4}f_4\!\rangle\! \!+\!\langle\! \, \overline{f_1}f_1\!\rangle\!  \!\left(\!\left(12 m_3\!+\!m_4\right)\!\!\langle\!\overline{f_2}f_2\!\rangle\!  \langle\! \, \overline{f_4}f_4\!\rangle\! \!+\!\langle\! \, \overline{f_3}f_3\!\rangle\!  \!\left(\!\!\left(m_1\!\!+\!12 m_2\right)\!\!\langle\!\overline{f_4}f_4\!\rangle\! \!+\!\!\left(m_3\!\!+\!12 m_4\right)\!\!\langle\!\overline{f_2}f_2\!\rangle\! \right)\!\right)\!\right)\nonumber \\
		&\qquad\qquad\times\left(\!\left(\delta _{2a} \delta _{1b} \delta _{4c} \delta _{3d}\!+\!\{\!c\!\!\leftrightarrow\!\! d\!\}\right)\!\!+\!\{\!1\!\!\leftrightarrow\!\! 2\!\}\right)\!\nonumber \\
		&\!-\!\frac{2}{q^2}\!\left(\!\left(\!\left(\!\left(\delta _{24} \delta _{1b} \delta _{3d} \delta _{ac} \langle\! \, \overline{f_2}f_2\!\rangle\!  \langle\! \, \overline{f_a}f_a\!\rangle\!  \!\left(\!\left(2 m_1\!+\!m_3\right)\!\!\langle\!\overline{f_3}f_3\!\rangle\! \!+\!\!\left(m_1\!\!+\!2 m_3\right)\!\!\langle\!\overline{f_1}f_1\!\rangle\! \right)\!\!+\!\{\!3\!\!\leftrightarrow\!\! 4\!\}\right)\!\!+\!\{\!c\!\!\leftrightarrow\!\! d\!\}\right)\!\!+\!\{\!a\!\!\leftrightarrow\!\! b\!\}\right)\!\!+\!\{\!1\!\!\leftrightarrow\!\! 2\!\}\right)
	\end{align}
}


{\tiny
	\begin{align}
		\Pi_{J^r_3}(q^2)\!=&\!\frac{q^8 g_s^2 \!\log\! \!\left(\!-\!\frac{q^2}{\mu ^2}\right)\!}{928972800 \pi ^8}\!\Big[\!1013 \!\left(\!\left(\!\left(\!\left(\delta _{24} \delta _{1b} \delta _{3d} \delta _{ac}\!-\!\{\!3\!\!\leftrightarrow\!\! 4,a\!\!\leftrightarrow\!\! c\!\}\right)\!\!-\!\{\!c\!\!\leftrightarrow\!\! d\!\}\right)\!\!-\!\{\!1\!\!\leftrightarrow\!\! 2\!\}\right)\!\!-\!\!\left(\!\left(\!\left(\delta _{24} \delta _{1a} \delta _{3d} \delta _{bc}\!-\!\{\!3\!\!\leftrightarrow\!\! 4,b\!\!\leftrightarrow\!\! c\!\}\right)\!\!-\!\{\!c\!\!\leftrightarrow\!\! d\!\}\right)\!\!-\!\{\!1\!\!\leftrightarrow\!\! 2\!\}\right)\!\right)\nonumber \\
		&\qquad\qquad\qquad\qquad+\!45972 \!\left(\!\left(\delta _{2a} \delta _{1b} \delta _{4c} \delta _{3d}\!-\!\{\!c\!\!\leftrightarrow\!\! d\!\}\right)\!\!-\!\{\!1\!\!\leftrightarrow\!\! 2\!\}\right)\!\Big]\!\nonumber \\
		&\!-\!\frac{q^8 g_s^2 \!\log\! ^2\!\left(\!-\!\frac{q^2}{\mu ^2}\right)\!}{8847360 \pi ^8}\!\Big[\!\left(\!\left(\!\left(\!\left(\delta _{24} \delta _{1b} \delta _{3d} \delta _{ac}\!-\!\{\!3\!\!\leftrightarrow\!\! 4,a\!\!\leftrightarrow\!\! c\!\}\right)\!\!-\!\{\!c\!\!\leftrightarrow\!\! d\!\}\right)\!\!-\!\{\!1\!\!\leftrightarrow\!\! 2\!\}\right)\!\!-\!\!\left(\!\left(\!\left(\delta _{24} \delta _{1a} \delta _{3d} \delta _{bc}\!-\!\{\!3\!\!\leftrightarrow\!\! 4,b\!\!\leftrightarrow\!\! c\!\}\right)\!\!-\!\{\!c\!\!\leftrightarrow\!\! d\!\}\right)\!\!-\!\{\!1\!\!\leftrightarrow\!\! 2\!\}\right)\!\right)\nonumber \\
		&\qquad\qquad\qquad\qquad+\!54 \!\left(\!\left(\delta _{2a} \delta _{1b} \delta _{4c} \delta _{3d}\!-\!\{\!c\!\!\leftrightarrow\!\! d\!\}\right)\!\!-\!\{\!1\!\!\leftrightarrow\!\! 2\!\}\right)\!\Big]\!\nonumber \\
		&\!-\!\frac{q^8 \!\log\! \!\left(\!-\!\frac{q^2}{\mu ^2}\right)\!}{10240 \pi ^6}\!\left(\!\left(\delta _{2a} \delta _{1b} \delta _{4c} \delta _{3d}\!-\!\{\!c\!\!\leftrightarrow\!\! d\!\}\right)\!\!-\!\{\!1\!\!\leftrightarrow\!\! 2\!\}\right)\!+\!\frac{q^4 \!\log\! \!\left(\!-\!\frac{q^2}{\mu ^2}\right)\!}{1024 \pi ^5}\langle\! \, GG\, \!\rangle\!  \!\left(\!\left(\delta _{2a} \delta _{1b} \delta _{4c} \delta _{3d}\!-\!\{\!c\!\!\leftrightarrow\!\! d\!\}\right)\!\!-\!\{\!1\!\!\leftrightarrow\!\! 2\!\}\right)\!\nonumber \\
		&\!-\!\frac{q^4 \!\log\! \!\left(\!-\!\frac{q^2}{\mu ^2}\right)\!}{192 \pi ^4}\!\left(\!\left(3 m_1\!+\!m_2\right)\!\!\langle\!\overline{f_1}f_1\!\rangle\! \!+\!\!\left(m_1\!\!+\!3 m_2\right)\!\!\langle\!\overline{f_2}f_2\!\rangle\! \!+\!\!\left(3 m_3\!+\!m_4\right)\!\!\langle\!\overline{f_3}f_3\!\rangle\! \!+\!\!\left(m_3\!\!+\!3 m_4\right)\!\!\langle\!\overline{f_4}f_4\!\rangle\! \right)\! \!\left(\!\left(\delta _{2a} \delta _{1b} \delta _{4c} \delta _{3d}\!-\!\{\!c\!\!\leftrightarrow\!\! d\!\}\right)\!\!-\!\{\!1\!\!\leftrightarrow\!\! 2\!\}\right)\!\nonumber \\
		&\!-\!\frac{13 q^2 \!\log\! \!\left(\!-\!\frac{q^2}{\mu ^2}\right)\!}{18432 \pi ^6}\langle\! G^3\!\rangle\!  \!\left(\!\left(\delta _{2a} \delta _{1b} \delta _{4c} \delta _{3d}\!-\!\{\!c\!\!\leftrightarrow\!\! d\!\}\right)\!\!-\!\{\!1\!\!\leftrightarrow\!\! 2\!\}\right)\!+\!\frac{q^2 \!\log\! \!\left(\!-\!\frac{q^2}{\mu ^2}\right)\!}{24 \pi ^2}\!\left(\!\!\langle\!\overline{f_1}f_1\!\rangle\!  \langle\! \, \overline{f_2}f_2\!\rangle\! \!+\!\langle\! \, \overline{f_3}f_3\!\rangle\!  \langle\! \, \overline{f_4}f_4\!\rangle\! \right)\! \!\left(\!\left(\delta _{2a} \delta _{1b} \delta _{4c} \delta _{3d}\!-\!\{\!c\!\!\leftrightarrow\!\! d\!\}\right)\!\!-\!\{\!1\!\!\leftrightarrow\!\! 2\!\}\right)\!\nonumber \\
		&\!-\!\frac{5 \!\log\! \!\left(\!-\!\frac{q^2}{\mu ^2}\right)\!}{192 \pi ^2}\!\left(\!\!\langle\!\overline{f_2}f_2\!\rangle\!  \langle\! \, \overline{f_1}\text{G}f_1\!\rangle\! \!+\!\langle\! \, \overline{f_1}f_1\!\rangle\!  \langle\! \, \overline{f_2}\text{G}f_2\!\rangle\! \!+\!\langle\! \, \overline{f_4}f_4\!\rangle\!  \langle\! \, \overline{f_3}\text{G}f_3\!\rangle\! \!+\!\langle\! \, \overline{f_3}f_3\!\rangle\!  \langle\! \, \overline{f_4}\text{G}f_4\!\rangle\! \right)\! \!\left(\!\left(\delta _{2a} \delta _{1b} \delta _{4c} \delta _{3d}\!-\!\{\!c\!\!\leftrightarrow\!\! d\!\}\right)\!\!-\!\{\!1\!\!\leftrightarrow\!\! 2\!\}\right)\!\nonumber \\
		&\!+\!\frac{49}{6144 \pi ^2 q^2}\!\left(\!\!\langle\!\overline{f_1}\text{G}f_1\!\rangle\!  \langle\! \, \overline{f_2}\text{G}f_2\!\rangle\! \!+\!\langle\! \, \overline{f_3}\text{G}f_3\!\rangle\!  \langle\! \, \overline{f_4}\text{G}f_4\!\rangle\! \right)\! \!\left(\!\left(\delta _{2a} \delta _{1b} \delta _{4c} \delta _{3d}\!-\!\{\!c\!\!\leftrightarrow\!\! d\!\}\right)\!\!-\!\{\!1\!\!\leftrightarrow\!\! 2\!\}\right)\!\nonumber \\
		&\!-\!\frac{5}{288 \pi  q^2}\langle\! \, GG\, \!\rangle\!  \!\left(\!\!\langle\!\overline{f_1}f_1\!\rangle\!  \langle\! \, \overline{f_2}f_2\!\rangle\! \!+\!\langle\! \, \overline{f_3}f_3\!\rangle\!  \langle\! \, \overline{f_4}f_4\!\rangle\! \right)\! \!\left(\!\left(\delta _{2a} \delta _{1b} \delta _{4c} \delta _{3d}\!-\!\{\!c\!\!\leftrightarrow\!\! d\!\}\right)\!\!-\!\{\!1\!\!\leftrightarrow\!\! 2\!\}\right)\!\nonumber \\
		&\!+\!\frac{1}{18 q^2}\!\left(\!\left(12 m_1\!+\!m_2\right)\!\!\langle\!\overline{f_2}f_2\!\rangle\!  \langle\! \, \overline{f_3}f_3\!\rangle\!  \langle\! \, \overline{f_4}f_4\!\rangle\! \!+\!\langle\! \, \overline{f_1}f_1\!\rangle\!  \!\left(\!\left(12 m_3\!+\!m_4\right)\!\!\langle\!\overline{f_2}f_2\!\rangle\!  \langle\! \, \overline{f_4}f_4\!\rangle\! \!+\!\langle\! \, \overline{f_3}f_3\!\rangle\!  \!\left(\!\!\left(m_1\!\!+\!12 m_2\right)\!\!\langle\!\overline{f_4}f_4\!\rangle\! \!+\!\!\left(m_3\!\!+\!12 m_4\right)\!\!\langle\!\overline{f_2}f_2\!\rangle\! \right)\!\right)\!\right)\nonumber \\
		&\qquad\qquad\times\left(\!\left(\delta _{2a} \delta _{1b} \delta _{4c} \delta _{3d}\!-\!\{\!c\!\!\leftrightarrow\!\! d\!\}\right)\!\!-\!\{\!1\!\!\leftrightarrow\!\! 2\!\}\right)\!
	\end{align}
}


{\tiny
	\begin{align}
		\Pi_{J^r_4}(q^2)\!=&\!\frac{q^8 g_s^2 \!\log\! \!\left(\!-\!\frac{q^2}{\mu ^2}\right)\!}{232243200 \pi ^8}\!\Big[2988 \!\left(\!\left(\delta _{2a} \delta _{1b} \delta _{4c} \delta _{3d}\!+\!\{\!c\!\!\leftrightarrow\!\! d\!\}\right)\!\!+\!\{\!1\!\!\leftrightarrow\!\! 2\!\}\right)\nonumber \\
		&\qquad\qquad\qquad\qquad-\!563 \!\left(\!\left(\!\left(\!\left(\delta _{24} \delta _{1b} \delta _{3d} \delta _{ac}\!+\!\{\!3\!\!\leftrightarrow\!\! 4,a\!\!\leftrightarrow\!\! c\!\}\right)\!\!+\!\{\!c\!\!\leftrightarrow\!\! d\!\}\right)\!\!+\!\{\!1\!\!\leftrightarrow\!\! 2\!\}\right)\!\!+\!\!\left(\!\left(\!\left(\delta _{24} \delta _{1a} \delta _{3d} \delta _{bc}\!+\!\{\!3\!\!\leftrightarrow\!\! 4,b\!\!\leftrightarrow\!\! c\!\}\right)\!\!+\!\{\!c\!\!\leftrightarrow\!\! d\!\}\right)\!\!+\!\{\!1\!\!\leftrightarrow\!\! 2\!\}\right)\!\right)\!\Big]\!\nonumber \\
		&\!-\!\frac{q^8 g_s^2 \!\log\! ^2\!\left(\!-\!\frac{q^2}{\mu ^2}\right)\!}{2211840 \pi ^8}\!\Big[\!\left(\!\left(\!\left(\!\left(\delta _{24} \delta _{1b} \delta _{3d} \delta _{ac}\!+\!\{\!3\!\!\leftrightarrow\!\! 4,a\!\!\leftrightarrow\!\! c\!\}\right)\!\!+\!\{\!c\!\!\leftrightarrow\!\! d\!\}\right)\!\!+\!\{\!1\!\!\leftrightarrow\!\! 2\!\}\right)\!\!+\!\!\left(\!\left(\!\left(\delta _{24} \delta _{1a} \delta _{3d} \delta _{bc}\!+\!\{\!3\!\!\leftrightarrow\!\! 4,b\!\!\leftrightarrow\!\! c\!\}\right)\!\!+\!\{\!c\!\!\leftrightarrow\!\! d\!\}\right)\!\!+\!\{\!1\!\!\leftrightarrow\!\! 2\!\}\right)\!\right)\nonumber \\
		&\qquad\qquad\qquad\qquad+\!54 \!\left(\!\left(\delta _{2a} \delta _{1b} \delta _{4c} \delta _{3d}\!+\!\{\!c\!\!\leftrightarrow\!\! d\!\}\right)\!\!+\!\{\!1\!\!\leftrightarrow\!\! 2\!\}\right)\!\Big]\!\nonumber \\
		&\!-\!\frac{q^8 \!\log\! \!\left(\!-\!\frac{q^2}{\mu ^2}\right)\!}{2560 \pi ^6}\!\left(\!\left(\delta _{2a} \delta _{1b} \delta _{4c} \delta _{3d}\!+\!\{\!c\!\!\leftrightarrow\!\! d\!\}\right)\!\!+\!\{\!1\!\!\leftrightarrow\!\! 2\!\}\right)\!+\!\frac{q^4 \!\log\! \!\left(\!-\!\frac{q^2}{\mu ^2}\right)\!}{256 \pi ^5}\langle\! \, GG\, \!\rangle\!  \!\left(\!\left(\delta _{2a} \delta _{1b} \delta _{4c} \delta _{3d}\!+\!\{\!c\!\!\leftrightarrow\!\! d\!\}\right)\!\!+\!\{\!1\!\!\leftrightarrow\!\! 2\!\}\right)\!\nonumber \\
		&\!+\!\frac{q^4 \!\log\! \!\left(\!-\!\frac{q^2}{\mu ^2}\right)\!}{48 \pi ^4}\!\left(\!\!\left(m_1\!\!-\!3 m_2\right)\!\!\langle\!\overline{f_2}f_2\!\rangle\! \!+\!\!\left(m_2\!\!-\!3 m_1\right)\!\!\langle\!\overline{f_1}f_1\!\rangle\! \!+\!\!\left(m_3\!\!-\!3 m_4\right)\!\!\langle\!\overline{f_4}f_4\!\rangle\! \!+\!\!\left(m_4\!\!-\!3 m_3\right)\!\!\langle\!\overline{f_3}f_3\!\rangle\! \right)\! \!\left(\!\left(\delta _{2a} \delta _{1b} \delta _{4c} \delta _{3d}\!+\!\{\!c\!\!\leftrightarrow\!\! d\!\}\right)\!\!+\!\{\!1\!\!\leftrightarrow\!\! 2\!\}\right)\!\nonumber \\
		&\!-\!\frac{13 q^2 \!\log\! \!\left(\!-\!\frac{q^2}{\mu ^2}\right)\!}{4608 \pi ^6}\langle\! G^3\!\rangle\!  \!\left(\!\left(\delta _{2a} \delta _{1b} \delta _{4c} \delta _{3d}\!+\!\{\!c\!\!\leftrightarrow\!\! d\!\}\right)\!\!+\!\{\!1\!\!\leftrightarrow\!\! 2\!\}\right)\!-\!\frac{q^2 \!\log\! \!\left(\!-\!\frac{q^2}{\mu ^2}\right)\!}{6 \pi ^2}\!\left(\!\!\langle\!\overline{f_1}f_1\!\rangle\!  \langle\! \, \overline{f_2}f_2\!\rangle\! \!+\!\langle\! \, \overline{f_3}f_3\!\rangle\!  \langle\! \, \overline{f_4}f_4\!\rangle\! \right)\! \!\left(\!\left(\delta _{2a} \delta _{1b} \delta _{4c} \delta _{3d}\!+\!\{\!c\!\!\leftrightarrow\!\! d\!\}\right)\!\!+\!\{\!1\!\!\leftrightarrow\!\! 2\!\}\right)\!\nonumber \\
		&\!+\!\frac{5 \!\log\! \!\left(\!-\!\frac{q^2}{\mu ^2}\right)\!}{48 \pi ^2}\!\left(\!\!\langle\!\overline{f_2}f_2\!\rangle\!  \langle\! \, \overline{f_1}\text{G}f_1\!\rangle\! \!+\!\langle\! \, \overline{f_1}f_1\!\rangle\!  \langle\! \, \overline{f_2}\text{G}f_2\!\rangle\! \!+\!\langle\! \, \overline{f_4}f_4\!\rangle\!  \langle\! \, \overline{f_3}\text{G}f_3\!\rangle\! \!+\!\langle\! \, \overline{f_3}f_3\!\rangle\!  \langle\! \, \overline{f_4}\text{G}f_4\!\rangle\! \right)\! \!\left(\!\left(\delta _{2a} \delta _{1b} \delta _{4c} \delta _{3d}\!+\!\{\!c\!\!\leftrightarrow\!\! d\!\}\right)\!\!+\!\{\!1\!\!\leftrightarrow\!\! 2\!\}\right)\!\nonumber \\
		&\!-\!\frac{49}{1536 \pi ^2 q^2}\!\left(\!\!\langle\!\overline{f_1}\text{G}f_1\!\rangle\!  \langle\! \, \overline{f_2}\text{G}f_2\!\rangle\! \!+\!\langle\! \, \overline{f_3}\text{G}f_3\!\rangle\!  \langle\! \, \overline{f_4}\text{G}f_4\!\rangle\! \right)\! \!\left(\!\left(\delta _{2a} \delta _{1b} \delta _{4c} \delta _{3d}\!+\!\{\!c\!\!\leftrightarrow\!\! d\!\}\right)\!\!+\!\{\!1\!\!\leftrightarrow\!\! 2\!\}\right)\!\nonumber \\
		&\!+\!\frac{5}{72 \pi  q^2}\langle\! \, GG\, \!\rangle\!  \!\left(\!\!\langle\!\overline{f_1}f_1\!\rangle\!  \langle\! \, \overline{f_2}f_2\!\rangle\! \!+\!\langle\! \, \overline{f_3}f_3\!\rangle\!  \langle\! \, \overline{f_4}f_4\!\rangle\! \right)\! \!\left(\!\left(\delta _{2a} \delta _{1b} \delta _{4c} \delta _{3d}\!+\!\{\!c\!\!\leftrightarrow\!\! d\!\}\right)\!\!+\!\{\!1\!\!\leftrightarrow\!\! 2\!\}\right)\!\nonumber \\
		&\!-\!\frac{2}{9 q^2}\!\left(\!\!\left(m_2\!\!-\!12 m_1\right)\!\!\langle\!\overline{f_2}f_2\!\rangle\!  \langle\! \, \overline{f_3}f_3\!\rangle\!  \langle\! \, \overline{f_4}f_4\!\rangle\! \!+\!\langle\! \, \overline{f_1}f_1\!\rangle\!  \!\left(\!\!\langle\!\overline{f_3}f_3\!\rangle\!  \!\left(\!\!\left(m_1\!\!-\!12 m_2\right)\!\!\langle\!\overline{f_4}f_4\!\rangle\! \!+\!\!\left(m_3\!\!-\!12 m_4\right)\!\!\langle\!\overline{f_2}f_2\!\rangle\! \right)\!\!+\!\!\left(m_4\!\!-\!12 m_3\right)\!\!\langle\!\overline{f_2}f_2\!\rangle\!  \langle\! \, \overline{f_4}f_4\!\rangle\! \right)\!\right)\nonumber \\
		&\qquad\qquad\times\left(\!\left(\delta _{2a} \delta _{1b} \delta _{4c} \delta _{3d}\!+\!\{\!c\!\!\leftrightarrow\!\! d\!\}\right)\!\!+\!\{\!1\!\!\leftrightarrow\!\! 2\!\}\right)\!
	\end{align}
}


{\tiny
	\begin{align}
		\Pi_{J^r_a}(q^2)\!=&\!\frac{\!\left(241\!+\!5 \sqrt{241}\right)\! q^8 \!\log\! \!\left(\!-\!\frac{q^2}{\mu ^2}\right)\!}{1280 \pi ^6}\!\left(\!\left(\delta _{2a} \delta _{1b} \delta _{4c} \delta _{3d}\!+\!\{\!c\!\!\leftrightarrow\!\! d\!\}\right)\!\!+\!\{\!1\!\!\leftrightarrow\!\! 2\!\}\right)\!+\!\frac{\!\left(241\!+\!59 \sqrt{241}\right)\! q^8 g_s^2 \!\log\! ^2\!\left(\!-\!\frac{q^2}{\mu ^2}\right)\!}{15360 \pi ^8}\!\left(\!\left(\delta _{2a} \delta _{1b} \delta _{4c} \delta _{3d}\!+\!\{\!c\!\!\leftrightarrow\!\! d\!\}\right)\!\!+\!\{\!1\!\!\leftrightarrow\!\! 2\!\}\right)\!\nonumber \\
		&\!-\!\frac{q^8 g_s^2 \!\log\! \!\left(\!-\!\frac{q^2}{\mu ^2}\right)\!}{1612800 \pi ^8}\!\Big[675 \!\left(\!\left(\!\left(\!\left(\delta _{24} \delta _{1b} \delta _{3d} \delta _{ac}\!+\!\{\!3\!\!\leftrightarrow\!\! 4,a\!\!\leftrightarrow\!\! c\!\}\right)\!\!+\!\{\!c\!\!\leftrightarrow\!\! d\!\}\right)\!\!+\!\{\!1\!\!\leftrightarrow\!\! 2\!\}\right)\!\!+\!\!\left(\!\left(\!\left(\delta _{24} \delta _{1a} \delta _{3d} \delta _{bc}\!+\!\{\!3\!\!\leftrightarrow\!\! 4,b\!\!\leftrightarrow\!\! c\!\}\right)\!\!+\!\{\!c\!\!\leftrightarrow\!\! d\!\}\right)\!\!+\!\{\!1\!\!\leftrightarrow\!\! 2\!\}\right)\!\right)\nonumber \\
		&\qquad\qquad\qquad\qquad+\!\!\left(486197\!+\!63403 \sqrt{241}\right)\! \!\left(\!\left(\delta _{2a} \delta _{1b} \delta _{4c} \delta _{3d}\!+\!\{\!c\!\!\leftrightarrow\!\! d\!\}\right)\!\!+\!\{\!1\!\!\leftrightarrow\!\! 2\!\}\right)\!\Big]\!\nonumber \\
		&\!-\!\frac{q^4 \!\log\! \!\left(\!-\!\frac{q^2}{\mu ^2}\right)\!}{8 \pi ^4}\!\left(241\!+\!5 \sqrt{241}\right)\! \!\left(m_1\! \langle\! \, \overline{f_1}f_1\!\rangle\! \!+\!m_2 \langle\! \, \overline{f_2}f_2\!\rangle\! \!+\!m_3 \langle\! \, \overline{f_3}f_3\!\rangle\! \!+\!m_4 \langle\! \, \overline{f_4}f_4\!\rangle\! \right)\! \!\left(\!\left(\delta _{2a} \delta _{1b} \delta _{4c} \delta _{3d}\!+\!\{\!c\!\!\leftrightarrow\!\! d\!\}\right)\!\!+\!\{\!1\!\!\leftrightarrow\!\! 2\!\}\right)\!\nonumber \\
		&\!-\!\frac{q^4 \!\log\! \!\left(\!-\!\frac{q^2}{\mu ^2}\right)\!}{96 \pi ^5}\!\left(241\!+\!59 \sqrt{241}\right)\!\!\langle \, GG\, \!\rangle\!  \!\left(\!\left(\delta _{2a} \delta _{1b} \delta _{4c} \delta _{3d}\!+\!\{\!c\!\!\leftrightarrow\!\! d\!\}\right)\!\!+\!\{\!1\!\!\leftrightarrow\!\! 2\!\}\right)\nonumber\nonumber \\
		&\!+\!\frac{q^2 \!\log\! \!\left(\!-\!\frac{q^2}{\mu ^2}\right)\!}{32 \pi ^6}\!\left(313\!+\!35 \sqrt{241}\right)\!\!\langle G^3\!\rangle\!  \!\left(\!\left(\delta _{2a} \delta _{1b} \delta _{4c} \delta _{3d}\!+\!\{\!c\!\!\leftrightarrow\!\! d\!\}\right)\!\!+\!\{\!1\!\!\leftrightarrow\!\! 2\!\}\right)\!\nonumber \\
		&\!-\!\frac{3 q^2 \!\log\! \!\left(\!-\!\frac{q^2}{\mu ^2}\right)\!}{\pi ^2}\!\left(265\!+\!17 \sqrt{241}\right)\! \!\left(\!\left(\!\left(\!\left(\delta _{14} \delta _{2b} \delta _{3d} \delta _{ac} \langle\! \, \overline{f_1}f_1\!\rangle\!  \langle\! \, \overline{f_b}f_b\!\rangle\! \!+\!\{\!1\!\!\leftrightarrow\!\! 2\!\}\right)\!\!+\!\{\!1\!\!\leftrightarrow\!\! 4,a\!\!\leftrightarrow\!\! b\!\}\right)\!\!+\!\{\!c\!\!\leftrightarrow\!\! d\!\}\right)\!\!+\!\{\!3\!\!\leftrightarrow\!\! 4\!\}\right)\!\nonumber \\
		&\!+\!\frac{2 \!\log\! \!\left(\!-\!\frac{q^2}{\mu ^2}\right)\!}{\pi ^2}\!\left(79\!+\!5 \sqrt{241}\right)\! \!\left(\!\left(\!\left(\!\left(\delta _{14} \delta _{2b} \delta _{3d} \delta _{ac} \!\left(\!\!\langle\!\overline{f_b}f_b\!\rangle\!  \langle\! \, \overline{f_1}\text{G}f_1\!\rangle\! \!+\!\langle\! \, \overline{f_1}f_1\!\rangle\!  \langle\! \, \overline{f_b}\text{G}f_b\!\rangle\! \right)\!\!+\!\{\!1\!\!\leftrightarrow\!\! 2\!\}\right)\!\!+\!\{\!1\!\!\leftrightarrow\!\! 4,a\!\!\leftrightarrow\!\! b\!\}\right)\!\!+\!\{\!c\!\!\leftrightarrow\!\! d\!\}\right)\!\!+\!\{\!3\!\!\leftrightarrow\!\! 4\!\}\right)\!\nonumber \\
		&\!-\!\frac{1}{48 \pi ^2 q^2}\!\left(385\!+\!23 \sqrt{241}\right)\! \!\left(\!\left(\!\left(\!\left(\delta _{14} \delta _{2b} \delta _{3d} \delta _{ac} \langle\! \, \overline{f_1}\text{G}f_1\!\rangle\!  \langle\! \, \overline{f_b}\text{G}f_b\!\rangle\! \!+\!\{\!1\!\!\leftrightarrow\!\! 2\!\}\right)\!\!+\!\{\!1\!\!\leftrightarrow\!\! 4,a\!\!\leftrightarrow\!\! b\!\}\right)\!\!+\!\{\!c\!\!\leftrightarrow\!\! d\!\}\right)\!\!+\!\{\!3\!\!\leftrightarrow\!\! 4\!\}\right)\!\nonumber \\
		&\!-\!\frac{1}{\pi  q^2}\!\left(265\!+\!17 \sqrt{241}\right)\! \!\left(\!\left(\!\left(\!\left(\delta _{14} \delta _{2b} \delta _{3d} \delta _{ac} \langle\! \, GG\, \!\rangle\!  \langle\! \, \overline{f_1}f_1\!\rangle\!  \langle\! \, \overline{f_b}f_b\!\rangle\! \!+\!\{\!1\!\!\leftrightarrow\!\! 2\!\}\right)\!\!+\!\{\!1\!\!\leftrightarrow\!\! 4,a\!\!\leftrightarrow\!\! b\!\}\right)\!\!+\!\{\!c\!\!\leftrightarrow\!\! d\!\}\right)\!\!+\!\{\!3\!\!\leftrightarrow\!\! 4\!\}\right)\!\nonumber \\
		&\!+\!\frac{16}{3 q^2}\!\left(241\!+\!5 \sqrt{241}\right)\! \!\left(m_2\! \langle\! \, \overline{f_1}f_1\!\rangle\!  \langle\! \, \overline{f_3}f_3\!\rangle\!  \langle\! \, \overline{f_4}f_4\!\rangle\! \!+\!\langle\! \, \overline{f_2}f_2\!\rangle\!  \!\left(m_3\! \langle\! \, \overline{f_1}f_1\!\rangle\!  \langle\! \, \overline{f_4}f_4\!\rangle\! \!+\!\langle\! \, \overline{f_3}f_3\!\rangle\!  \!\left(m_1\! \langle\! \, \overline{f_4}f_4\!\rangle\! \!+\!m_4 \langle\! \, \overline{f_1}f_1\!\rangle\! \right)\!\right)\!\right)\nonumber\\ &\qquad\qquad\qquad\qquad\times\left(\!\left(\delta _{2a} \delta _{1b} \delta _{4c} \delta _{3d}\!+\!\{\!c\!\!\leftrightarrow\!\! d\!\}\right)\!\!+\!\{\!1\!\!\leftrightarrow\!\! 2\!\}\right)\!\nonumber \\
		&\!-\!\frac{4}{q^2}\!\left(265\!+\!17 \sqrt{241}\right)\! \!\left(\!\left(\!\left(\!\left(\delta _{24} \delta _{1b} \delta _{3d} \delta _{ac} \langle\! \, \overline{f_2}f_2\!\rangle\!  \langle\! \, \overline{f_a}f_a\!\rangle\!  \!\left(\!\left(2 m_1\!+\!m_3\right)\!\!\langle\!\overline{f_3}f_3\!\rangle\! \!+\!\!\left(m_1\!\!+\!2 m_3\right)\!\!\langle\!\overline{f_1}f_1\!\rangle\! \right)\!\!+\!\{\!3\!\!\leftrightarrow\!\! 4\!\}\right)\!\!+\!\{\!c\!\!\leftrightarrow\!\! d\!\}\right)\!\!+\!\{\!a\!\!\leftrightarrow\!\! b\!\}\right)\!\!+\!\{\!1\!\!\leftrightarrow\!\! 2\!\}\right)
	\end{align}
}


{\tiny
	\begin{align}
		\Pi_{J^r_b}(q^2)\!=&\!\frac{\!\left(241\!+\!5 \sqrt{241}\right)\! q^8 \!\log\! \!\left(\!-\!\frac{q^2}{\mu ^2}\right)\!}{640 \pi ^6}\!\left(\!\left(\delta _{2a} \delta _{1b} \delta _{4c} \delta _{3d}\!-\!\{\!c\!\!\leftrightarrow\!\! d\!\}\right)\!\!-\!\{\!1\!\!\leftrightarrow\!\! 2\!\}\right)\!+\!\frac{\!\left(2651\!+\!163 \sqrt{241}\right)\! q^8 g_s^2 \!\log\! ^2\!\left(\!-\!\frac{q^2}{\mu ^2}\right)\!}{15360 \pi ^8}\!\left(\!\left(\delta _{2a} \delta _{1b} \delta _{4c} \delta _{3d}\!-\!\{\!c\!\!\leftrightarrow\!\! d\!\}\right)\!\!-\!\{\!1\!\!\leftrightarrow\!\! 2\!\}\right)\!\nonumber \\
		&\!-\!\frac{q^8 g_s^2 \!\log\! \!\left(\!-\!\frac{q^2}{\mu ^2}\right)\!}{1612800 \pi ^8}\!\Big[\!\left(\!3023377\!+\!164921 \sqrt{241}\right)\! \!\left(\!\left(\delta _{2a} \delta _{1b} \delta _{4c} \delta _{3d}\!-\!\{\!c\!\!\leftrightarrow\!\! d\!\}\right)\!\!-\!\{\!1\!\!\leftrightarrow\!\! 2\!\}\right)\nonumber \\
		&\qquad\qquad\qquad\qquad+\!2700 \!\left(\!\left(\!\left(\!\left(\delta _{24} \delta _{1b} \delta _{3d} \delta _{ac}\!-\!\{\!3\!\!\leftrightarrow\!\! 4,a\!\!\leftrightarrow\!\! c\!\}\right)\!\!-\!\{\!c\!\!\leftrightarrow\!\! d\!\}\right)\!\!-\!\{\!1\!\!\leftrightarrow\!\! 2\!\}\right)\!\!-\!\!\left(\!\left(\!\left(\delta _{24} \delta _{1a} \delta _{3d} \delta _{bc}\!-\!\{\!3\!\!\leftrightarrow\!\! 4,b\!\!\leftrightarrow\!\! c\!\}\right)\!\!-\!\{\!c\!\!\leftrightarrow\!\! d\!\}\right)\!\!-\!\{\!1\!\!\leftrightarrow\!\! 2\!\}\right)\!\right)\!\Big]\!\nonumber \\
		&\!-\!\frac{q^4 \!\log\! \!\left(\!-\!\frac{q^2}{\mu ^2}\right)\!}{4 \pi ^4}\!\left(241\!+\!5 \sqrt{241}\right)\! \!\left(m_1\! \langle\! \, \overline{f_1}f_1\!\rangle\! \!+\!m_2 \langle\! \, \overline{f_2}f_2\!\rangle\! \!+\!m_3 \langle\! \, \overline{f_3}f_3\!\rangle\! \!+\!m_4 \langle\! \, \overline{f_4}f_4\!\rangle\! \right)\! \!\left(\!\left(\delta _{2a} \delta _{1b} \delta _{4c} \delta _{3d}\!-\!\{\!c\!\!\leftrightarrow\!\! d\!\}\right)\!\!-\!\{\!1\!\!\leftrightarrow\!\! 2\!\}\right)\!\nonumber \\
		&\!-\!\frac{q^4 \!\log\! \!\left(\!-\!\frac{q^2}{\mu ^2}\right)\!}{96 \pi ^5}\!\left(2651\!+\!163 \sqrt{241}\right)\!\!\langle \, GG\, \!\rangle\!  \!\left(\!\left(\delta _{2a} \delta _{1b} \delta _{4c} \delta _{3d}\!-\!\{\!c\!\!\leftrightarrow\!\! d\!\}\right)\!\!-\!\{\!1\!\!\leftrightarrow\!\! 2\!\}\right)\nonumber\nonumber \\
		&\!+\!\frac{q^2 \!\log\! \!\left(\!-\!\frac{q^2}{\mu ^2}\right)\!}{8 \pi ^6}\!\left(151\!+\!11 \sqrt{241}\right)\!\!\langle G^3\!\rangle\!  \!\left(\!\left(\delta _{2a} \delta _{1b} \delta _{4c} \delta _{3d}\!-\!\{\!c\!\!\leftrightarrow\!\! d\!\}\right)\!\!-\!\{\!1\!\!\leftrightarrow\!\! 2\!\}\right)\!\nonumber \\
		&\!-\!\frac{12 q^2 \!\log\! \!\left(\!-\!\frac{q^2}{\mu ^2}\right)\!}{\pi ^2}\!\left(181\!+\!11 \sqrt{241}\right)\! \!\left(\!\left(\!\left(\!\left(\delta _{14} \delta _{2b} \delta _{3d} \delta _{ac} \langle\! \, \overline{f_1}f_1\!\rangle\!  \langle\! \, \overline{f_b}f_b\!\rangle\! \!-\!\{\!1\!\!\leftrightarrow\!\! 2\!\}\right)\!\!+\!\{\!1\!\!\leftrightarrow\!\! 4,a\!\!\leftrightarrow\!\! b\!\}\right)\!\!-\!\{\!c\!\!\leftrightarrow\!\! d\!\}\right)\!\!-\!\{\!3\!\!\leftrightarrow\!\! 4\!\}\right)\!\nonumber \\
		&\!-\!\frac{2 \!\log\! \!\left(\!-\!\frac{q^2}{\mu ^2}\right)\!}{\pi ^2}\!\left(173\!+\!13 \sqrt{241}\right)\! \!\left(\!\left(\!\left(\!\left(\delta _{14} \delta _{2b} \delta _{3d} \delta _{ac} \!\left(\!\!\langle\!\overline{f_b}f_b\!\rangle\!  \langle\! \, \overline{f_1}\text{G}f_1\!\rangle\! \!+\!\langle\! \, \overline{f_1}f_1\!\rangle\!  \langle\! \, \overline{f_b}\text{G}f_b\!\rangle\! \right)\!\!-\!\{\!1\!\!\leftrightarrow\!\! 2\!\}\right)\!\!+\!\{\!1\!\!\leftrightarrow\!\! 4,a\!\!\leftrightarrow\!\! b\!\}\right)\!\!-\!\{\!c\!\!\leftrightarrow\!\! d\!\}\right)\!\!-\!\{\!3\!\!\leftrightarrow\!\! 4\!\}\right)\!\nonumber \\
		&\!-\!\frac{1}{48 \pi ^2 q^2}\!\left(961\!+\!41 \sqrt{241}\right)\! \!\left(\!\left(\!\left(\!\left(\delta _{14} \delta _{2b} \delta _{3d} \delta _{ac} \langle\! \, \overline{f_1}\text{G}f_1\!\rangle\!  \langle\! \, \overline{f_b}\text{G}f_b\!\rangle\! \!-\!\{\!1\!\!\leftrightarrow\!\! 2\!\}\right)\!\!+\!\{\!1\!\!\leftrightarrow\!\! 4,a\!\!\leftrightarrow\!\! b\!\}\right)\!\!-\!\{\!c\!\!\leftrightarrow\!\! d\!\}\right)\!\!-\!\{\!3\!\!\leftrightarrow\!\! 4\!\}\right)\!\nonumber \\
		&\!-\!\frac{4}{\pi  q^2}\!\left(181\!+\!11 \sqrt{241}\right)\! \!\left(\!\left(\!\left(\!\left(\delta _{14} \delta _{2b} \delta _{3d} \delta _{ac} \langle\! \, GG\, \!\rangle\!  \langle\! \, \overline{f_1}f_1\!\rangle\!  \langle\! \, \overline{f_b}f_b\!\rangle\! \!-\!\{\!1\!\!\leftrightarrow\!\! 2\!\}\right)\!\!+\!\{\!1\!\!\leftrightarrow\!\! 4,a\!\!\leftrightarrow\!\! b\!\}\right)\!\!-\!\{\!c\!\!\leftrightarrow\!\! d\!\}\right)\!\!-\!\{\!3\!\!\leftrightarrow\!\! 4\!\}\right)\!\nonumber \\
		&\!+\!\frac{32}{3 q^2}\!\left(241\!+\!5 \sqrt{241}\right)\! \!\left(m_2\! \langle\! \, \overline{f_1}f_1\!\rangle\!  \langle\! \, \overline{f_3}f_3\!\rangle\!  \langle\! \, \overline{f_4}f_4\!\rangle\! \!+\!\langle\! \, \overline{f_2}f_2\!\rangle\!  \!\left(m_3\! \langle\! \, \overline{f_1}f_1\!\rangle\!  \langle\! \, \overline{f_4}f_4\!\rangle\! \!+\!\langle\! \, \overline{f_3}f_3\!\rangle\!  \!\left(m_1\! \langle\! \, \overline{f_4}f_4\!\rangle\! \!+\!m_4 \langle\! \, \overline{f_1}f_1\!\rangle\! \right)\!\right)\!\right)\nonumber\\ &\qquad\qquad\qquad\qquad\times\left(\!\left(\delta _{2a} \delta _{1b} \delta _{4c} \delta _{3d}\!-\!\{\!c\!\!\leftrightarrow\!\! d\!\}\right)\!\!-\!\{\!1\!\!\leftrightarrow\!\! 2\!\}\right)\!\nonumber \\
		&\!-\!\frac{16}{q^2}\!\left(181\!+\!11 \sqrt{241}\right)\! \!\left(\!\left(\!\left(\!\left(\delta _{24} \delta _{1b} \delta _{3d} \delta _{ac} \langle\! \, \overline{f_2}f_2\!\rangle\!  \langle\! \, \overline{f_a}f_a\!\rangle\!  \!\left(\!\left(2 m_1\!+\!m_3\right)\!\!\langle\!\overline{f_3}f_3\!\rangle\! \!+\!\!\left(m_1\!\!+\!2 m_3\right)\!\!\langle\!\overline{f_1}f_1\!\rangle\! \right)\!\!-\!\{\!3\!\!\leftrightarrow\!\! 4\!\}\right)\!\!-\!\{\!c\!\!\leftrightarrow\!\! d\!\}\right)\!\!-\!\{\!a\!\!\leftrightarrow\!\! b\!\}\right)\!\!-\!\{\!1\!\!\leftrightarrow\!\! 2\!\}\right)
	\end{align}
}

{\tiny
	\begin{align}
		\Pi_{J^r_c}(q^2)\!=&\!\frac{q^8 g_s^2 \!\log\! \!\left(\!-\!\frac{q^2}{\mu ^2}\right)\!}{232243200 \pi ^8}\!\Big[877 \!\left(\!\left(\!\left(\!\left(\delta _{24} \delta _{1b} \delta _{3d} \delta _{ac}\!-\!\{\!3\!\!\leftrightarrow\!\! 4,a\!\!\leftrightarrow\!\! c\!\}\right)\!\!-\!\{\!c\!\!\leftrightarrow\!\! d\!\}\right)\!\!-\!\{\!1\!\!\leftrightarrow\!\! 2\!\}\right)\!\!-\!\!\left(\!\left(\!\left(\delta _{24} \delta _{1a} \delta _{3d} \delta _{bc}\!-\!\{\!3\!\!\leftrightarrow\!\! 4,b\!\!\leftrightarrow\!\! c\!\}\right)\!\!-\!\{\!c\!\!\leftrightarrow\!\! d\!\}\right)\!\!-\!\{\!1\!\!\leftrightarrow\!\! 2\!\}\right)\!\right)\nonumber \\
		&\qquad\qquad\qquad\qquad-\!26406 \!\left(\!\left(\delta _{2a} \delta _{1b} \delta _{4c} \delta _{3d}\!-\!\{\!c\!\!\leftrightarrow\!\! d\!\}\right)\!\!-\!\{\!1\!\!\leftrightarrow\!\! 2\!\}\right)\!\Big]\!\nonumber \\
		&\!+\!\frac{q^8 g_s^2 \!\log\! ^2\!\left(\!-\!\frac{q^2}{\mu ^2}\right)\!}{2211840 \pi ^8}\!\Big[18 \!\left(\!\left(\delta _{2a} \delta _{1b} \delta _{4c} \delta _{3d}\!-\!\{\!c\!\!\leftrightarrow\!\! d\!\}\right)\!\!-\!\{\!1\!\!\leftrightarrow\!\! 2\!\}\right)\nonumber \\
		&\qquad\qquad\qquad\qquad-\!\!\left(\!\left(\!\left(\!\left(\delta _{24} \delta _{1b} \delta _{3d} \delta _{ac}\!-\!\{\!3\!\!\leftrightarrow\!\! 4,a\!\!\leftrightarrow\!\! c\!\}\right)\!\!-\!\{\!c\!\!\leftrightarrow\!\! d\!\}\right)\!\!-\!\{\!1\!\!\leftrightarrow\!\! 2\!\}\right)\!\!-\!\!\left(\!\left(\!\left(\delta _{24} \delta _{1a} \delta _{3d} \delta _{bc}\!-\!\{\!3\!\!\leftrightarrow\!\! 4,b\!\!\leftrightarrow\!\! c\!\}\right)\!\!-\!\{\!c\!\!\leftrightarrow\!\! d\!\}\right)\!\!-\!\{\!1\!\!\leftrightarrow\!\! 2\!\}\right)\!\right)\!\Big]\!\nonumber \\
		&\!-\!\frac{q^8 \!\log\! \!\left(\!-\!\frac{q^2}{\mu ^2}\right)\!}{30720 \pi ^6}\!\left(\!\left(\delta _{2a} \delta _{1b} \delta _{4c} \delta _{3d}\!-\!\{\!c\!\!\leftrightarrow\!\! d\!\}\right)\!\!-\!\{\!1\!\!\leftrightarrow\!\! 2\!\}\right)\!-\!\frac{q^4 \!\log\! \!\left(\!-\!\frac{q^2}{\mu ^2}\right)\!}{768 \pi ^5}\langle\! \, GG\, \!\rangle\!  \!\left(\!\left(\delta _{2a} \delta _{1b} \delta _{4c} \delta _{3d}\!-\!\{\!c\!\!\leftrightarrow\!\! d\!\}\right)\!\!-\!\{\!1\!\!\leftrightarrow\!\! 2\!\}\right)\!\nonumber \\
		&\!-\!\frac{q^4 \!\log\! \!\left(\!-\!\frac{q^2}{\mu ^2}\right)\!}{192 \pi ^4}\!\left(m_1\! \langle\! \, \overline{f_1}f_1\!\rangle\! \!+\!m_2 \langle\! \, \overline{f_2}f_2\!\rangle\! \!+\!m_3 \langle\! \, \overline{f_3}f_3\!\rangle\! \!+\!m_4 \langle\! \, \overline{f_4}f_4\!\rangle\! \right)\! \!\left(\!\left(\delta _{2a} \delta _{1b} \delta _{4c} \delta _{3d}\!-\!\{\!c\!\!\leftrightarrow\!\! d\!\}\right)\!\!-\!\{\!1\!\!\leftrightarrow\!\! 2\!\}\right)\!\nonumber \\
		&\!-\!\frac{q^2 \!\log\! \!\left(\!-\!\frac{q^2}{\mu ^2}\right)\!}{3456 \pi ^6}\langle\! G^3\!\rangle\!  \!\left(\!\left(\delta _{2a} \delta _{1b} \delta _{4c} \delta _{3d}\!-\!\{\!c\!\!\leftrightarrow\!\! d\!\}\right)\!\!-\!\{\!1\!\!\leftrightarrow\!\! 2\!\}\right)\!\nonumber \\
		&\!+\!\frac{2}{9 q^2}\!\left(m_2\! \langle\! \, \overline{f_1}f_1\!\rangle\!  \langle\! \, \overline{f_3}f_3\!\rangle\!  \langle\! \, \overline{f_4}f_4\!\rangle\! \!+\!\langle\! \, \overline{f_2}f_2\!\rangle\!  \!\left(m_3\! \langle\! \, \overline{f_1}f_1\!\rangle\!  \langle\! \, \overline{f_4}f_4\!\rangle\! \!+\!\langle\! \, \overline{f_3}f_3\!\rangle\!  \!\left(m_1\! \langle\! \, \overline{f_4}f_4\!\rangle\! \!+\!m_4 \langle\! \, \overline{f_1}f_1\!\rangle\! \right)\!\right)\!\right)\! \!\left(\!\left(\delta _{2a} \delta _{1b} \delta _{4c} \delta _{3d}\!-\!\{\!c\!\!\leftrightarrow\!\! d\!\}\right)\!\!-\!\{\!1\!\!\leftrightarrow\!\! 2\!\}\right)\!
	\end{align}
}

{\tiny
	\begin{align}
		\Pi_{J^r_d}(q^2)\!=&\!\frac{q^8 g_s^2 \!\log\! \!\left(\!-\!\frac{q^2}{\mu ^2}\right)\!}{232243200 \pi ^8}\!\Big[877 \!\left(\!\left(\!\left(\!\left(\delta _{24} \delta _{1b} \delta _{3d} \delta _{ac}\!+\!\{\!3\!\!\leftrightarrow\!\! 4,a\!\!\leftrightarrow\!\! c\!\}\right)\!\!+\!\{\!c\!\!\leftrightarrow\!\! d\!\}\right)\!\!+\!\{\!1\!\!\leftrightarrow\!\! 2\!\}\right)\!\!+\!\!\left(\!\left(\!\left(\delta _{24} \delta _{1a} \delta _{3d} \delta _{bc}\!+\!\{\!3\!\!\leftrightarrow\!\! 4,b\!\!\leftrightarrow\!\! c\!\}\right)\!\!+\!\{\!c\!\!\leftrightarrow\!\! d\!\}\right)\!\!+\!\{\!1\!\!\leftrightarrow\!\! 2\!\}\right)\!\right)\nonumber \\
		&\qquad\qquad\qquad\qquad+\!4806 \!\left(\!\left(\delta _{2a} \delta _{1b} \delta _{4c} \delta _{3d}\!+\!\{\!c\!\!\leftrightarrow\!\! d\!\}\right)\!\!+\!\{\!1\!\!\leftrightarrow\!\! 2\!\}\right)\!\Big]\!\nonumber \\
		&\!-\!\frac{q^8 g_s^2 \!\log\! ^2\!\left(\!-\!\frac{q^2}{\mu ^2}\right)\!}{2211840 \pi ^8}\!\Big[\!\left(\!\left(\!\left(\!\left(\delta _{24} \delta _{1b} \delta _{3d} \delta _{ac}\!+\!\{\!3\!\!\leftrightarrow\!\! 4,a\!\!\leftrightarrow\!\! c\!\}\right)\!\!+\!\{\!c\!\!\leftrightarrow\!\! d\!\}\right)\!\!+\!\{\!1\!\!\leftrightarrow\!\! 2\!\}\right)\!\!+\!\!\left(\!\left(\!\left(\delta _{24} \delta _{1a} \delta _{3d} \delta _{bc}\!+\!\{\!3\!\!\leftrightarrow\!\! 4,b\!\!\leftrightarrow\!\! c\!\}\right)\!\!+\!\{\!c\!\!\leftrightarrow\!\! d\!\}\right)\!\!+\!\{\!1\!\!\leftrightarrow\!\! 2\!\}\right)\!\right)\nonumber \\
		&\qquad\qquad\qquad\qquad+\!18 \!\left(\!\left(\delta _{2a} \delta _{1b} \delta _{4c} \delta _{3d}\!+\!\{\!c\!\!\leftrightarrow\!\! d\!\}\right)\!\!+\!\{\!1\!\!\leftrightarrow\!\! 2\!\}\right)\!\Big]\!\nonumber \\
		&\!-\!\frac{q^8 \!\log\! \!\left(\!-\!\frac{q^2}{\mu ^2}\right)\!}{15360 \pi ^6}\!\left(\!\left(\delta _{2a} \delta _{1b} \delta _{4c} \delta _{3d}\!+\!\{\!c\!\!\leftrightarrow\!\! d\!\}\right)\!\!+\!\{\!1\!\!\leftrightarrow\!\! 2\!\}\right)\!+\!\frac{q^4 \!\log\! \!\left(\!-\!\frac{q^2}{\mu ^2}\right)\!}{768 \pi ^5}\langle\! \, GG\, \!\rangle\!  \!\left(\!\left(\delta _{2a} \delta _{1b} \delta _{4c} \delta _{3d}\!+\!\{\!c\!\!\leftrightarrow\!\! d\!\}\right)\!\!+\!\{\!1\!\!\leftrightarrow\!\! 2\!\}\right)\!\nonumber \\
		&\!-\!\frac{q^4 \!\log\! \!\left(\!-\!\frac{q^2}{\mu ^2}\right)\!}{96 \pi ^4}\!\left(m_1\! \langle\! \, \overline{f_1}f_1\!\rangle\! \!+\!m_2 \langle\! \, \overline{f_2}f_2\!\rangle\! \!+\!m_3 \langle\! \, \overline{f_3}f_3\!\rangle\! \!+\!m_4 \langle\! \, \overline{f_4}f_4\!\rangle\! \right)\! \!\left(\!\left(\delta _{2a} \delta _{1b} \delta _{4c} \delta _{3d}\!+\!\{\!c\!\!\leftrightarrow\!\! d\!\}\right)\!\!+\!\{\!1\!\!\leftrightarrow\!\! 2\!\}\right)\!\nonumber \\
		&\!+\!\frac{5 q^2 \!\log\! \!\left(\!-\!\frac{q^2}{\mu ^2}\right)\!}{6912 \pi ^6}\langle\! G^3\!\rangle\!  \!\left(\!\left(\delta _{2a} \delta _{1b} \delta _{4c} \delta _{3d}\!+\!\{\!c\!\!\leftrightarrow\!\! d\!\}\right)\!\!+\!\{\!1\!\!\leftrightarrow\!\! 2\!\}\right)\!\nonumber \\
		&\!+\!\frac{4}{9 q^2}\!\left(m_2\! \langle\! \, \overline{f_1}f_1\!\rangle\!  \langle\! \, \overline{f_3}f_3\!\rangle\!  \langle\! \, \overline{f_4}f_4\!\rangle\! \!+\!\langle\! \, \overline{f_2}f_2\!\rangle\!  \!\left(m_3\! \langle\! \, \overline{f_1}f_1\!\rangle\!  \langle\! \, \overline{f_4}f_4\!\rangle\! \!+\!\langle\! \, \overline{f_3}f_3\!\rangle\!  \!\left(m_1\! \langle\! \, \overline{f_4}f_4\!\rangle\! \!+\!m_4 \langle\! \, \overline{f_1}f_1\!\rangle\! \right)\!\right)\!\right)\! \!\left(\!\left(\delta _{2a} \delta _{1b} \delta _{4c} \delta _{3d}\!+\!\{\!c\!\!\leftrightarrow\!\! d\!\}\right)\!\!+\!\{\!1\!\!\leftrightarrow\!\! 2\!\}\right)\!
	\end{align}
}

{\tiny
	\begin{align}
		\Pi_{J^r_e}(q^2)\!=&\!\frac{\!\left(5 \sqrt{241}\!-\!241\right)\! q^8 \!\log\! \!\left(\!-\!\frac{q^2}{\mu ^2}\right)\!}{640 \pi ^6}\!\left(\!\left(\delta _{2a} \delta _{1b} \delta _{4c} \delta _{3d}\!-\!\{\!c\!\!\leftrightarrow\!\! d\!\}\right)\!\!-\!\{\!1\!\!\leftrightarrow\!\! 2\!\}\right)\!-\!\frac{\!\left(163 \sqrt{241}\!-\!2651\right)\! q^8 g_s^2 \!\log\! ^2\!\left(\!-\!\frac{q^2}{\mu ^2}\right)\!}{15360 \pi ^8}\!\left(\!\left(\delta _{2a} \delta _{1b} \delta _{4c} \delta _{3d}\!-\!\{\!c\!\!\leftrightarrow\!\! d\!\}\right)\!\!-\!\{\!1\!\!\leftrightarrow\!\! 2\!\}\right)\!\nonumber \\
		&\!+\!\frac{q^8 g_s^2 \!\log\! \!\left(\!-\!\frac{q^2}{\mu ^2}\right)\!}{1612800 \pi ^8}\!\Big[\!\left(164921 \sqrt{241}\!-\!3023377\right)\! \!\left(\!\left(\delta _{2a} \delta _{1b} \delta _{4c} \delta _{3d}\!-\!\{\!c\!\!\leftrightarrow\!\! d\!\}\right)\!\!-\!\{\!1\!\!\leftrightarrow\!\! 2\!\}\right)\nonumber \\
		&\qquad\qquad\qquad\qquad-\!2700 \!\left(\!\left(\!\left(\!\left(\delta _{24} \delta _{1b} \delta _{3d} \delta _{ac}\!-\!\{\!3\!\!\leftrightarrow\!\! 4,a\!\!\leftrightarrow\!\! c\!\}\right)\!\!-\!\{\!c\!\!\leftrightarrow\!\! d\!\}\right)\!\!-\!\{\!1\!\!\leftrightarrow\!\! 2\!\}\right)\!\!-\!\!\left(\!\left(\!\left(\delta _{24} \delta _{1a} \delta _{3d} \delta _{bc}\!-\!\{\!3\!\!\leftrightarrow\!\! 4,b\!\!\leftrightarrow\!\! c\!\}\right)\!\!-\!\{\!c\!\!\leftrightarrow\!\! d\!\}\right)\!\!-\!\{\!1\!\!\leftrightarrow\!\! 2\!\}\right)\!\right)\!\Big]\!\nonumber \\
		&\!+\!\frac{q^4 \!\log\! \!\left(\!-\!\frac{q^2}{\mu ^2}\right)\!}{4 \pi ^4}\!\left(5 \sqrt{241}\!-\!241\right)\! \!\left(m_1\! \langle\! \, \overline{f_1}f_1\!\rangle\! \!+\!m_2 \langle\! \, \overline{f_2}f_2\!\rangle\! \!+\!m_3 \langle\! \, \overline{f_3}f_3\!\rangle\! \!+\!m_4 \langle\! \, \overline{f_4}f_4\!\rangle\! \right)\! \!\left(\!\left(\delta _{2a} \delta _{1b} \delta _{4c} \delta _{3d}\!-\!\{\!c\!\!\leftrightarrow\!\! d\!\}\right)\!\!-\!\{\!1\!\!\leftrightarrow\!\! 2\!\}\right)\!\nonumber \\
		&\!+\!\frac{q^4 \!\log\! \!\left(\!-\!\frac{q^2}{\mu ^2}\right)\!}{96 \pi ^5}\!\left(163 \sqrt{241}\!-\!2651\right)\!\!\langle \, GG\, \!\rangle\!  \!\left(\!\left(\delta _{2a} \delta _{1b} \delta _{4c} \delta _{3d}\!-\!\{\!c\!\!\leftrightarrow\!\! d\!\}\right)\!\!-\!\{\!1\!\!\leftrightarrow\!\! 2\!\}\right)\nonumber\nonumber \\
		&\!-\!\frac{q^2 \!\log\! \!\left(\!-\!\frac{q^2}{\mu ^2}\right)\!}{8 \pi ^6}\!\left(11 \sqrt{241}\!-\!151\right)\!\!\langle G^3\!\rangle\!  \!\left(\!\left(\delta _{2a} \delta _{1b} \delta _{4c} \delta _{3d}\!-\!\{\!c\!\!\leftrightarrow\!\! d\!\}\right)\!\!-\!\{\!1\!\!\leftrightarrow\!\! 2\!\}\right)\!\nonumber \\
		&\!+\!\frac{12 q^2 \!\log\! \!\left(\!-\!\frac{q^2}{\mu ^2}\right)\!}{\pi ^2}\!\left(11 \sqrt{241}\!-\!181\right)\! \!\left(\!\left(\!\left(\!\left(\delta _{14} \delta _{2b} \delta _{3d} \delta _{ac} \langle\! \, \overline{f_1}f_1\!\rangle\!  \langle\! \, \overline{f_b}f_b\!\rangle\! \!-\!\{\!1\!\!\leftrightarrow\!\! 2\!\}\right)\!\!+\!\{\!1\!\!\leftrightarrow\!\! 4,a\!\!\leftrightarrow\!\! b\!\}\right)\!\!-\!\{\!c\!\!\leftrightarrow\!\! d\!\}\right)\!\!-\!\{\!3\!\!\leftrightarrow\!\! 4\!\}\right)\!\nonumber \\
		&\!+\!\frac{2 \!\log\! \!\left(\!-\!\frac{q^2}{\mu ^2}\right)\!}{\pi ^2}\!\left(13 \sqrt{241}\!-\!173\right)\! \!\left(\!\left(\!\left(\!\left(\delta _{14} \delta _{2b} \delta _{3d} \delta _{ac} \!\left(\!\!\langle\!\overline{f_b}f_b\!\rangle\!  \langle\! \, \overline{f_1}\text{G}f_1\!\rangle\! \!+\!\langle\! \, \overline{f_1}f_1\!\rangle\!  \langle\! \, \overline{f_b}\text{G}f_b\!\rangle\! \right)\!\!-\!\{\!1\!\!\leftrightarrow\!\! 2\!\}\right)\!\!+\!\{\!1\!\!\leftrightarrow\!\! 4,a\!\!\leftrightarrow\!\! b\!\}\right)\!\!-\!\{\!c\!\!\leftrightarrow\!\! d\!\}\right)\!\!-\!\{\!3\!\!\leftrightarrow\!\! 4\!\}\right)\!\nonumber \\
		&\!+\!\frac{1}{48 \pi ^2 q^2}\!\left(41 \sqrt{241}\!-\!961\right)\! \!\left(\!\left(\!\left(\!\left(\delta _{14} \delta _{2b} \delta _{3d} \delta _{ac} \langle\! \, \overline{f_1}\text{G}f_1\!\rangle\!  \langle\! \, \overline{f_b}\text{G}f_b\!\rangle\! \!-\!\{\!1\!\!\leftrightarrow\!\! 2\!\}\right)\!\!+\!\{\!1\!\!\leftrightarrow\!\! 4,a\!\!\leftrightarrow\!\! b\!\}\right)\!\!-\!\{\!c\!\!\leftrightarrow\!\! d\!\}\right)\!\!-\!\{\!3\!\!\leftrightarrow\!\! 4\!\}\right)\!\nonumber \\
		&\!+\!\frac{4}{\pi  q^2}\!\left(11 \sqrt{241}\!-\!181\right)\! \!\left(\!\left(\!\left(\!\left(\delta _{14} \delta _{2b} \delta _{3d} \delta _{ac} \langle\! \, GG\, \!\rangle\!  \langle\! \, \overline{f_1}f_1\!\rangle\!  \langle\! \, \overline{f_b}f_b\!\rangle\! \!-\!\{\!1\!\!\leftrightarrow\!\! 2\!\}\right)\!\!+\!\{\!1\!\!\leftrightarrow\!\! 4,a\!\!\leftrightarrow\!\! b\!\}\right)\!\!-\!\{\!c\!\!\leftrightarrow\!\! d\!\}\right)\!\!-\!\{\!3\!\!\leftrightarrow\!\! 4\!\}\right)\!\nonumber \\
		&\!-\!\frac{32}{3 q^2}\!\left(5 \sqrt{241}\!-\!241\right)\! \!\left(m_2\! \langle\! \, \overline{f_1}f_1\!\rangle\!  \langle\! \, \overline{f_3}f_3\!\rangle\!  \langle\! \, \overline{f_4}f_4\!\rangle\! \!+\!\langle\! \, \overline{f_2}f_2\!\rangle\!  \!\left(m_3\! \langle\! \, \overline{f_1}f_1\!\rangle\!  \langle\! \, \overline{f_4}f_4\!\rangle\! \!+\!\langle\! \, \overline{f_3}f_3\!\rangle\!  \!\left(m_1\! \langle\! \, \overline{f_4}f_4\!\rangle\! \!+\!m_4 \langle\! \, \overline{f_1}f_1\!\rangle\! \right)\!\right)\!\right)\nonumber\\ &\qquad\qquad\qquad\qquad\times\left(\!\left(\delta _{2a} \delta _{1b} \delta _{4c} \delta _{3d}\!-\!\{\!c\!\!\leftrightarrow\!\! d\!\}\right)\!\!-\!\{\!1\!\!\leftrightarrow\!\! 2\!\}\right)\!\nonumber \\
		&\!+\!\frac{16}{q^2}\!\left(11 \sqrt{241}\!-\!181\right)\! \!\left(\!\left(\!\left(\!\left(\delta _{24} \delta _{1b} \delta _{3d} \delta _{ac} \langle\! \, \overline{f_2}f_2\!\rangle\!  \langle\! \, \overline{f_a}f_a\!\rangle\!  \!\left(\!\left(2 m_1\!+\!m_3\right)\!\!\langle\!\overline{f_3}f_3\!\rangle\! \!+\!\!\left(m_1\!\!+\!2 m_3\right)\!\!\langle\!\overline{f_1}f_1\!\rangle\! \right)\!\!-\!\{\!3\!\!\leftrightarrow\!\! 4\!\}\right)\!\!-\!\{\!c\!\!\leftrightarrow\!\! d\!\}\right)\!\!-\!\{\!a\!\!\leftrightarrow\!\! b\!\}\right)\!\!-\!\{\!1\!\!\leftrightarrow\!\! 2\!\}\right)
	\end{align}
}

{\tiny
	\begin{align}
		\Pi_{J^r_f}(q^2)\!=&\!\frac{\!\left(5 \sqrt{241}\!-\!241\right)\! q^8 \!\log\! \!\left(\!-\!\frac{q^2}{\mu ^2}\right)\!}{1280 \pi ^6}\!\left(\!\left(\delta _{2a} \delta _{1b} \delta _{4c} \delta _{3d}\!+\!\{\!c\!\!\leftrightarrow\!\! d\!\}\right)\!\!+\!\{\!1\!\!\leftrightarrow\!\! 2\!\}\right)\!-\!\frac{\!\left(59 \sqrt{241}\!-\!241\right)\! q^8 g_s^2 \!\log\! ^2\!\left(\!-\!\frac{q^2}{\mu ^2}\right)\!}{15360 \pi ^8}\!\left(\!\left(\delta _{2a} \delta _{1b} \delta _{4c} \delta _{3d}\!+\!\{\!c\!\!\leftrightarrow\!\! d\!\}\right)\!\!+\!\{\!1\!\!\leftrightarrow\!\! 2\!\}\right)\!\nonumber \\
		&\!+\!\frac{q^8 g_s^2 \!\log\! \!\left(\!-\!\frac{q^2}{\mu ^2}\right)\!}{1612800 \pi ^8}\!\Big[\!\left(63403 \sqrt{241}\!-\!486197\right)\! \!\left(\!\left(\delta _{2a} \delta _{1b} \delta _{4c} \delta _{3d}\!+\!\{\!c\!\!\leftrightarrow\!\! d\!\}\right)\!\!+\!\{\!1\!\!\leftrightarrow\!\! 2\!\}\right)\nonumber \\
		&\qquad\qquad\qquad\qquad-\!675 \!\left(\!\left(\!\left(\!\left(\delta _{24} \delta _{1b} \delta _{3d} \delta _{ac}\!+\!\{\!3\!\!\leftrightarrow\!\! 4,a\!\!\leftrightarrow\!\! c\!\}\right)\!\!+\!\{\!c\!\!\leftrightarrow\!\! d\!\}\right)\!\!+\!\{\!1\!\!\leftrightarrow\!\! 2\!\}\right)\!\!+\!\!\left(\!\left(\!\left(\delta _{24} \delta _{1a} \delta _{3d} \delta _{bc}\!+\!\{\!3\!\!\leftrightarrow\!\! 4,b\!\!\leftrightarrow\!\! c\!\}\right)\!\!+\!\{\!c\!\!\leftrightarrow\!\! d\!\}\right)\!\!+\!\{\!1\!\!\leftrightarrow\!\! 2\!\}\right)\!\right)\!\Big]\!\nonumber \\
		&\!+\!\frac{q^4 \!\log\! \!\left(\!-\!\frac{q^2}{\mu ^2}\right)\!}{8 \pi ^4}\!\left(5 \sqrt{241}\!-\!241\right)\! \!\left(m_1\! \langle\! \, \overline{f_1}f_1\!\rangle\! \!+\!m_2 \langle\! \, \overline{f_2}f_2\!\rangle\! \!+\!m_3 \langle\! \, \overline{f_3}f_3\!\rangle\! \!+\!m_4 \langle\! \, \overline{f_4}f_4\!\rangle\! \right)\! \!\left(\!\left(\delta _{2a} \delta _{1b} \delta _{4c} \delta _{3d}\!+\!\{\!c\!\!\leftrightarrow\!\! d\!\}\right)\!\!+\!\{\!1\!\!\leftrightarrow\!\! 2\!\}\right)\!\nonumber \\
		&\!+\!\frac{q^4 \!\log\! \!\left(\!-\!\frac{q^2}{\mu ^2}\right)\!}{96 \pi ^5}\!\left(59 \sqrt{241}\!-\!241\right)\!\!\langle \, GG\, \!\rangle\!  \!\left(\!\left(\delta _{2a} \delta _{1b} \delta _{4c} \delta _{3d}\!+\!\{\!c\!\!\leftrightarrow\!\! d\!\}\right)\!\!+\!\{\!1\!\!\leftrightarrow\!\! 2\!\}\right)\nonumber\nonumber \\
		&\!-\!\frac{q^2 \!\log\! \!\left(\!-\!\frac{q^2}{\mu ^2}\right)\!}{32 \pi ^6}\!\left(35 \sqrt{241}\!-\!313\right)\!\!\langle G^3\!\rangle\!  \!\left(\!\left(\delta _{2a} \delta _{1b} \delta _{4c} \delta _{3d}\!+\!\{\!c\!\!\leftrightarrow\!\! d\!\}\right)\!\!+\!\{\!1\!\!\leftrightarrow\!\! 2\!\}\right)\!\nonumber \\
		&\!+\!\frac{3 q^2 \!\log\! \!\left(\!-\!\frac{q^2}{\mu ^2}\right)\!}{\pi ^2}\!\left(17 \sqrt{241}\!-\!265\right)\! \!\left(\!\left(\!\left(\!\left(\delta _{14} \delta _{2b} \delta _{3d} \delta _{ac} \langle\! \, \overline{f_1}f_1\!\rangle\!  \langle\! \, \overline{f_b}f_b\!\rangle\! \!+\!\{\!1\!\!\leftrightarrow\!\! 2\!\}\right)\!\!+\!\{\!1\!\!\leftrightarrow\!\! 4,a\!\!\leftrightarrow\!\! b\!\}\right)\!\!+\!\{\!c\!\!\leftrightarrow\!\! d\!\}\right)\!\!+\!\{\!3\!\!\leftrightarrow\!\! 4\!\}\right)\!\nonumber \\
		&\!-\!\frac{2 \!\log\! \!\left(\!-\!\frac{q^2}{\mu ^2}\right)\!}{\pi ^2}\!\left(5 \sqrt{241}\!-\!79\right)\! \!\left(\!\left(\!\left(\!\left(\delta _{14} \delta _{2b} \delta _{3d} \delta _{ac} \!\left(\!\!\langle\!\overline{f_b}f_b\!\rangle\!  \langle\! \, \overline{f_1}\text{G}f_1\!\rangle\! \!+\!\langle\! \, \overline{f_1}f_1\!\rangle\!  \langle\! \, \overline{f_b}\text{G}f_b\!\rangle\! \right)\!\!+\!\{\!1\!\!\leftrightarrow\!\! 2\!\}\right)\!\!+\!\{\!1\!\!\leftrightarrow\!\! 4,a\!\!\leftrightarrow\!\! b\!\}\right)\!\!+\!\{\!c\!\!\leftrightarrow\!\! d\!\}\right)\!\!+\!\{\!3\!\!\leftrightarrow\!\! 4\!\}\right)\!	\nonumber \\
		&\!+\!\frac{1}{48 \pi ^2 q^2}\!\left(23 \sqrt{241}\!-\!385\right)\! \!\left(\!\left(\!\left(\!\left(\delta _{14} \delta _{2b} \delta _{3d} \delta _{ac} \langle\! \, \overline{f_1}\text{G}f_1\!\rangle\!  \langle\! \, \overline{f_b}\text{G}f_b\!\rangle\! \!+\!\{\!1\!\!\leftrightarrow\!\! 2\!\}\right)\!\!+\!\{\!1\!\!\leftrightarrow\!\! 4,a\!\!\leftrightarrow\!\! b\!\}\right)\!\!+\!\{\!c\!\!\leftrightarrow\!\! d\!\}\right)\!\!+\!\{\!3\!\!\leftrightarrow\!\! 4\!\}\right)\!\nonumber \\
		&\!+\!\frac{1}{\pi  q^2}\!\left(17 \sqrt{241}\!-\!265\right)\! \!\left(\!\left(\!\left(\!\left(\delta _{14} \delta _{2b} \delta _{3d} \delta _{ac} \langle\! \, GG\, \!\rangle\!  \langle\! \, \overline{f_1}f_1\!\rangle\!  \langle\! \, \overline{f_b}f_b\!\rangle\! \!+\!\{\!1\!\!\leftrightarrow\!\! 2\!\}\right)\!\!+\!\{\!1\!\!\leftrightarrow\!\! 4,a\!\!\leftrightarrow\!\! b\!\}\right)\!\!+\!\{\!c\!\!\leftrightarrow\!\! d\!\}\right)\!\!+\!\{\!3\!\!\leftrightarrow\!\! 4\!\}\right)\!\nonumber \\
		&\!-\!\frac{16}{3 q^2}\!\left(5 \sqrt{241}\!-\!241\right)\! \!\left(m_2\! \langle\! \, \overline{f_1}f_1\!\rangle\!  \langle\! \, \overline{f_3}f_3\!\rangle\!  \langle\! \, \overline{f_4}f_4\!\rangle\! \!+\!\langle\! \, \overline{f_2}f_2\!\rangle\!  \!\left(m_3\! \langle\! \, \overline{f_1}f_1\!\rangle\!  \langle\! \, \overline{f_4}f_4\!\rangle\! \!+\!\langle\! \, \overline{f_3}f_3\!\rangle\!  \!\left(m_1\! \langle\! \, \overline{f_4}f_4\!\rangle\! \!+\!m_4 \langle\! \, \overline{f_1}f_1\!\rangle\! \right)\!\right)\!\right)\nonumber\\
		&\qquad\qquad\qquad\qquad\times\left(\!\left(\delta _{2a} \delta _{1b} \delta _{4c} \delta _{3d}\!+\!\{\!c\!\!\leftrightarrow\!\! d\!\}\right)\!\!+\!\{\!1\!\!\leftrightarrow\!\! 2\!\}\right)\!\nonumber \\
		&\!+\!\frac{4}{q^2}\!\left(17 \sqrt{241}\!-\!265\right)\! \!\left(\!\left(\!\left(\!\left(\delta _{24} \delta _{1b} \delta _{3d} \delta _{ac} \langle\! \, \overline{f_2}f_2\!\rangle\!  \langle\! \, \overline{f_a}f_a\!\rangle\!  \!\left(\!\left(2 m_1\!+\!m_3\right)\!\!\langle\!\overline{f_3}f_3\!\rangle\! \!+\!\!\left(m_1\!\!+\!2 m_3\right)\!\!\langle\!\overline{f_1}f_1\!\rangle\! \right)\!\!+\!\{\!3\!\!\leftrightarrow\!\! 4\!\}\right)\!\!+\!\{\!c\!\!\leftrightarrow\!\! d\!\}\right)\!\!+\!\{\!a\!\!\leftrightarrow\!\! b\!\}\right)\!\!+\!\{\!1\!\!\leftrightarrow\!\! 2\!\}\right)
	\end{align}
}

{\tiny
	\begin{align}
		\Pi_{I^r_3}(q^2)\!=&\!\frac{q^8 g_s^2 \!\log\! \!\left(\!-\!\frac{q^2}{\mu ^2}\right)\!}{464486400 \pi ^8}\!\Big[4586 \!\left(\!\left(\!\left(\!\left(\delta _{34} \delta _{1c} \delta _{2d} \delta _{ab}\!+\!\{\!2\!\!\leftrightarrow\!\! 4,a\!\!\leftrightarrow\!\! b\!\}\right)\!\!+\!\{\!b\!\!\leftrightarrow\!\! d\!\}\right)\!\!+\!\{\!1\!\!\leftrightarrow\!\! 3\!\}\right)\!\!+\!\!\left(\!\left(\!\left(\delta _{34} \delta _{1a} \delta _{2d} \delta _{bc}\!+\!\{\!2\!\!\leftrightarrow\!\! 4,b\!\!\leftrightarrow\!\! c\!\}\right)\!\!+\!\{\!b\!\!\leftrightarrow\!\! d\!\}\right)\!\!+\!\{\!1\!\!\leftrightarrow\!\! 3\!\}\right)\!\right)\nonumber\nonumber \\
		&\qquad\qquad\qquad\qquad-\!27477 \!\left(\!\left(\delta _{3a} \delta _{4b} \delta _{1c} \delta _{2d}\!+\!\{\!b\!\!\leftrightarrow\!\! d\!\}\right)\!\!+\!\{\!1\!\!\leftrightarrow\!\! 3\!\}\right)\!\Big]\!\nonumber\nonumber \\
		&\!-\!\frac{q^8 g_s^2 \!\log\! ^2\!\left(\!-\!\frac{q^2}{\mu ^2}\right)\!}{552960 \pi ^8}\!\Big[\!\left(\!\left(\!\left(\!\left(\delta _{34} \delta _{1c} \delta _{2d} \delta _{ab}\!+\!\{\!2\!\!\leftrightarrow\!\! 4,a\!\!\leftrightarrow\!\! b\!\}\right)\!\!+\!\{\!b\!\!\leftrightarrow\!\! d\!\}\right)\!\!+\!\{\!1\!\!\leftrightarrow\!\! 3\!\}\right)\!\!+\!\!\left(\!\left(\!\left(\delta _{34} \delta _{1a} \delta _{2d} \delta _{bc}\!+\!\{\!2\!\!\leftrightarrow\!\! 4,b\!\!\leftrightarrow\!\! c\!\}\right)\!\!+\!\{\!b\!\!\leftrightarrow\!\! d\!\}\right)\!\!+\!\{\!1\!\!\leftrightarrow\!\! 3\!\}\right)\!\right)\nonumber\nonumber \\
		&\qquad\qquad\qquad\qquad+\!18 \!\left(\!\left(\delta _{3a} \delta _{4b} \delta _{1c} \delta _{2d}\!+\!\{\!b\!\!\leftrightarrow\!\! d\!\}\right)\!\!+\!\{\!1\!\!\leftrightarrow\!\! 3\!\}\right)\!\Big]\!\nonumber\nonumber \\
		&\!-\!\frac{q^8 \!\log\! \!\left(\!-\!\frac{q^2}{\mu ^2}\right)\!}{3840 \pi ^6}\!\left(\!\left(\delta _{3a} \delta _{4b} \delta _{1c} \delta _{2d}\!+\!\{\!b\!\!\leftrightarrow\!\! d\!\}\right)\!\!+\!\{\!1\!\!\leftrightarrow\!\! 3\!\}\right)\!+\!\frac{q^4 \!\log\! \!\left(\!-\!\frac{q^2}{\mu ^2}\right)\!}{192 \pi ^5}\langle\! \, GG\, \!\rangle\!  \!\left(\!\left(\delta _{3a} \delta _{4b} \delta _{1c} \delta _{2d}\!+\!\{\!b\!\!\leftrightarrow\!\! d\!\}\right)\!\!+\!\{\!1\!\!\leftrightarrow\!\! 3\!\}\right)\!\nonumber\nonumber \\
		&\!-\!\frac{q^4 \!\log\! \!\left(\!-\!\frac{q^2}{\mu ^2}\right)\!}{24 \pi ^4}\!\left(m_1\! \langle\! \, \overline{f_1}f_1\!\rangle\! \!+\!m_2 \langle\! \, \overline{f_2}f_2\!\rangle\! \!+\!m_3 \langle\! \, \overline{f_3}f_3\!\rangle\! \!+\!m_4 \langle\! \, \overline{f_4}f_4\!\rangle\! \right)\! \!\left(\!\left(\delta _{3a} \delta _{4b} \delta _{1c} \delta _{2d}\!+\!\{\!b\!\!\leftrightarrow\!\! d\!\}\right)\!\!+\!\{\!1\!\!\leftrightarrow\!\! 3\!\}\right)\!\nonumber\nonumber \\
		&\!+\!\frac{5 q^2 \!\log\! \!\left(\!-\!\frac{q^2}{\mu ^2}\right)\!}{1728 \pi ^6}\langle\! G^3\!\rangle\!  \!\left(\!\left(\delta _{3a} \delta _{4b} \delta _{1c} \delta _{2d}\!+\!\{\!b\!\!\leftrightarrow\!\! d\!\}\right)\!\!+\!\{\!1\!\!\leftrightarrow\!\! 3\!\}\right)\!\nonumber \\
		&\!+\!\frac{16}{9 q^2}\!\left(m_2\! \langle\! \, \overline{f_1}f_1\!\rangle\!  \langle\! \, \overline{f_3}f_3\!\rangle\!  \langle\! \, \overline{f_4}f_4\!\rangle\! \!+\!\langle\! \, \overline{f_2}f_2\!\rangle\!  \!\left(m_3\! \langle\! \, \overline{f_1}f_1\!\rangle\!  \langle\! \, \overline{f_4}f_4\!\rangle\! \!+\!\langle\! \, \overline{f_3}f_3\!\rangle\!  \!\left(m_1\! \langle\! \, \overline{f_4}f_4\!\rangle\! \!+\!m_4 \langle\! \, \overline{f_1}f_1\!\rangle\! \right)\!\right)\!\right)\! \!\left(\!\left(\delta _{3a} \delta _{4b} \delta _{1c} \delta _{2d}\!+\!\{\!b\!\!\leftrightarrow\!\! d\!\}\right)\!\!+\!\{\!1\!\!\leftrightarrow\!\! 3\!\}\right)\!
	\end{align}
}

}

\endgroup

\acknowledgments
This work is supported by NSFC (No.11175153, No.11775187)

\bibliographystyle{JHEP}
\bibliography{refs}
\end{document}